%% file: rmp_main.tex
\documentclass[rmp,aps,nofootinbib,twocolumn,floatfix]{revtex4}

\usepackage{graphics}
\usepackage{epsfig}
\usepackage{calligra}
\DeclareMathAlphabet{\mathcalligra}{T1}{calligra}{m}{n}
\DeclareFontShape{T1}{calligra}{m}{n}{<->s*[2.2]callig15}{}
\usepackage{color}
\usepackage{amssymb,amsmath}
\usepackage{multirow}
\usepackage{nicefrac}

\begin{document}
\title{Spinor Bose gases: symmetries, magnetism and quantum dynamics}

\author{Dan M. Stamper-Kurn}
\email{dmsk@berkeley.edu}
\affiliation{Department of Physics, University of California, Berkeley CA 94720, USA}
\affiliation{Materials Sciences Division, Lawrence Berkeley National Laboratory, Berkeley, CA 94720, USA}
\author{Masahito Ueda}
\email{ueda@phys.s.u-tokyo.ac.jp}
\affiliation{Department of Physics, University of Tokyo, 7-3-1 Hongo, Bunkyo-ku, Tokyo 113-0033, Japan}
\affiliation{ERATO Macroscopic Quantum Control Project, JST, 7-3-1 Hongo, Bunkyo-ku, Tokyo 113-0033, Japan}

\begin{abstract}
Spinor Bose gases form a family of quantum fluids manifesting both magnetic order and superfluidity.  This article reviews experimental and theoretical progress in understanding the static and dynamic properties of these fluids.  The connection between system properties and the rotational symmetry properties of the atomic states and their interactions are investigated.  Following a review of the experimental techniques used for characterizing spinor gases, their mean-field and many-body ground states, both in isolation and under the application of symmetry-breaking external fields, are discussed. These states serve as the starting point for understanding low-energy dynamics, spin textures and topological defects, effects of magnetic dipole interactions, and various non-equilibrium collective spin-mixing phenomena.  The paper aims to form connections and establish coherence among the vast range of works on spinor Bose gases, so as to point to open questions and future research opportunities.
\end{abstract}

\maketitle
\tableofcontents

\newcommand{\hami}{H}
\newcommand{\scripty}[1]{\ensuremath{\mathcalligra{#1}}}
\newcommand{\littler}{\scripty{r}}
\newcommand{\littlervec}{\mathbf{\scripty{r}}}
\newcommand{\littlerhat}{\hat{\mathbf{\scripty{r}}}}
\newcommand{\comment}[1]{}
\newcommand{\markup}[1]{\textbf{#1}}
\newcommand{\bmu}{\boldsymbol{\mu}}
\newcommand{\brho}{\boldsymbol{\rho}}
\newcommand{\bpi}{\boldsymbol{\pi}}
\newcommand{\bsigma}{\boldsymbol{\sigma}}

\input{rmp_ch_intro} 
\input{rmp_ch_quantumstates}

\input{rmp_ch_rotsymm} 
\input{rmp_ch_exprealities} 
\input{rmp_ch_groundstates} 
\input{rmp_ch_textures} 
\input{rmp_ch_magnetization} 
\input{rmp_ch_dipolar} 
\input{rmp_ch_spinmixing} 
\input{rmp_ch_equilibration} 
\input{rmp_ch_future} 
\input{rmp_conclusions}

\section*{Acknowledgments}

The authors are indebted to our close colleagues within our research groups with whom we have explored the physics of spinor Bose gases.  At Berkeley, we acknowledge the experimental contributions of A.\ Chikkatur, J.\ Guzman, J.\ Higbie, S.\ Inouye, G.-B.\ Jo, S.\ Leslie, K.\ Murch, L.\ Sadler, V.\ Savalli, F.\ Serwane, C.\ Thomas, M.\ Vengalattore, A.\ Wenz, and also the theoretical inputs of J.\ Sau and M.\ Cohen.  M.U.\ acknowledges his scientific collaborators including M.\ Kobayashi, Y.\ Kawaguchi, M.\ Koashi, M.\ Nitta, H.\ Saito, and S.\ Uchino.
We are also grateful to A.\ Lamacraft, W.V.\ Liu, J. Moore and E. Mueller for valuable discussions during the preparation of this work, and to C.\ Raman and Y.\ Liu for assistance with several figures.  D.S.-K.\ thanks the NSF, the ARO with funds from the DARPA OLE program, the Division of Materials Sciences and Engineering at the  Office of Basic Energy Sciences (DOE), the Sloan and Packard Foundations, and the Hellman Faculty Fund and the Miller Institute at the University of California, for their support.  M.U.\ acknowledge support by Grants-in-Aid for Scientific Research (Grant Nos.\ 22103005 and 22340114), a Global COE Program ``the Physical Sciences Frontier,'' and the Photon Frontier Network Program, from MEXT of Japan.


\bibliographystyle{apsrmp4-1}
\bibliography{allrefs_x2,spinor_notes}

\end{document}

%% file: rmp_ch_intro.tex
\section{Introduction}
\label{sec:intro}

Nature has provided us with few quantum fluids, in which macroscopic characteristics of the fluid derive directly from quantum coherences.  Among such fluids, ones with a non-trivial internal degree of freedom are even rarer, examples being the d-wave and p-wave superconductors and certain phases of superfluid $^3$He.  Ultracold atomic physics has supplied us with a new family of such fluids, degenerate Bose gases with a spin degree of freedom.  These so-called ``spinor gases'' are subject to the interplay of magnetism and superfluidity, both of which involve quantum phase coherence, long-range order and symmetry breaking.  These fluids are interesting in their own right, as a newly conceived material now available for experimental investigation.  In addition, since these atomic gases may be described starting from a simple theoretical framework, and since their properties may be readily manipulated and measured in experiments, their study promises to shed insight on a range of topics such as the role of symmetry and topology in quantum-ordered materials, quantum phase transitions, non-equilibrium quantum dynamics, and the entanglement and squeezing of quantum fields.

The study of degenerate spinor Bose gases was sparked in 1998 by experiments on ultracold rubidium \cite{hall98dyn,hall98phas,matt98} and sodium \cite{sten98spin}, followed quickly by theoretical investigations of Bose-Einstein condensation in an interacting spin-1 gas \cite{ho98,ohmi98}.  A significant opus of literature has amassed since those works covering an extremely wide range of topics.  The present review is intended to coalesce the understandings presented in that work and to focus attention on core physical concepts brought up in the study of spinor gases.  We hope such a discussion will broaden the impact of spinor-gas research and draw new participants and perspective to this topic.

\subsection{Internal degrees of freedom of Bose gases}

By now, a wide selection of atoms and a few simple molecules have been produced in gaseous form in the ultracold regime\footnote{The accepted definition of ``ultracold'' is time varying.  Spurred by the quest for correlated atomic systems with ever-fewer defects and ever-subtler types of correlation, physicists are devising new methods to reach ever-lower temperature regimes.  In this Review we will define relevant temperature regimes more specifically with respect to the properties of atomic interactions (Sec.\ \ref{sec:collisions}), the onset of quantum degeneracy (Sec.\ \ref{sec:nonzeroT}), and the expected appearance of magnetic order in optical lattices (Sec.\ \ref{sec:lattices}).}.  Such objects carry internal degrees of freedom, so that the state description of a single molecule or atom or of a system assembled of many such particles must describe both the external degrees of freedom, i.e.\ the center-of-mass motional state, and also the internal degrees of freedom.  Quantum fluids composed of such particles may be described by a multi-component order parameter.  In comparison, liquid $^4$He in its superfluid state is described by a scalar order parameter.  The interplay between the external and internal degrees of freedom in the multi-component systems leads to a range of phenomena unfamiliar from studies of scalar quantum fluids.

At first glance, it would seem that one has an enormous selection of multi-component quantum gases from which to choose.  In fact, this plethora is severely depleted by the instability of most atomic and molecular internal states.  Electronic excited states will typically decay due to spontaneous emission on a time scale much faster than that required for a gas to equilibrate kinetically through collisions. Short lifetimes are also expected (or observed) in many instances even within the electronic ground-state family of states, due to inelastic collisions that release large amounts of energy.

However, in several cases, one does find one or more sets of internal states that may coexist for sufficiently long times to allow experimental studies of equilibrated multi-component gases.  Let us discuss three examples:

\subsubsection{Hyperfine spin manifolds of hydrogen-like atoms}

\begin{figure*}[tb]
\begin{center}
\includegraphics[width=1.5 \columnwidth]{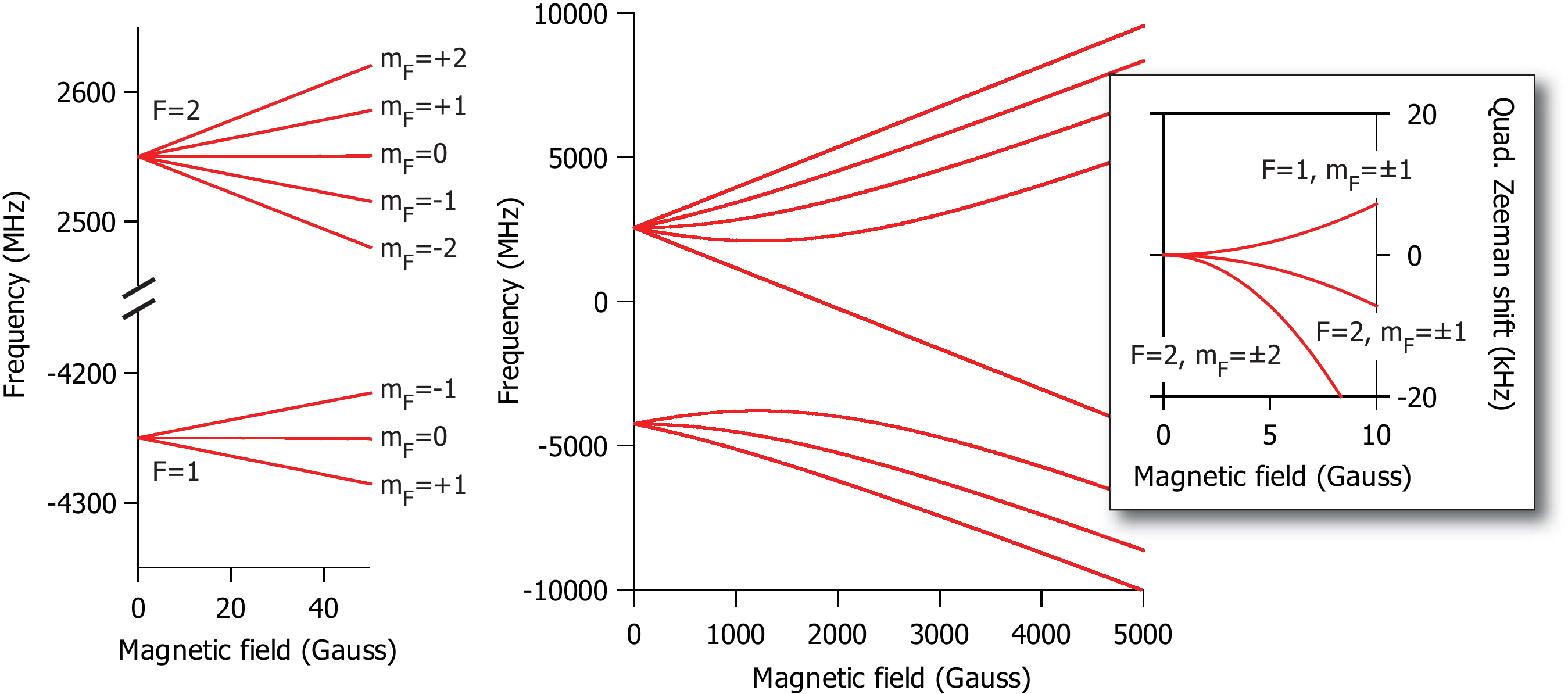}
\caption{(Color online) Ground-state hyperfine structure of $^{87}$Rb.  With a nuclear spin of $I=3/2$, $^{87}$Rb has a total angular momentum of either $F=1$ or $F=2$ in the $J=1/2$ electronic ground state.  At low magnetic fields, $F$ is a good quantum number due to rotational invariance.  The $F=1$ and $F=2$ spin manifolds are separated by the hyperfine interaction.  The Zeeman sublevels, labeled by the magnetic quantum number $m_F$, have a Zeeman energy that shows a linear variation with field at low magnetic fields (left), but deviates from the linear dependence strongly when the linear Zeeman shift is comparable to the hyperfine interaction energy (center).  Right inset: the \emph{net} quadratic Zeeman energy shift is exhibited after subtracting off the zero-field hyperfine splitting, the linear Zeeman term, and the energy of the $|m_z=0\rangle$ state; thus, the net quadratic shift of the $|m_z=\rangle$ states is zero by definition.  The quadratic shift due to a dc magnetic field is positive for the $F=1$ levels and negative for the $F=2$ levels.  The hyperfine structure of other hydrogen-like atoms shows similar features.} \label{fig:rbhyperfine}
\end{center}
\end{figure*}

The electronic ground states of hydrogen and alkali-metal atoms correspond to different orientations of the $J=1/2$ electron and of the nuclear spin $I$.  At low magnetic fields, the hyperfine interaction between these spins dominates, and the ground-state subspace breaks into manifolds of states with definite total spin $F = I \pm 1/2$.  At higher magnetic fields, the dominance of the electron magnetic moment causes the eigenstate structure to reorganize, as shown in Fig.\ \ref{fig:rbhyperfine}.

The collision properties of mixtures of different atomic hyperfine states are not always favorable to experiments on quantum gases.  Atoms in the higher hyperfine spin manifold may undergo hyperfine-relaxation collisions \cite{gorl03}, leading to significant heating and atom loss.  In the case of $^{133}$Cs, atoms in the lower hyperfine spin manifold also suffer from exothermic collisions \cite{guer98isBECpossible,sodi98} due to scattering resonances \cite{arnd97,kokk98prospects}.  In other gases, such as $^7$Li and $^{85}$Rb, the s-wave scattering lengths for several collision channels are negative, implying that a Bose-condensed gas with large atom number acquires a negative compressibility and collapses, leading to rapid decay \cite{sack98,donl01collapse}.

Several compatible mixtures of internal states do remain.  The main focus of this Review is mixtures containing all magnetic sublevels of a manifold of states with a single value of the hyperfine spin $F$.  The internal state of an atom in such a manifold of states is represented as a spherical tensor of rank $2 F$ in spin space; thus, such gases are called \emph{spinor} gases.  This spherical-tensor character endows the spinor gas and the quantum fluid that it attains with an elegant phenomenology, as we describe in the remainder of this work.

Experimentally, long-lived alkali spinor gases have been explored in three spin manifolds (Table \ref{tab:spinorlist}): the lower-energy $F=1$ hyperfine spin manifolds of both $^{23}$Na \cite{stam98odt} and $^{87}$Rb \cite{barr01}, and also the higher-energy $F=2$ manifold of $^{87}$Rb \cite{schm04,chan04,kuwa04spin2}.  The latter gas is moderately stable against hyperfine relaxation collisions on account of fortuitous properties of the molecular potential of the rubidium dimer \cite{burk97}.  In comparison, the spinor gas in the $F=2$ upper hyperfine levels of $^{23}$Na was found to decay rapidly \cite{gorl03}.  Several other promising systems still await experimental exploration, such as the $F=1$ spinor gases of $^7$Li and $^{41}$K, which represent the lower of the two hyperfine spin manifolds for each atom, and perhaps also the higher hyperfine spin manifolds of $^{41}$K and of hydrogen.  Experiments with radioactive alkali isotopes are also possible\footnote{In atoms with lifetimes longer than the typical experimental cycle time of around 10 s, the lower-spin manifolds of the electronic ground state offer $F=1$ gases of $^{21}$Na, $^{43,45}$K, $^{77, 81, 89, 91}$Rb, $^{121}$Cs, and $^{223,225}$Fr; $F=2$ gases of $^{25}$Na, $^{79,83}$Rb, $^{131}$Cs, and $^{221}$Fr; $F=3$ gases of $^{135, 137, 139, 141}$Cs; and $F=4$ gases of $^{119}$Cs and $^{207,209,211,213}$Fr.}.

\renewcommand{\multirowsetup}{\centering}
\begin{table}[tb]
\begin{center}
\begin{tabular}{| c | c | c|}
\hline \hline
\multicolumn{2}{|c|}{Stable} & \multirow{2}{2cm}{Unstable} \\
\cline{1-2} $\langle F_z \rangle$ conserved & $\langle F_z \rangle$  not conserved &  \\
\hline  \hline
$^7$Li, $F=1$ (f) & $^{52}$Cr, $F = 3$ (not f) & $^7$Li, $F=2$ \\
$^{23}$Na, $F=1$ (af) & Dy, $F=8$ (?) & $^{23}$Na, $F=2$ \\
$^{41}$K, $F=1$ (f) & Er, $F=6$ (?) & $^{39}$K \\
$^{87}$Rb, $F=1$ (f)&  &  $^{85}$Rb \\
$^{87}$Rb, $F=2$ (af or cyc) & & $^{133}$Cs \\
\hline
$^{87}$Rb pseudo-spin: & \multicolumn{2}{ c|}{Tm, $F=4$ (?) } \\
$|1,0\rangle$, $|2,0\rangle$ & \multicolumn{2}{c|}{} \\
$|1, \pm 1\rangle$, $|2, \mp 1\rangle$ & \multicolumn{2}{c|}{} \\
\hline
\end{tabular}
\caption{Experimental candidates for the study of ultracold spinor Bose gases.  Species are divided according to whether they are stable at zero magnetic field (information on thulium is lacking), and whether the dipolar relaxation rate is small enough to allow the longitudinal magnetization ($\langle F_z\rangle$) to be conserved in an experiment.  The nature of the spin-dependent contact interactions is indicated in parentheses (f: ferromagnetic, af: antiferromagnetic, cyc:cyclic or tetrahedral, ?:unknown).  Stable pseudo-spin-1/2 gases of $^{87}$Rb are indicated, with states labeled with quantum numbers $|F, m_F\rangle$ having the same low-field magnetic moment.}
\label{tab:spinorlist}
\end{center}
\end{table}

The collisional stability of $^{87}$Rb also permits experiments on mixtures of atoms from both the $F=1$ and $F=2$ hyperfine spin manifolds \cite{myat97}.  In particular, the $|F=1, m_z=-1\rangle$ and $|F=2, m_z=+1\rangle$ states possess magnetic moments that are nearly equal at low magnetic field (identical at a field of 3.23 G).  Both species can be confined in magnetic traps, in which the external potential is nearly state-independent.  The magnetic-field insensitivity of the internal energy difference between these states allows the coherence between them to be probed over seconds-long evolution times \cite{harb02pra,treu04,deut10self}.  Similarly, mixtures of the magnetic-field-insensitive $|F=1, m_z = 0\rangle$ and $|F=2, m_z=0\rangle$ states of $^{87}$Rb are stable in optical traps, and show long coherence times \cite{klei11coherence}. In contrast, for the spinor gases described above, the linear Zeeman energy shift between sublevels makes this coherence difficult to access, as discussed further in Sec.\ \ref{sec:probes}.  The two internal states of such pseudo-spin-1/2 Bose gases are not related by rotational symmetry in spin space, and so one expects phenomena associated with ground-state degeneracy and symmetry breaking to differ from those of spinor gases.  These differences are explored in Sec.\ \ref{sec:pseudospin}.

Other species of ultracold atoms comprise the metastable $^3\mbox{S}_1$ states of helium and the noble gases, and the $^3\mbox{P}_J$ states of two-electron atoms such as the alkali-earth elements and Yb.  For bosonic isotopes of these elements, all of which have a nuclear spin $I=0$, the electron angular momentum may be non-zero.  Thus, held in optical traps, such atoms offer another realization of spinor Bose gases.  However, the metastability of these atoms relies on the suppression of collisional decay to the electronic ground state due to spin polarization.  Experiments on metastable $^4\mbox{He}$ demonstrated that s-wave collisions between a pair of atoms with total projected angular momentum of $M_{\rm pair}=0$ lead to strong inelastic losses \cite{part10}.  Thus, it appears that stable spinor quantum fluids of such gases are precluded.

\subsubsection{Many-electron atoms: Spinor gases with strong dipolar interactions}

In condensed-matter systems, magnetic order is established both by local interactions, as in the Heisenberg-Dirac model of magnetism, and also by long-range interactions, such as the magnetic dipole interaction.  Similarly, as we discuss below, magnetic order in degenerate spinor Bose gases is influenced both by the local spin-dependent contact interactions and by the long-range magnetic dipole-dipole interaction (MDDI).  The strength of dipolar interactions increases with the square of the magnetic moment $\mu$ of the atoms forming the spinor gas.  For the ground-state alkali gases discussed above, neglecting the small nuclear contribution, the atomic magnetic moment is $\mu = \mu_B F / (I+1/2)$, i.e.\ no larger than one Bohr magneton $\mu_B$.  Dipolar effects are visible in such gases \cite{yi06texture,kawa06dipolar,kawa07observe,veng08helix}, although the local magnetic order is still selected predominantly by the contact interaction.

Both the long-range and the local spin-dependent interactions are varied strongly in gases with larger magnetic moments. For example, quantum degenerate gases of $^{52}$Cr display strong dipolar effects such as density deformations \cite{laha07ferrofluid} and instability \cite{laha08dwave} due to the anisotropic long-range dipolar interaction.  The magnetic moment of this spin-3 gas is $6 \mu_B$, yielding dipolar energies that are at least 36 times higher than those in the alkali gases (144 times larger than in $F=1$ $^{87}$Rb).  Several lanthanide elements have been laser cooled, presenting new candidates for high-spin and highly magnetic spinor Bose gases.  Thulium \cite{suka10thmot,suka10subdoppler} carries a total spin of $F=4$ in its ground state, with the magnetic moment of $4 \mu_B$.  Stronger dipolar interactions are realized in Er \cite{mccl06erbium,berg07erbium} and Dy \cite{youn10dy}, in which the atomic angular momenta of bosonic species are $J=6$ and $J=8$, leading to magnetic moments of $7 \mu_B$ and $10 \mu_B$, respectively.  Bose-Einstein condensates of both Er and Dy have been produced, and they show signs of strong magnetic dipolar interactions \cite{lu11dybec,aika12erbium}.

The large electron spin of these gases also influences the short-range interactions between atoms.  The presence of several molecular potentials, split by the fine structure for different configurations of the total electronic spin of the colliding atoms, and the strong dipolar interaction at molecular distances, altogether lead to much higher rates for dipolar relaxation among the Zeeman sublevels \cite{wein98cr,hens03relaxation} than in the alkali gases.  The phenomenology of spinor gases for these highly magnetic elements is thus expected to be different.  In the alkali gases, the total spin of the gas along the direction of an applied magnetic field is typically constant over the duration of an experiment. In contrast, the longitudinal magnetization of high-magnetic-moment spinor gases is dynamic.  The enhanced dipolar relaxation couples the spin and mass currents of spinor Bose gases, leading to phenomena such as the Einstein-de Haas effect \cite{kawa06dehaas,sant06spin3,garw07res}.  The anisotropy of the dipolar interaction also favors low-energy states in which the spin-state of the condensate is not spatially uniform; rather, the spin order parameter varies spatially, producing what is known as a spin texture.  This spin texture spontaneously breaks chiral symmetry \cite{yi06texture,kawa06dipolar}.  Strong dipolar relaxation also requires that spinor-gas experiments with these elements occur at precisely controlled, weak magnetic fields.  Such experimental conditions have recently been achieved, allowing for the first observation of spinor gas physics with $^{52}$Cr \cite{pasq11crspinor}.  Dipolar effects in spinor gases are discussed in Sec.\ \ref{sec:dipolar}.

\subsubsection{Orbital degrees of freedom in optical lattices}

The physics of spinor Bose gases may also find realization in systems of atoms in optical lattice potentials.  Similar to the orbital states of electrons within the unit cell of a periodic crystal, the quantum states of the center-of-mass motion within a unit cell of a periodic optical lattice can be regarded as an internal state space for a lattice-bound atomic gas.  These motional states may have non-zero angular momentum.  In deep optical lattices where the lattice site potentials tend toward spherical symmetry, the system closely resembles a spinor Bose gas insofar as the spherical symmetry implies the presence of ground-state degeneracies and a particular structure to the interactions between atoms.  As the lattice depth is reduced, the orbital degeneracy is strongly influenced by the structure and discrete symmetries of the lattice.

A degenerate Bose gas in the superfluid state in such a shallow lattice may be regarded as analogous to a bulk spinor Bose-Einstein condensate. Both the kinetic and interaction energy in the lattice can depend strongly on the orbital quantum state.  A simple example of this dependence arises from the anisotropic tunneling of p-orbital particles in a primitive lattice, with the tunneling rate being higher for the p-orbital aligned along the direction of tunneling \cite{isac05multiflavor,liu06nonzero}.  In terms of orbital magnetism in a bosonic superfluid, such tunneling gives anisotropic stiffness of the magnetic order parameter, which may lead to ground-state ``orbital textures'' with circulating orbital currents, reminiscent of the spin textures that are expected to arise from the anisotropic spin interactions of dipolar spinor condensates.  Such lattice-gas systems have been realized experimentally \cite{mull07pband,wirt11pband}.  The interplay of orbital dynamics and spin dynamics in non-Bravais lattices has also been considered \cite{wagn11spin1}.  However, one must note that orbital magnetism comes about from the excitation of high-energy bands, and, at least for Bose gases, is generally not protected against decay to the ground band; in contrast, at low magnetic fields or in the absence of dipolar relaxation, non-trivial spin ordering can persist for long times.

\subsection{General properties of a quantum Bose liquid with internal symmetry}

Consider a scalar Bose-Einstein gas in a box potential.  As the temperature of the gas is lowered into the regime of quantum degeneracy, why does the condensate form in the motional ground state of the box?

The answer lies in the influence of bosonic particle-exchange symmetry on the enumeration of microstates available to the system.  In particular, in comparison with Maxwell-Boltzmann statistics of distinguishable particles, Bose-Einstein statistics greatly reduces, by a factor of $N!$, the number of nondegenerate quantum states available to an $N$-body system. Thus, Bose-Einstein statistics makes states with macroscopic populations in single-particle states more likely.  Once the system temperature $T$ is reduced so that the number of single-particle states $M$ accessible to the system (roughly the number of states with energies below $k_B T$) becomes smaller than $N$, multiply occupied single-particle states become the norm.  In a $d$-dimensional box potential, this degeneracy occurs at $k_B T_c \sim E_1 N^{2/d}$ where $E_1$ is the energy of the first excited single-particle state in the box. Note that a huge number of single-particle states exist below the energy scale $k_BT_c$, yet a macroscopic number of particles condense into the lowest single-particle state due to bosonic enhancement.

To address the issue posed above, we consider now the population of atoms in the first excited single-particle state once the system is cooled below $T_c$.  This population is given as $N_1 \lesssim k_B T_c / E_1 = N^{2/d}$, so that, in three dimensions, with the scaling exponent $2/d<1$, only the ground state is macroscopically occupied.  In contrast, for a Maxwell-Boltzmann gas, the dominant macroscopic occupation of the ground state occurs only at the temperature $T_{MB} = E_1/k_B$.  Comparing the two situations, we see that Bose-Einstein statistics have the effect of greatly enhancing the effects of extremely small energy differences (here, $E_1$) so as to allow the ground state to be distinguished already at temperatures that are $N^{2/d}$ higher than $E_1/k_B$.

In degenerate spinor Bose gases, one finds that the spin-dependent contact and dipolar interactions distinguish between different spin states of the spinor order parameter.  The energy differences per particle are often minute, far smaller than the temperatures accessed in experiments; e.g.\ for the $F=1$ gas of $^{87}$Rb, the spin-dependent contact interaction contributes roughly 1 nK of energy per particle, far lower than the several $100$ nK critical temperature for Bose-Einstein condensation.  And yet, due to the quantum-statistical Bose enhancement effect described above, the gas becomes magnetically ordered just below the Bose-Einstein condensation temperature.  Such Bose-enhanced magnetization is a novel mechanism for magnetism, distinct from the more conventional mechanisms for localized or itinerant electrons in solids.

However, such Bose enhancement still does not determine uniquely the state of a quantum degenerate spinor Bose gas.  In the absence of explicit symmetry breaking, e.g.\ due to applied magnetic fields or non-spherically symmetric trapping potentials, the internal rotational symmetry of the spinor Bose gas guarantees that several single-particle spinor states achieve the same lowest energy of the interacting system.  The set of spinor ground states for the spin-$F$ gas may not retain the full $SU(2F+1)$ symmetry owing to spin-dependent interactions, but the lower-symmetry manifold of spinor ground states is still guaranteed to be degenerate over a continuous symmetry group.

In condensed-matter magnetic systems, such degeneracies are often resolved by experimental actualities: extraneous magnetic bias fields, magnetic defects, stress anisotropy, etc.  In spinor Bose gases, many of these artifacts are absent.  Does the Bose condensed system now break symmetry spontaneously and choose a conventional Bose-Einstein condensate, with just a single macroscopically occupied single-particle state \cite{penr56} but with random magnetization?  Or does the system abide by the postulate of molecular chaos -- assuming there are no constraints that bar the system from thermalizing -- and form an unbroken symmetry state with macroscopic populations in several single-particle spin states \cite{nozi82}?  This question is taken up in Sec.\ \ref{sec:fragmentation}.

The presence of a degenerate ground-state manifold also influences the dynamics and structure of the degenerate spinor Bose gas.  In a broken-symmetry state, the gas will possess one or more gapless Goldstone modes.  The symmetry group of the ground-state manifold determines the range of topological defects that may be found in low-energy spin textures, as discussed in Sec.\ \ref{sec:topological}

\subsection{Outline of this review}

The central characteristic of spinor Bose gases is their response to geometric transformations such as rotations and inversions.  To visualize the response of a high-spin object to such transformations, in Sec.\ \ref{sec:quantumstates} we outline a method, first developed by Majorana and subsequently re-introduced for the description of spinor gases, in which quantum states are represented by the positions of several points on the unit sphere; for a spin-1/2 state, this method reproduces the familiar Bloch-sphere representation.  We also introduce a family of high-symmetry spin states, known as inert states for the reason that they represent extremal energy states for generic rotationally symmetric spin Hamiltonians.

We then begin the physical description of spinor Bose gases.  Section\ \ref{sec:rotsymm} illustrates how rotational symmetry is used to simplify enormously the characterization of interparticle interactions.  After specifying several approximations, we find that the spin-dependent interactions within the spinor gas are determined by the relative strength of the s-wave scattering lengths for collisions among particles with a given value of the total spin $F_{\rm pair}$ of the colliding pair.  This simplification sets the spinor gas apart from other multi-component Bose gases, as we discuss for the experimentally relevant pseudo-spin-1/2 system.

The description of atomic interactions in Sec.\ \ref{sec:rotsymm} is strictly valid only when the full rotational symmetry of the gas is preserved.  In fact, the true experimental conditions under which spinor gases are prepared and studied depart from such ideal conditions.  Thus, before continuing with our exploration of the physics of spinor Bose gases, we use Sec.\ \ref{sec:exprealities} to describe these experimental conditions and techniques.

In Sec.\ \ref{sec:groundstates}, we describe the spatially uniform ground states of spinor Bose gases as determined by their spin-dependent s-wave interaction energies.  This discussion includes a review of experimental observations that characterize the ground states of the $F=1$ $^{23}$Na spinor gas (polar), the $F=1$ $^{87}$Rb spinor gas (ferromagnetic), and the $F=2$ $^{87}$Rb spinor gas (either antiferromagnetic or cyclic/tetrahedral).  Ground states are determined not only via the mean-field (Hartree) approximation, but also via a many-body quantum treatment that admits ground states that are fragmented.  

For a non-uniform degenerate spinor Bose gas, long length-scale variations of the local magnetic order parameters produce low-energy spin textures.  In Sec.\ \ref{sec:textures}, we discuss salient properties of these textures.  We highlight the connections between spatial variation of the spin-state (which may represent spin currents) and the superfluid velocity (i.e.\ mass currents), which are required due to the invariance of certain order-parameter manifolds to combinations of geometric rotations and gauge transformations.  We also summarize our knowledge of topological excitations that are either observed or are predicted to occur in such spin textures.  The coupling of spin- and mass-currents is also inherent to several hydrodynamic descriptions that have been developed to describe the low-energy dynamics of spin textures.

Another key characteristic of spinor Bose gases is the onset of magnetic ordering due to Bose-Einstein statistics and the enhancement of small single-particle energy differences in the quantum degenerate gas.   What happens to such magnetic ordering when the proclivity to Bose-Einstein condensation is weakened, such as at non-zero temperature, in two- or one-dimensional systems where finite temperature Bose-Einstein condensation does not occur, or in periodic potentials where the role of particle interactions is enhanced?  These questions are considered in Sec.\ \ref{sec:magneticorder}.

Next, in Sec.\ \ref{sec:dipolar}, we consider the influence of the magnetic dipole-dipole interaction on the low-temperature spinor Bose gas.  Such interactions add new features to the spinor gas: a new source of spin-dependent interactions, which can compete with the contact interactions to change the local magnetic order of the gas; and a new long-range interaction that competes with the kinetic-energy-inducted magnetic stiffness of the spinor gas.  Intricate spin textures are predicted to develop in the ground state to reduce the dipolar energy.  Dipolar effects are expected to strongly influence the nature of spinor gases composed of atoms with large magnetic moments.  However, even for the less magnetic alkali spinor gases, the long-range nature of the dipolar interaction is expected to be significant, and we describe experimental evidence for its effects.

In the next two sections, we discuss the dynamics of degenerate spinor Bose gases.  Sec.\ \ref{sec:spinmixing} is devoted to coherent spin mixing.  We summarize the observations of spin-mixing oscillations in both microscopic and macroscopic spinor gas systems.  Spin mixing collisions also induce dynamical instabilities that can be controlled by varying the single-particle energies (via the quadratic Zeeman shift) and the initial state of the degenerate spinor gas.  These instabilities provide access to a rich variety of physical phenomena, such as symmetry breaking, quantum atom optics, and spin squeezing.

Whereas the early-time dynamics of a non-equilibrium spinor condensate may exhibit well-controlled coherent spin mixing, at longer times one expects the gas to evolve toward thermal equilibrium via phase-ordering kinetics. In Sec.\ \ref{sec:equilibration}, we summarize several areas where coarsening and equilibration have been examined.

Following Sec.\ \ref{sec:magnetometry}, in which we demonstrate how the properties of ultracold spinor Bose gases may be useful for applications in magnetic sensing and imaging, we conclude by suggesting several open questions to be addressed in future research (Sec.\ \ref{sec:conclusions}).

Several previous reviews are complementary to this work.  The early round of experiments on spinor Bose-Einstein condensates of sodium is summarized by \citet{stam01leshouches}.  A more up-to-date review is presented by \citet{kawa12review}, and also briefly by \citet{ueda12annrev}, with a focus on theoretical frameworks.  Spinor gases are also treated briefly in \citet{lewe07review}, with a focus on gases in optical lattices.  In addition, several of the topics covered in this Review have been discussed separately.  Several reviews on properties of Bose gases with strong dipolar interactions focus on gases with uniformly oriented dipole moments \cite{bara02dipolarreview,bara08,laha09review}.  In contrast, the present focus is on dipolar interactions in spinor gases with a variable dipole orientation (Sec.\ \ref{sec:dipolar}).  Aspects of topological defects, particularly of vortices, in spinor condensates are discussed in \citet{kasa05}.  Several reviews have summarized the explorations of the Kibble-Zurek mechanism in condensed-matter systems, both for thermal phase transitions \cite{zure96review,arod02book} and, recently, for quantum phase transitions \cite{dzia10review}.  A broader range of non-equilibrium phenomena is discussed in the Colloquium by \citet{polk11colloq}.

%% file: rmp_ch_quantumstates.tex
\section{Quantum states of a spin-$F$ object}
\label{sec:quantumstates}

We begin by introducing and relating several methods for characterizing the states of a spin-$F$ object.

\subsection{Geometric representation of spinor wavefunctions}
\label{sec:representations}

Majorana invented a geometric representation of a general spin-$F$ state \cite{majo32}, which helps illustrate the symmetries retained by a spinor wavefunction independent of the special coordinate system chosen.  This representation can be used to picture the mean-field ground states, dynamics, and topological structures of degenerate spinor Bose gases \cite{barn06class,turn07order,make07inert,barn07classvortex,barn08lattice,barn09geometrical,turn09vortexmolecule,lama10lowenergy,kawa11symmetry,marc12spin56}.

First, let us recall that spin-1/2 quantum states may be represented pictorially as a point on the unit sphere (the Bloch sphere).  That is, given a spin-1/2 state $|\psi\rangle$, one can find a projection of the Pauli spin vector, $\hat{\sigma}_\pi = \hat{\bsigma} \cdot \bpi$ where $\bpi$ is a unit vector, so that $|\psi\rangle$ is the $+1$ eigenstate of $\hat{\sigma}_\pi$.  Apart from an overall phase, the state $|\psi\rangle$ is then represented by a point on the Bloch sphere at the position $\bpi$.  Alternately, one can instead consider the family of fully magnetized states, defined by the relation $\hat{\boldsymbol{\sigma}} \cdot \boldsymbol{\pi}^\prime |\psi^\prime\rangle = + |\psi^\prime\rangle$.  We then identify the fully magnetized state that is orthogonal to $|\psi\rangle$, thereby identifying a unit vector $\boldsymbol{\pi}^\prime$ that equivalently represents the state $|\psi\rangle$ on the Bloch sphere.  In this case we see trivially that $\boldsymbol{\pi} = - \boldsymbol{\pi}^\prime$.

Now we extend this pictorial scheme to states of higher spin.  First, let us imagine that a spin-$F$ particle is actually composed of $2F$ spin-1/2 sub-particles, and that $F$ is the quantum number of the total spin of these sub-particles.  A spin-$F$ wavefunction, written in the tensor space of these spin-1/2 sub-particles, is symmetric under the exchange of any two of the sub-particles.  The state of each of these sub-particles defines a unit vector $\bpi_r$ ($r \in \{1, 2, \cdots 2F\}$) in the manner described above for spin-1/2 states.  This set of vectors corresponds to $2 F$ points on the unit sphere, providing a pictorial representation of the spin-$F$ state apart from an overall phase.

Alternately, we ask whether the spin-$F$ state $|\psi\rangle$ is orthogonal to the maximum-value eigenstate of $\hat{\mathbf{F}} \cdot \bpi^\prime$, the projection of the spin vector along the unit vector $\bpi^\prime$ whose angular coordinates are $(\theta^\prime, \phi^\prime)$.  Expressing the state $|\psi\rangle$ in the $\hat{F}_z$ eigenbasis as $(\psi_F,\psi_{F-1},\cdots,\psi_{-F})^T$ ($T$ denotes the transpose), the orthogonality condition produces the following polynomial equation:
\begin{eqnarray}
b_0\zeta^{2F}+b_1\zeta^{2F-1}+\cdots+b_{2F}=0,
\label{eq:majoranapoly}
\end{eqnarray}
where $b_j=\psi_{F-j}^*/\sqrt{(2F-j)!j!}$ and where $\zeta$ is the stereographic projection on the equatorial complex plane of the point $(\theta^\prime,\phi^\prime)$ from the south pole:
\begin{eqnarray}
\tan\left(\theta^\prime/2\right) \, e^{i\phi^\prime}=\zeta
\label{eq:definepoint}
\end{eqnarray}
The $2 F$ complex roots of the polynomial equation, $\zeta_1,\zeta_2,\cdots,\zeta_{2F}$, identify the set of unit vectors $\boldsymbol{\pi}^\prime_r$.  As in the spin-1/2 case, these vectors are antipodal to the unit vectors obtained from the spin-1/2 sub-particle approach, i.e.\ $\bpi_r = - \bpi_r^\prime$.

In this Review, we adopt the convention that a spinor $|\psi\rangle$ is represented by the set $\bpi_r$.  The state described by the unit vectors $\bpi_r^\prime$ is generated from $|\psi\rangle$ by time reversal, by which one reverses the orientation of each of the spin-1/2 sub-particles of which $|\psi\rangle$ is composed.

\subsection{Spin moments}

The state of a spin-$F$ object, and of the magnetic ordering produced in an ensemble of such objects, may also be described by its spin moments.  Since the single-particle density matrix of a spin-$F$ system is a $(2 F + 1) \times (2 F+1)$ square matrix, its full description requires the specification of spin moments up to rank $2F$.  The rank-1 moment is the expectation value of the spin vector operator $\hat{\mathbf{F}}$.  The rank-2 moment is the expectation value of the spin quadrupole tensor, defined as
\begin{equation}
\hat{N}_{\mu \nu} = \frac{1}{2} \left( \hat{F}_\mu \hat{F}_\nu + \hat{F}_\nu \hat{F}_\mu\right) - \frac{1}{3} \hat{\mathbf{F}}^2\delta_{\mu \nu}
\end{equation}
where $\mu$ and $\nu$ specify Cartesian axes.  This tensor, and also its density as defined below, are sometimes referred to as the nematicity. Throughout this Review, both the vector spin operator $\hat{\mathbf{F}}$ and the spin quadrupole tensor $\hat{N}_{\mu \nu}$ are taken to be dimensionless, with the quantum of angular momentum $\hbar$ absorbed into the definitions of other terms.

We shall also use the symbols $\hat{\cal{F}}_\mu(\mathbf{r})$ and $\hat{\cal N}_{\mu \nu}(\mathbf{r})$, written now as functions of the spatial position $\mathbf{r}$, to define the \emph{densities} of the spin vector and spin quadrupole moments.  That is, let us define the bosonic field operator $\hat{\psi}_{m_F}(\mathbf{r})$ for particles in the eigenstate of the $\hat{F}_z$ operator with magnetic quantum number $m_F$.  The $\mu$ component of the spin vector density is then given as
\begin{equation}
\hat{\cal{F}}_\mu(\mathbf{r}) = \sum_{m_1, m_2} \hat{\psi}^\dagger_{m_1}(\mathbf{r}) \left(F_\mu\right)_{m_1, m_2} \hat{\psi}(\mathbf{r})
\end{equation}
where $F_\mu$ is the matrix representation of the corresponding projection of the spin vector operator in the $\hat{F}_z$ eigenbasis.  The components of the spin quadrupole tensor density are written similarly.  Both quantities have the units of number density.  The magnetization $\mathbf{M}(\mathbf{r})$ is found by multiplying the spin vector density by the maximum magnetic moment of a spin-$F$ atom.

The spin moments may be expressed in terms of the unit vectors $\bpi_r$ defined in the Majorana representation, although such expressions become complicated for higher spin and higher-rank moments \cite{barn09geometrical}.  For the simple case of $F=1$, the vector spin is found to be
\begin{equation}
\langle \hat{\mathbf{F}} \rangle = \frac{2 (\bpi_1 + \bpi_2)}{3 + \bpi_1 \cdot \bpi_2}
\end{equation}
and the spin-quadrupole tensor as
\begin{eqnarray}
\langle \hat{\mathbf{N}} \rangle & = & - \frac{\mathbf{I}}{6} + \frac{1}{3 + \bpi_1 \cdot \bpi_2} \Bigg[ \bpi_1 \bpi_2 + \bpi_2 \bpi_1  \\
& + &\frac{(\bpi_1 + \bpi_2)(\bpi_1 + \bpi_2) + (\bpi_1 \times \bpi_2)(\bpi_1 \times \bpi_2)}{1 + \bpi_1 \cdot \bpi_2} \Bigg] \nonumber
\end{eqnarray}
where the matrices in the square brackets are written as dyadic tensors, and $\mathbf{I}$ is the identity matrix.

Let us consider two examples of Majorana representations for the simple case of $F=1$.  For the spinor $(1/2, 1/\sqrt{2}, 1/2)^T$, where $T$ denotes the transpose, the polynomial $\zeta^2 + 2 \zeta + 1 = 0$ has a double root at $\zeta=-1$, so that the state is represented by two overlapping points at the location $(\theta = \pi/2, \phi= 0)$ and has the $SO(2)$ symmetry about the $\mathbf{x}$ axis.  Geometric rotations of this state are represented by two overlapping points at the corresponding positions on the sphere.  Majorana representations of longitudinal ferromagnetic states are shown in Fig.\ \ref{fig:spin1phasediagrams}.  States within this manifold maximize the vector spin $|\langle \hat{\mathbf{F}} \rangle|$, and so they are known as ``ferromagnetic'' states.  Note that the time-reversed spinor state is obtained by moving the points representing the state to their antipodal positions; indeed, the ferromagnetic states reverse their orientation under time reversal.

Another manifold of states is obtained starting from the spinor $\psi = (1/\sqrt{2}, 0, -1/\sqrt{2})^T$, for which Eq.~(\ref{eq:majoranapoly}) gives roots at $\zeta = 1$ and $\zeta = -1$.  The corresponding points are located at antipodal points on the unit sphere; an example is shown in Fig.\ \ref{fig:spin1phasediagrams}.  For this state, we identify $\mathbf{x}$ as the nematic director $\mathbf{n}$, i.e.\ the axis along which the spinor is the eigenstate of the spin projection $\mathbf{F} \cdot \mathbf{n}$ with eigenvalue zero.  The state $|\psi\rangle$ has both $SO(2)$ symmetry about the $\mathbf{x}$ axis and also $\pi$ rotation symmetry about any axis on the $\mathbf{y}$-$\mathbf{z}$ plane.  The states obtained from this starting state by rotations are known as ``polar'' states.  They are characterized by zero magnetization, while their inversion symmetry is reflected in their nematicity.
%

Representations for higher-spin states can be similarly obtained.  Spin-2 states are represented by four points on the unit sphere.  Examples of such states relevant to the zero-field mean-field ground states of $F=2$ condensates (Fig.\ \ref{fig:spin2phasediagram}) include the ferromagnetic state, represented by four overlapping dots; the uniaxial nematic state, represented by two doubly degenerate dots at antipodal points on the sphere; the biaxial nematic state, represented by points lying on a great circle at the vertices of a regular square; and the cyclic/tetrahedral phase\footnote{The term ``cyclic'' was introduced by \citet{ciob00}, referring to earlier classifications of the states of d-wave superconductors.  The term may also refer to the 3-fold rotational symmetry of the state.  The Majorana representation of the state clarifies that its symmetry group is that of the tetrahedron, so a better name for the state may be ``tetrahedral.''  We use both names in this Review in keeping with previous usage.}, represented by points at the corners of a regular tetrahedron.

Spin-3 states are represented by six points on the unit sphere \cite{barn06class,kawa11symmetry}. The possible symmetries constructed from six vertices are octahedron, tetrahedron, dihedral-2,3,4,5,6 and $C_{2,3,4,5,6}$, where $C_n$ represents the cyclic symmetry of order $n$. The icosahedron symmetry is expected to appear for $F\geq 6$.

\subsection{Inert states}
\label{sec:inert}

There is a special class of states, called ``inert states,'' which is defined to be stationary for generic rotationally symmetric, spin-dependent energy functionals and independent of the strengths of the interactions \cite{bart74,bart75,voll90he3,yip07inert,make07inert}.  There is a close relationship between the symmetry of the order parameter and a stationary point of the free-energy functional~\cite{mich80}. This relationship greatly simplifies the otherwise rather involved calculation of stationary states.  Let $g\in G$ be an element of the group of operations on the order parameter $\psi$ of the system that leave the free-energy functional invariant: $f[g\psi]=f[\psi]$. Let $h\in H$ be an element of the isotropy group $H$  that leaves $\psi$ unaltered: $h\psi=\psi$. The order parameter manifold $M$ is then given by a coset $M=G/H$. Now take one $\psi$ and operate every $g\in G$ on it. The resulting set of $g\psi$ is said to constitute an orbit. Now take another $\psi'$ and find the corresponding orbit by operating every element of $G$. If two orbits share the same isotropy group, they are said to belong to the same stratum which is defined as the union of orbits with the same isotropy group.  \citet{mich80} asserts that if an orbit is isolated in its stratum, i.e., if it does not have another orbit in its neighborhood, the corresponding order parameter is a stationary function of the free-energy functional. This theorem is useful to find inert states because they depend only on the symmetry of the system and are independent of the details of the system's parameters.
As an illustration, let us consider the following eigenvalue problem: $e^{i\phi}e^{-i f_z \phi}\psi=\psi$.
For a spin-1 BEC, the solution is $\psi=(1,0,0)^T$, which implies that the ferromagnetic state is an extremum of the mean-field energy.

The ground states of spinor Bose condensates in the absence of externally imposed symmetry-breaking fields are inert, as are the A, A${}_1$, and B phases of superfluid $^3$He \cite{voll90he3}.  It can be shown that the ferromagnetic and polar states are the only two inert states of the spin-1 condensate. For the spin-2 condensate, the ferromagnetic, uniaxial nematic, biaxial nematic, and cyclic/tetrahedral states are all inert states.

%% file: rmp_ch_rotsymm.tex
\section{Bose gases with internal rotational symmetry}
\label{sec:rotsymm}

In high-energy, condensed-matter, atomic and molecular physics alike, the starting point for understanding a new system is the recognition of its underlying symmetries, by which one can simplify its energetics, dynamics, topological structures, and so on.  In this spirit, we elucidate how rotational symmetry constrains and simplifies the interactions present in a spinor Bose gas.

\subsection{Quantum scattering under rotational symmetry}
\label{sec:collisions}

At the low densities of atomic gases, interactions result almost exclusively from binary collisions and, even in a many-body system, such interactions can be understood through the treatment of the two-body quantum scattering problem.  Neglecting the effects of particle confinement, we treat such scattering in three spatial dimensions.  Collisions are then characterized by the scattering matrix that connects the incoming and outgoing asymptotic states.  Such states are chosen as product states of the orbital motion and internal states.  We specify the former in the spherical harmonic basis by the relative wave number $k$, and the quantum numbers for the total orbital angular momentum $L_{\rm pair}$ and its projection on the quantization axis $m_{L,{\rm pair}}$.  We specify the latter using basis states of spinor wavefunctions $\psi$ for the colliding particles.  Time-reversal symmetry allows us to interchange the incident and outgoing quantum states.  Energy conservation (invariance under translations in time) ensures that differences between the incident and outgoing kinetic energies, determined by $k_i$ and $k_f$, must be compensated by changes in the internal energies of the spin states in the collision; e.g.\ dipolar relaxation in an applied magnetic field converts the Zeeman energy into the kinetic energy of the colliding particles.

The next step in simplifying the scattering problem is to consider the character of the two-body interaction, which separates fairly cleanly into short-range and long-range parts, with the divide between these parts occurring at an internuclear distance $r$ longer than the effective range of the potential $r_0$.  The long-range part is ultimately dominated by the magnetic dipole-dipole interaction (MDDI), which falls off as $1/r^3$.  This functional dependence requires that the dipole interaction be retained explicitly in the interaction Hamiltonian.  For the sake of simplicity, we neglect the long-range portion at this point, deferring the treatment of dipolar interactions to Sec. \ref{sec:dipolar}.

The short-range portion of the potential is complicated as it comes about from the interaction among all the constituent particles within the colliding atoms.  This portion also contains terms that mix the spin and orbital angular momenta of the colliding particles, such as through second-order spin-orbit coupling.  The short-range interaction also influences the strength of the MDDI; for example, a scattering resonance causes the two-body wavefunction to be enhanced at short internuclear distances, enhancing the effect of dipolar interactions.

To treat the quantum scattering due to the short-range potential, several approximations are commonly made.  First, we assume that the \emph{incident} collision energy is very low, in the sense that the thermal de Broglie length of the relative motion of the atoms obeys $\lambda_{dB} \gg r_0$.  In this cold-collision approximation, only the lowest-order incident partial waves ($L_{{\rm pair},i}=0$) undergo collisions (subscript $i$ indicates ``initial"); all other partial waves can be ignored.

Second, we make the ``spinor gas collision'' approximation that the short-range potential is rotationally invariant -- such rotations affect both the internal spin degrees of freedom and also the orbital spatial degrees of freedom.  This approximation is exact in the absence of any external source of rotational symmetry breaking such as applied magnetic fields, non-spherical trapping potentials, vector or tensor ac Stark shifts in the optical dipole force, etc.  In practice, such influences are rarely completely absent; nevertheless, the spinor gas collision approximation still applies at low magnetic fields, away from magnetic or optical Feshbach resonances and in the normal Zeeman regime of the ground-state hyperfine structure.  This approximation guarantees the conservation of the total angular momentum of the colliding pair, which is the sum of the total orbital angular momentum $\hat{\mathbf{L}}_{\rm pair}$ and the internal angular momentum (nuclear and electronic) $\hat{\mathbf{F}}_{\rm pair}$.  By the cold-collision approximation, the total angular momentum of a colliding pair is $F_{{\rm pair}, i}$.  However, each source of angular momentum need not be separately conserved.  Notably, dipolar relaxation can convert internal to orbital angular momentum; to first order, such a collision promotes the colliding pair from an incident orbital s-wave ($L_{{\rm pair},i} = 0$) to an outgoing d-wave ($L_{{\rm pair},f} = 2$) wavefunction, with a concomitant change in $F_{{\rm pair},f}$, where subscript $f$ indicates ``final."

To simplify further, one commonly makes a third assumption, the ``weak-dipolar approximation,'' whereby one neglects the spin-orbit coupling through the short-range molecular potential.  Now, the orbital and internal angular momenta are separately conserved:  for a colliding atom pair, $L_{{\rm pair},i} = L_{{\rm pair},f}=0$ and $F_{{\rm pair},i} = F_{{\rm pair},f} = F_{\rm pair}$.

A final approximation is that s-wave collisions do not mix total hyperfine states of the colliding atoms.  For example, in considering the properties of a spinor gas in an upper hyperfine manifold of the electronic ground state (say the $F=2$ spinor gas of $^{87}$Rb), we neglect collisions by which atoms undergo hyperfine relaxation to the lower manifold ($F=1$).

Only after all these approximations can we conclude that the myriad collisions among spinor-gas atoms are characterized simply by the s-wave scattering lengths $a_{F,{\rm pair}}$ between two particles in the collision channel $F_{\rm pair}$ \cite{ho98,ohmi98}.  The parity of $F_{\rm pair}$ is constrained further by quantum statistics.  The many-body wave function of identical spin-$F$ particles changes by a factor of $(-1)^{2F}$ under exchange of any two particles. By the same exchange, the spin and orbital parts of the wave function change by $(-1)^{F_{\rm pair}+2F}$ and $(-1)^{L_{\rm pair}}$, respectively.  To be consistent, we must have $(-1)^{2 F}=(-1)^{F_{\rm pair}+2F}\times(-1)^{L_{\rm pair}}$; hence $F_{\rm pair}+L_{\rm pair}$ must be even. For s-wave collisions, $F_{\rm pair}$ must be even.  This conclusion holds for both bosons and fermions.


Let us now construct the Hamiltonian describing the rotationally symmetric s-wave interaction under the approximations listed above.  By the pseudopotential method \cite{huan57}, we obtain \cite{ho98,ohmi98}
\begin{eqnarray}
V=\frac{1}{2} \sum_{i,j} \delta^3\left(\mathbf{r}_i -\mathbf{r}_j\right) \sum_{{\mathrm{even}}\,  F_{\rm pair}} \frac{4\pi\hbar^2 a_{F,{\rm pair}}}{M} \hat{P}_{F,{\rm pair}},
\label{eq:vinteraction} 
\end{eqnarray}
where $M$ is the atomic mass and the sum is taken over particle pairs, labeled $i$ and $j$.  The operator $\hat{P}_{F,{\rm pair}}$ projects a pair of atoms -- implicitly the pair $(i,j)$ -- into the total spin-$F_{\rm pair}$ state, where only even values of $F_{\rm pair}$ are now considered.

Table \ref{tab:scatlengths} summarizes the known scattering lengths $a_{F, {\rm pair}}$ for several spinor Bose gases that are experimentally accessible.  As discussed below, and in Sec.\ \ref{sec:meanfieldground}, the different relations among the scattering lengths for different $F_{\rm pair}$ yield different predictions for the spin ordering of the ground state.

\renewcommand{\multirowsetup}{\centering}
\begin{table}[tb]
\begin{center}
\begin{tabular}{| c | c |}
\hline \hline
Spinor gas & Scattering lengths [$a_B$]\\
\hline  \hline
$^7$Li, $F=1$ & $a_0 = 23.9$, $a_2 = 6.8$ \\
$^{23}$Na, $F=1$ & $a_0 = 50.0 \pm 1.6$, $a_2 = 55.0 \pm 1.7 $ \\
$^{41}$K, $F=1$ & $a_0 = 68.5 \pm 0.7$, $a_2 = 63.5 \pm 0.6$ \\
$^{87}$Rb, $F=1$ & $a_0 = 101.8 \pm 0.2$, $a_2 = 100.4 \pm 0.1 $ \\
\multirow{2}{2 cm}{$^{87}$Rb, $F=2$} & \multirow{2}{5cm}{$a_0 = 87.4 \pm 1.0 $, $a_2 = 92.4 \pm 1.0 $, $a_4 = 100.5 \pm 1.0 $}\\
 & \\
\multirow{2}{2 cm}{$^{52}$Cr, $F=3$} & \multirow{2}{5cm}{$a_0 =30-50$, $a_2 = -7 \pm 20$, $a_4 = 58 \pm 6$, $a_6 = 102.5 \pm 0.4$}\\
 & \\
\hline
\end{tabular}
\caption{Scattering lengths $a_{F, {\rm pair}}$ for several realizations of spinor Bose gases.  These scattering lengths are calculated theoretically, using numerical models for the molecular potential with input parameters that are determined by comparison with experimental measurements of the positions of Feshbach resonances, of the line strengths of photoassociation resonances, and of scattering length differences surmised from the dynamics of quantum gases.  Values for lithium are from \citet{note:lithium}, for sodium from \citet{crub99}, for potassium from \citet{falk08} and \citet{lyse10} (we guess errors at the level of 1\%), for rubidium from \citet{klau01rbspin} and \citet{vankemp02}, and for chromium from \citet{wern05feshbach}, \citet{pasq10dipolar} and the value of $a_0$ is due to H. Sadeghpour (private communication).  Values and standard error estimates (where they are known) are given in units of the Bohr radius $a_B$.}
\label{tab:scatlengths}
\end{center}
\end{table}

\subsection{Spin-dependent s-wave interaction energy}

It is convenient to express the s-wave interaction energy (Eq.\ (\ref{eq:vinteraction})) in less opaque nomenclature involving one-particle and two-particle spin operators, as we demonstrate below for the spin-1 and spin-2 cases.

\subsubsection{Spin-1 case}

As a starting point, we consider the identity operator for a system of two spin-$F$ particles:
\begin{eqnarray}
\hat{I}_1 \otimes \hat{I}_2 = \sum_{\mathrm{all}\, F_{\rm pair}}\hat{P}_{F,{\rm pair}},
\label{eq:spinidentity}
\end{eqnarray}
where we sum over all $F_{\rm pair} \in \{0, 1, 2, \cdots, 2 F\}$.  Here, $\hat{I}_i$ is the identity operator of the $i$th particle, and $\otimes$ denotes the tensor product.  From the composition law of spin operators, we obtain also
\begin{eqnarray}
 \hat{\mathbf{F}}_1 \cdot \hat{\mathbf{F}}_2 \! = \! \sum_{\mathrm{all}\, F_{\rm pair}} \!\! \left[ \frac{1}{2}F_{\rm pair}(F_{\rm pair}+1)-F(F+1)\right]\! \hat{P}_{F,{\rm pair}}.
\label{eq:spincomposition}
\end{eqnarray}
where $\mathbf{\hat{F}}_i$ is the one-particle vector spin for particle $i$.

We recall that the restriction to s-wave interactions allows us to consider only the two-particle spin states that are symmetric under the exchange of two particles, i.e.\ for all two-particle spin operators multiplied by the spatial $\delta$-function in Eq.\ (\ref{eq:vinteraction}), we may eliminate terms with odd $F_{\rm pair}$ from the summations.  Restricting to such states, we then identify, for $F=1$, $\left[\hat{I}_1\otimes \hat{I}_2\right]_S = \hat{P}_0 + \hat{P}_2$ and $\left[\hat{\mathbf{F}}_1 \cdot \hat{\mathbf{F}}_2\right]_S=\hat{P}_2-2 \hat{P}_0$, where the $S$ subscript reminds us of the symmetric-state restriction.  Combined with Eq.\ (\ref{eq:vinteraction}), we obtain the interaction Hamiltonian of spin-1 bosons as
\begin{equation}
\hat{V}^{(1)} =  \frac{c_0^{(1)}}{2} \left[\hat{I}_1\otimes \hat{I}_2\right]_S+\frac{c_1^{(1)}}{2} \left[\hat{\mathbf{F}}_1 \cdot \hat{\mathbf{F}}_2 \right]_{S}
\label{eq:spin1interaction}
\end{equation}
with
\begin{eqnarray}
c_0^{(1)}\equiv \frac{4 \pi \hbar^2}{M} \frac{a_0 + 2 a_2}{3}, \\
c_1^{(1)}\equiv \frac{4 \pi \hbar^2}{M} \frac{a_2 - a_0}{3}.
\label{eq:definespin1cs}
\end{eqnarray}
In second quantization, each identity operator becomes the number-density operator $\hat{n}(\mathbf{r}) =\sum_{m=-F}^F\hat{\psi}_m^\dagger(\mathbf{r}) \hat{\psi}_{m}(\mathbf{r})$ and the $\nu$ component of the spin vector becomes $\hat{F}_\nu=\sum_{m,n=-F}^F\hat{\psi}_m^\dagger(\mathbf{r}) (F_\nu)_{mn} \hat{\psi}_n(\mathbf{r})$, where $\hat{\psi}_{m_F}(\mathbf{r})$ is the Bose field operator for the Zeeman state labeled by $m_F$. Thus, the second-quantized interaction operator is given by
\begin{eqnarray}
\hat{V}^{(1)} & = & \frac{1}{2}\int d{\bf r} \,
\left[  c_0^{(1)}: \hat{n}^2 : + c_1^{(1)} : \hat{\mathbf{F}}_1\cdot \hat{\mathbf{F}}_2 : \right],
\label{2.9}
\end{eqnarray}
where $::$ denotes normal ordering.  The density and spin operators are evaluated locally, enforcing the restriction to symmetric spin states and allowing us to drop the subscript $S$.

%

\subsubsection{Spin-2 case}

For $F=2$, Eq.~(\ref{eq:spincomposition}) gives $\left[\hat{\mathbf{F}}_1\otimes\hat{\mathbf{F}}_2\right]_S=-6 \hat{P}_0-3\hat{P}_2+4\hat{P}_4$.  Combined with Eqs.\ (\ref{eq:vinteraction}) and (\ref{eq:spinidentity}) leads to
\begin{eqnarray}
V^{(2)}
&=& \frac{c_0^{(2)}}{2}\left[\hat{I}_1\otimes \hat{I}_2\right] + \frac{c_1^{(2)}}{2} \hat{\mathbf{F}}_1 \cdot \hat{\mathbf{F}}_2 + \frac{5c_2^{(2)}}{2} \hat{P}_0,
\label{eq:vforspin2}
\end{eqnarray}
where
\begin{eqnarray}
\lefteqn{\left\{ c_0^{(2)},c_1^{(2)}, c_2^{(2)}  \right\}  \equiv } \\
&  & \frac{4 \pi \hbar^2}{M} \times \left\{ \frac{4a_2+3a_4}{7}, \frac{a_4-a_2}{7}, \frac{7a_0 - 10a_2 + 3a_4}{35} \right\}. \nonumber
\label{eq:definespin2cs}
\end{eqnarray}
The second-quantized form of Eq.~(\ref{eq:vforspin2}) is given by
\begin{eqnarray}
\hat{V}^{(2)}
=\frac{1}{2} \int d{\bf r} \;
[ c_0^{(2)} :\!\hat{n}^2\!:+c_1^{(2)} :\!\hat{\mathbf{F}}_1 \cdot \hat{\mathbf{F}}_2\!:+c_2^{(2)} \hat{A}_2^\dagger\hat{A}_2]
\label{2.12}
\end{eqnarray}
where $\hat{A}_2\equiv\sum_{m=-2}^2(-1)^m\hat{\psi}_m\hat{\psi}_{-m}/\sqrt{5}$ is the annihilation operator of a spin-singlet pair of $F=2$ atoms.
\subsection{Pseudo-spin systems}
\label{sec:pseudospin}

Any mixture of $N$ condensates may, in principle, be described as a fictitious spin-$(N-1)/2$ system. For example, a mixture of condensates occupying the two hyperfine states $|F=1, m_z = -1\rangle$ and $|F=2, m_z = +1\rangle$ of $^{87}$Rb \cite{myat97} constitutes a pseudo-spin-1/2 system. Unlike spinor gases, pseudo-spin systems do not possess rotational symmetry in spin space, and so many of the simplifications attained above for describing the interactions and structures of quantum gases do not apply.


Consider a Bose-Einstein condensate composed of two internal states of the same atom.  Its energetics are described by the following Gross-Pitaevskii (mean-field) energy functional:
\begin{eqnarray}
E &\! =\! & \int d\mathbf{r}\left[
\sum_{i=1,2}\Psi_i^*\left(-\frac{\hbar^2\nabla^2}{2 M}+U_{i}+\frac{2 \pi \hbar^2 a_{ii}}{M}|\Psi_i|^2\right)\Psi_i \right.\nonumber \\
& & \left. + \frac{4 \pi \hbar^2 a_{12}}{M}|\Psi_1|^2|\Psi_2|^2
\right],
\label{2.30}
\end{eqnarray}
where $M$ is the atomic mass, $U_{i}$ is the trapping potential for state $|i=1,2\rangle$, and $a_{ii}$ and $a_{12}$ are the intra- and inter-species s-wave scattering lengths.  To draw connection to the characteristics of spinor Bose-Einstein condensates, this energy functional can be expressed in the pseudo-spin representation as follows. For simplicity, we assume $U_{1}=U_{2}\equiv U$, and decompose the condensate pseudo-spinor wavefunction $\Psi_i$ as $\Psi_i(\mathbf{r})=\sqrt{n(\mathbf{r})}\chi_i(\mathbf{r})$, where $n(\mathbf{r})$ is the total density of atoms and $|\chi_1(\mathbf{r})|^2+|\chi_2(\mathbf{r})|^2=1$. Using $\chi_i$, we can introduce a unit spin vector $\mathbf{S}$ as
\begin{eqnarray}
\left(\begin{array}{c} S_x \\ S_y \\ S_z \end{array} \right) = \left( \begin{array}{c} 2 \, {\rm Re}(\chi_1\chi_2^*) \\ 2 \, {\rm Im}(\chi_1\chi_2^*) \\ |\chi_1|^2-|\chi_2|^2 \end{array} \right)
\label{2.31}
\end{eqnarray}
where Re and Im denote the real and imaginary parts, respectively.
Then Eq.~(\ref{2.30}) can be rewritten as
\begin{eqnarray}
E &= &\int d\mathbf{r}\left[
\frac{\hbar^2}{2M}(\nabla\sqrt{n})^2+\frac{\hbar^2 n}{8M}\sum_{i=x,y,z}(\nabla S_i)^2+\frac{M n}{2}\mathbf{v}_{\rm eff}^2 \right.\nonumber \\
& & +
\left. U+\frac{n^2}{2}(c_0+c_1S_z+c_2S_z^2)\right],
\label{2.32}
\end{eqnarray}
where $\mathbf{v}_{\rm eff}\equiv(\hbar/m){\rm Im}\sum_{i=1,2}(\chi_i^*\nabla\chi_i)$, and $c_0=4 \pi \hbar^2(a_1+a_2+2a_{12})/M$, $c_1=4 \pi \hbar^2(a_1-a_2)/(2M)$, and $c_2=\pi \hbar^2(a_1+a_2-2a_{12})/M$. This result shows that, even at zero applied magnetic field, the system experiences fictitious interaction-induced linear and quadratic Zeeman effects for nonzero $c_1$ and $c_2$, respectively.  Differing from the case of spinor Bose gases, these terms break the $SU(2)$ symmetry of the nonlinear sigma model~\cite{raja87solitons} in which only the $\sum_{i=x,y,z}(\nabla S_i)^2$ term exists.

Another generic feature of spinor Bose gases that may be absent in the pseudo-spin system is the dynamics of interaction-induced spin mixing.  For example, for the $F=1$ spinor gas the s-wave interaction energy (Eq.\ (\ref{2.9})) contains the terms
\begin{equation}
\frac{4 \pi \hbar^2 (a_2 - a_0)}{3 M} \int d^3 \mathbf{r} \, \left(\hat{\psi}^\dagger_0  \hat{\psi}^\dagger_0  \hat{\psi}_{+1}  \hat{\psi}_{-1}   + \hat{\psi}^\dagger_{+1}  \hat{\psi}^\dagger_{-1}  \hat{\psi}_0  \hat{\psi}_0  \right).
\label{eq:spinmixingterm}
\end{equation}
These terms describe spin-mixing collisions wherein two atoms in the $|m_z = 0\rangle$ state collide to yield atoms in the $|m_z = +1\rangle$ and $|m_z = -1\rangle$ states, and vice versa.  Such spin mixing is necessarily present if the s-wave interactions are spin dependent ($a_2 \neq a_0$), and leads to characteristic phenomena such as coherent spin oscillations, spontaneous magnetization and symmetry breaking, parametric amplification and spin squeezing, as discussed in Sec.\ \ref{sec:spinmixing}.  Such phenomena are physical consequences of the coherence between the different internal-state components of a Bose-condensed spinor gas.  In contrast, in pseudo-spin Bose systems, such internal-state dynamics may be absent.  However, the pseudo-spin Bose condensate may still show spontaneous phase coherence among its internal-state populations, as demonstrated experimentally by \citet{whee04spin}.  In other words, a spin-$F$ spinor Bose-Einstein condensate cannot be considered an incoherent overlap of $N=2 F$ individual condensates, whereas, in many regards, a pseudo-spin-$F$ condensate can.

\subsection{Exchange interactions and spin waves}
\label{sec:spinwaves}

As argued above, a pair of atoms in a spinor gas will only experience a contact interaction if its total spin $F_{\rm pair}$ is even.  For two atoms in distinguishable motional states, this particle-exchange effect leads to an interaction-energy splitting between states of even and odd $F_{\rm pair}$, even when the s-wave scattering lengths are all equal, similar to exchange-energy effects within multi-electron atoms.  The exchange-interaction effect is pronounced for a gas in the cold-collision regime, for which all odd partial wave collisions are frozen out.  Interactions among atoms in a Bose-Einstein condensate are not affected by this phenomenon, as the particles all occupy the same state of motion.  Rather, the exchange interaction affects the spin dynamics of nondegenerate gases.

Spin dynamics driven by the exchange effect in the cold-collision regime were observed in experiments on trapped hydrogen \cite{john84,bige89}.  The gas was comprised of atoms with a common electronic spin polarization, but differing nuclear spin orientations.  For this particular pseudo-spin $1/2$ system, because all collisions occur through the electronic-triplet molecular potential, all s-wave scattering lengths are expected to be the same.  However, nuclear magnetic resonances detected in a gas cell placed in an inhomogeneous magnetic field showed narrow resonances ascribed to standing spin-wave modes of this rarified gas.

In a pseudo-spin-$1/2$ gas, the exchange interaction yields a singlet-triplet splitting that can be understood simply as causing the spins of two distinguishable atoms to precess about one another.  Similarly, in a mean-field picture, the spin of an atom passing through a gas that is spin polarized in the $\mathbf{e}$ direction will precess about $\mathbf{e}$ at a rate proportional to the product of the gas density and the s-wave scattering length.  Such an interaction leads to a dynamical coupling between spin currents and spin polarizations.  Solving the gas-kinetic equations in the presence of such coupling reveals a spectrum of spin waves, which are damped by particle diffusion \cite{lhul82trans1,lhul82trans2}.

The physics of spin waves arose again in studies of pseudo-spin $1/2$ rubidium gases for which the s-wave scattering lengths are, accidentally, all nearly equal.  A nondegenerate gas was prepared in a superposition of two spin states, within a magnetic trapping potential that differed slightly for the two states.  The slight spin dependence of the trap led to ``anomalously'' rapid spin dynamics, seen as a spatial segregation of the two spin components \cite{lewa02}.  Such dynamics were soon explained as a real-space signature of spin waves \cite{mcgu02,will02spinwave,okte02spinwave,fuch02}.  Spin waves have also been imaged in two-component ultracold Fermi gases \cite{du09}.  Spin-wave dynamics, also called the identical spin rotation effect, have been shown to prevent the dephasing of spins (``spin-locking'') in tightly confined pseudo-spin-$1/2$ gases \cite{deut10self,klei11coherence}.  The interplay between the strong spin-wave dynamics of the thermal fraction and the otherwise languid spin dynamics of the condensed fraction of a partly condensed pseudo-spin-$1/2$ gas was shown to accelerate the condensate spin dynamics significantly \cite{mcgu03normal}.

These works on pseudo-spin gases suggest that exchange effects in nonzero temperature spinor Bose gases should be significant, particular in situations where the differences among s-wave scattering lengths $a_{F,{\rm pair}}$ are small, and also richer than those observed previously owing to the higher value of the spin $F$.  Several distinct spin-wave modes have been predicted, marked by damped oscillations of both the spin vector moments and also of the spin quadrupole moments due to exchange effects \cite{niku08spindynamics}.  An examination by \citet{natu10spinwave} suggests that spin waves should lead to spin-mixing instabilities in nondegenerate $F=1$ spinor Bose gases.  However, the full implications of exchange effects in nonzero-temperature spinor Bose gases remain unexplored.

%% file: rmp_ch_exprealities.tex
\newcommand{\Dfs}{\Delta_{\mbox{\scriptsize{fs}}}}
\newcommand{\Dhfs}{\Delta_{\mbox{\scriptsize{hfs}}}}

\section{Experimental realities}
\label{sec:exprealities}

\subsection{Optical trapping}
\label{sec:opticaltraps}


Experiments on spinor gases commence typically with the preparation of a spin polarized, and thus scalar, gas at low temperature.  This spin polarization is established in the preparation of the atomic gas, for example, by optical pumping and the selective trapping of a single spin state in a magnetic trap.  The spinor properties of the gas become relevant once the atoms are held in a trap that accommodates all spin states and the atomic spin is no longer constrained to the state of maximum magnetization.

A suitably spin-independent trap is provided by the optical dipole force of a focused beam of light.  An optical trap confines an atom by dint of the electric dipole moment that is induced on the polarizable atom by the optical electric field \cite{metc99book}.  The trapping potential is expressed as a dyad of dipole vector operators, and is, thus, composed generally of a scalar, vector, or rank-2 tensor term.  In the case of alkali gases, for light far detuned from atomic resonance, the relative strength of these three contributions scale as $1/\Delta_{eg}$, $\Dfs/\Delta_{eg}^2$ and $\Dhfs/\Delta_{eg}^2$, respectively, where $\Delta_{eg}$ is the detuning of the trapping light from the atomic resonance, $\Dfs$ is the excited-state fine-structure splitting and $\Dhfs$ is the excited-state hyperfine-structure splitting.  By using a detuning $\Delta_{eg}$ that greatly exceeds $\Dfs$ and $\Dhfs$, the scalar light shift can be made to dominate the vector and rank-2 tensor shifts\footnote{For example, for the $^{87}$Rb atom, $\Dhfs$ is on the order of $10^8$ Hz in the excited $5P$ states.  The fine-structure splitting of $\Dfs = 7 \times 10^{12}$ is determined from the difference in the resonant wavelengths of 795 and 780 nm for the D1 and D2 lines, respectively.  An optical trap with light at a wavelength of 1 $\mu$m, or $\Delta_{eg} = 8 \times 10^{13}$ Hz, satisfies the relation $\Delta_{eg} \gg (\Dfs, \Dhfs)$ and is therefore dominated by the scalar contribution.}.  Moreover, by time-reversal symmetry, the vector term is eliminated for linearly polarized trapping light.  For such conditions, the optical trap potential is regarded as being insensitive to the atomic spin.  Alternately, by reducing the laser detuning or using elliptically polarized light, the optical dipole potential becomes a tool for manipulating the atomic spin.

In optical traps, the confinement strength, and hence the extent of the trapped gas, along different dimensions can be varied.  An extreme elliptical focus produces pancake-shaped gases, with the diameter being far smaller along one axis than along the others \cite{sadl06symm}.  A single-beam red-detuned optical trap with a tight circular focus can produce prolate gases with high aspect ratios \cite{sten98spin,schm04}.   Forming light traps using two independent intersecting beams allowed for tightly confining and nearly isotropic traps \cite{chan04,sche10}.  Under such confinement, spatial variations of the spin and also the density of the degenerate spinor gas can be made energetically prohibited along one, two, or three dimensions, respectively.  Such a dimensional restriction will apply if the spatial dimension of the gas along a given direction is smaller than the healing lengths, $\xi_\mathrm{sp} \sim (8 \pi n \Delta a)^{-1/2}$, and $\xi_n \sim(8 \pi n \bar{a})^{-1/2}$, pertinent to the spin and density, respectively.  Here, we take into account just the contact interactions, letting $\Delta a$ characterize the relevant scattering-length difference that gives rise to the spin-dependent interaction term and $\bar{a}$ ($>0$) the scattering length that gives rise to the compressibility.  Under the typical condition $|\Delta a| < \bar{a}$, the spin dynamics may be restricted to a lower dimensionality than the density dynamics.  Similar considerations apply to determining whether thermal excitations or the MDDI can lead to spin variations in each spatial dimension.

\subsection{Effects of applied fields}

With the atoms in an optical trap, an experiment is conducted by setting the initial state of the spin distribution and then allowing this state to evolve, either freely or under continued perturbation.  The chief tool for manipulating atomic quantum gases, and spinor gases in particular, is the application of external electric and magnetic fields.  Below, we point out some uses and implications of these applied fields.

\subsubsection{Static magnetic fields: the constraint of spin conservation and spin-orbit coupling}
\label{sec:linearzeeman}

The spin-dependent interactions of spinor Bose gases, defined by characteristic energies on the order of $h \times 100$ Hz, become far smaller than the linear Zeeman energy splitting between spin states already at magnetic fields of tens of $\mu$G.  Under the application of such fields, exothermic dipolar relaxation collisions would allow a spinor Bose gas to evolve toward the trivial lowest-energy state with uniform magnetization along the field direction.  Recent experiments on Cr spinor gases \cite{pasq11crspinor}, for which the dipolar relaxation rate is substantial \cite{wein98cr,hens03relaxation}, confirm this evolution toward a single Zeeman state.  Only when the magnetic field is stabilized to very weak strengths, at which either the thermal energy $k_B T$ or the spin-dependent interaction energy of a condensate becomes comparable to the Zeeman energy, does the gas populate additional Zeeman states.  Studies of non-trivial spinor-gas physics in high-magnetic-moment gases will therefore require fine magnetic field control and shielding.

However, for alkali spinor gases \cite{boes96decay,gert99dipolar} (other than cesium \cite{sodi98}), dipolar relaxation collisions are rare, and thus, in a uniform bias field, the magnetization of the gas along the field direction can be considered a conserved quantity.  In light of the (approximate) conservation of the longitudinal spin, the theoretical treatments for the ground states of the spinor Bose gas must be reconsidered.

One approach is to introduce spin conservation into the many-body Hamiltonian via a Lagrange multiplier $\lambda$ \cite{sten98spin}.  Let the spinor gas in the absence of an applied magnetic field be described by the Hamiltonian $\hat{H}_0$.  Considering also the linear Zeeman energy from a uniform magnetic field $\mathbf{B} = B_z \mathbf{z}$, we substitute
\begin{equation}
\hat{H}_0 \rightarrow \hat{H}_0 - (\hbar \gamma B_z - \lambda) \sum_i \hat{F}_{z,i}
\label{eq:constrainhami1}
\end{equation}
where $\gamma$ is the atomic gyromagnetic ratio, and the sum is taken over the atoms in the gas.  Let us consider the case where $\langle \sum_i \hat{F}_{z,i}\rangle$ is constrained to be zero.  By symmetry, we expect $\lambda = \hbar \gamma B_z$: the effect of the applied field has disappeared altogether!  Under this global conservation constraint, the evolution of the spinor gas magnetization is still quite rich, so that meaningful studies of the spinor Bose gas are still possible under non-stringent experimental conditions.

The insertion of a Lagrange multiplier is equivalent to treating the evolution of the spinor Bose gas in a frame rotating at the Larmor precession frequency $\omega_L = - \gamma B_z$.  In that rotating frame, the magnetic field appears to be absent, and the effects of spin-dependent interactions and/or spontaneous symmetry breaking are free to determine the ground-state or other configurations of the quantum fluid.  Viewed in the lab frame, however, this configuration undergoes rapid Larmor precession, with $\omega_L$ typically greatly exceeding other dynamical frequencies of the gas.  This rapid precession must be taken into account, for example, in probing the magnetization of the gas (Sec.\ \ref{sec:dispersive}).

While the linear Zeeman effect of a uniform magnetic field is effectively eliminated through global spin conservation, that of a non-uniform field $\mathbf{B}(\mathbf{r})$ is not.  If we may neglect spatial variations in the orientation of the magnetic field, e.g.\ in the case that the spinor gas is confined tightly along some symmetry axis of the magnetic field, the constrained Hamiltonian of Eq.\ (\ref{eq:constrainhami1}) is simply modified as follows:
\begin{equation}
\hat{H}_0 \rightarrow \hat{H}_0 - (\hbar \gamma B_z({\mathbf{r}}) - \lambda) \sum_i \hat{F}_{z,i}
\label{eq:constrainhami2}
\end{equation}
The global constraint provides a uniform adjustment to the applied magnetic field, but does not cancel out the field inhomogeneity.

This remnant variation was utilized to reveal features of the mean-field phase diagram of sodium spinor condensates  \cite{sten98spin}.  Considering just the gradient $B'_z$ of the field magnitude along the long axis of their quasi-one-dimensional spinor gas and adopting the local density approximation, the spinor state at the axial position $z$ was taken to represent the constrained equilibrium state for different values of $p(z) = \lambda - \hbar \gamma B_z(z) = p_0 + p' z$, with $p_0$ determined by the longitudinal magnetization of the sample. By comparison to the theoretical phase diagram (Fig.\ \ref{fig:spin1phasediagrams}, in which one varies also the quadratic Zeeman shift; see below), the researchers determined that the $F=1$ spinor gas of sodium experiences antiferromagnetic interactions.  Their approach is among the first applications of the now-common method of measuring slices through a phase diagram through local measurements of a gas in an inhomogeneous potential \cite{shin08phasediagram,ho10phasediagram,navo10eqstate,hori10}.  This method relies on the validity of the local-density approximation.  For the spinor Bose gas, this approximation holds when the locally determined equilibrium spin order varies on length scales larger than the spin healing length (on the order of microns).

Accounting for the near-conservation of longitudinal spin in a magnetic field with inhomogeneous orientation is more subtle.  It is convenient to describe the spinor wavefunction in a spatially dependent basis of spin states, by applying a spatially dependent spin rotation operator $\hat{R}({\bf{r}}) = \exp[ -i {\bf{\Omega}}({\bf{r}}) \cdot {\hat{\bf{F}}}]$ so that the basis states correspond to local eigenstates of the spin projection along the magnetic field.  Applied in this basis, the constrained Hamiltonian has the form of Eq.\ (\ref{eq:constrainhami2}) and also acquires a gauge field, with the momentum operator transforming as
\begin{equation}
\hat{\mathbf{P}} \rightarrow \hat{\bf{P}} - \sum_\mu {\bf{A}}_\mu \hat{F}_{\mu}
\end{equation}
with the sum taken over components of the spin vector.  The gauge field is defined through the relation
\begin{equation}
\sum_\mu {\bf{A}}_\mu \hat{F}_{\mu} = i \hbar \nabla_{\bf{r}} \hat{R}|_{\bf{r}}\, \hat{R}^\dagger({\bf{r}})
\end{equation}
The imposition of additional rotations about the local field direction in $\hat{R}({\bf{r}})$ corresponds to a gauge transformation of the gauge field $\bf{A}$.  This gauge field generally provides a form of spin-orbit coupling to the spinor gas.  As described by \citet{ho96}, such spin-orbit coupling may introduce vorticity to the ground state of a spinor Bose condensate.

This means of synthesizing a gauge field for a neutral-atom gas is related to that realized experimentally by the Spielman group \cite{lin09bfield,lin11efield}.  In that work, a pair of intersecting laser beams drives a steady Raman transition between hyperfine spin states.  In a rotating frame, this Raman coupling provides a spatially periodic effective magnetic field, which then leads to an effective spin-dependent gauge field by the mechanism described above.

Even in the absence of dipolar relaxation collisions, the magnetization of the spinor gas is generally no longer conserved in such inhomogeneous fields.  For example, atoms in magnetic spherical quadrupole traps may undergo spin flips when they cannot adiabatically follow the rapidly varying field orientation near the location of zero field.  In the case of pure magnetic trapping, such spin flips are known as ``Majorana losses,'' \cite{majo32} since the spin-flipped atoms are expelled from the trap.  For an optically trapped spinor gas, such Majorana transitions would provide a means of Zeeman relaxation.  However, if the magnetic fields within the volume of the spinor gas do not vary rapidly in their orientation, the absence of dipolar relaxation collisions will still permit the longitudinal magnetization -- properly defined at each locale -- to be conserved.

\subsubsection{Quadratic shifts}

The Zeeman shift of hyperfine spin states at low magnetic field is not strictly linear.  For alkali gases, one can evaluate the next-order term in the Zeeman shift given the Breit-Rabi Hamiltonian \cite{brei31} (Fig.\ \ref{fig:rbhyperfine}).  Measuring energies with respect to the $|m_z = 0\rangle$ state, we obtain the net single-particle quadratic Zeeman energy
\begin{equation}
\hat{H}_q = \mp \frac{\left( g_s \mu_B - g_I \mu_N \right)^2}{\Delta W \left(1 + 2 I\right)^2} B_z^2 \hat{F}^2_z= q \hat{F}^2_z,
\end{equation}
where $g_s\simeq 2$ is the g-factor of the electron, $g_I \mu_N/\hbar$ is the nuclear gyromagnetic ratio, $\Delta W$ is the hyperfine energy splitting, $B_z$ is the magnitude of the magnetic field, assumed to be oriented along $\mathbf{z}$, and the hyperfine spin is $F = I \pm 1/2$.

A similar quadratic energy shift can be obtained by other means as well.  For instance, a quadratic energy shift for the $F=1$ spinor gas is obtained using the ac Zeeman effect from a microwave drive tuned near the $|F=1; m_z=0\rangle \rightarrow |F=0,2;  m_z = 0\rangle$ hyperfine transition \cite{gerb06rescontrol}.  With this method, the sign of $q$ can be varied by using either positive or negative detuning.  Additionally, off-resonance linear polarized light will exert a quadratic shift -- the tensor portion of the ac Stark shift -- along the polarization axis \cite{cct72,sant07quadratic,jens09cancel}.  This optically induced shift can be spatially tailored at high resolution and rapidly adjusted using electro-optics.

This quadratic shift (which we will call a ``Zeeman'' shift regardless of its provenance) breaks rotational symmetry by favoring one spatial axis, but still preserves the lower $SO(2)$ symmetry about that axis.  This shift gives a significant experimental handle on the ground-state phase diagram of spinor Bose-Einstein condensates (Sec.\ \ref{sec:meanfieldground}) and on spin-mixing dynamics (Sec.\ \ref{sec:spinmixing}).

\subsubsection{Effects of radio-frequency, microwave, and Raman coupling between spin states}

Magnetic fields are also applied to manipulate the spin of the gas dynamically.  Transitions among the magnetic sublevels of the spinor gas are driven with rf magnetic fields.  Given the strong Zeeman sensitivity of the transition frequencies, the radio-frequency (rf) excitation is usually either pulsed or frequency-chirped across the spin resonance frequency for the purpose of rapid adiabatic passage.  When the differences between the various $\Delta m = 1$ transition frequencies are small compared to the Rabi frequency and the inverse of the pulse duration, e.g.\ at low magnetic fields where the quadratic Zeeman shift is small, the rf excitation effects a rotation of the spin.  When the different $\Delta m = 1$ transition frequencies are resolved, e.g.\ at high magnetic fields where the quadratic Zeeman shift is large, rf fields can be used for arbitrary unitary transformations of the atomic spin \cite{gior03}.

The internal state of the spinor gas may also be manipulated by driving transitions among the different hyperfine spin manifolds.  Here, even at low magnetic fields, hyperfine transitions for the different magnetic sublevels of the spinor gas are readily resolved.  Thus, chosen unitary transformations may be imposed without applying large magnetic fields and contending with the resulting eddy currents or hysteresis in the experimental chamber.

Internal-state transitions of the spinor gas can also be achieved optically \cite{wrig08raman}.  This method has the benefit of allowing fine spatial control over the coupling between states.  A striking use of this capability was the generation of spin-vortex-like structures in a sodium spinor condensate using Gauss-Laguerre light beams to impose the necessary phase variation across the gas \cite{lesl09skyrmion,wrig09sculpting}.

\subsection{Probes}
\label{sec:probes}
Here, we highlight measurement techniques tailored for studies of magnetism in spinor gases.

\subsubsection{Stern-Gerlach and time-of-flight analysis}

\begin{figure}[tb]
\begin{center}
\includegraphics[width=\columnwidth]{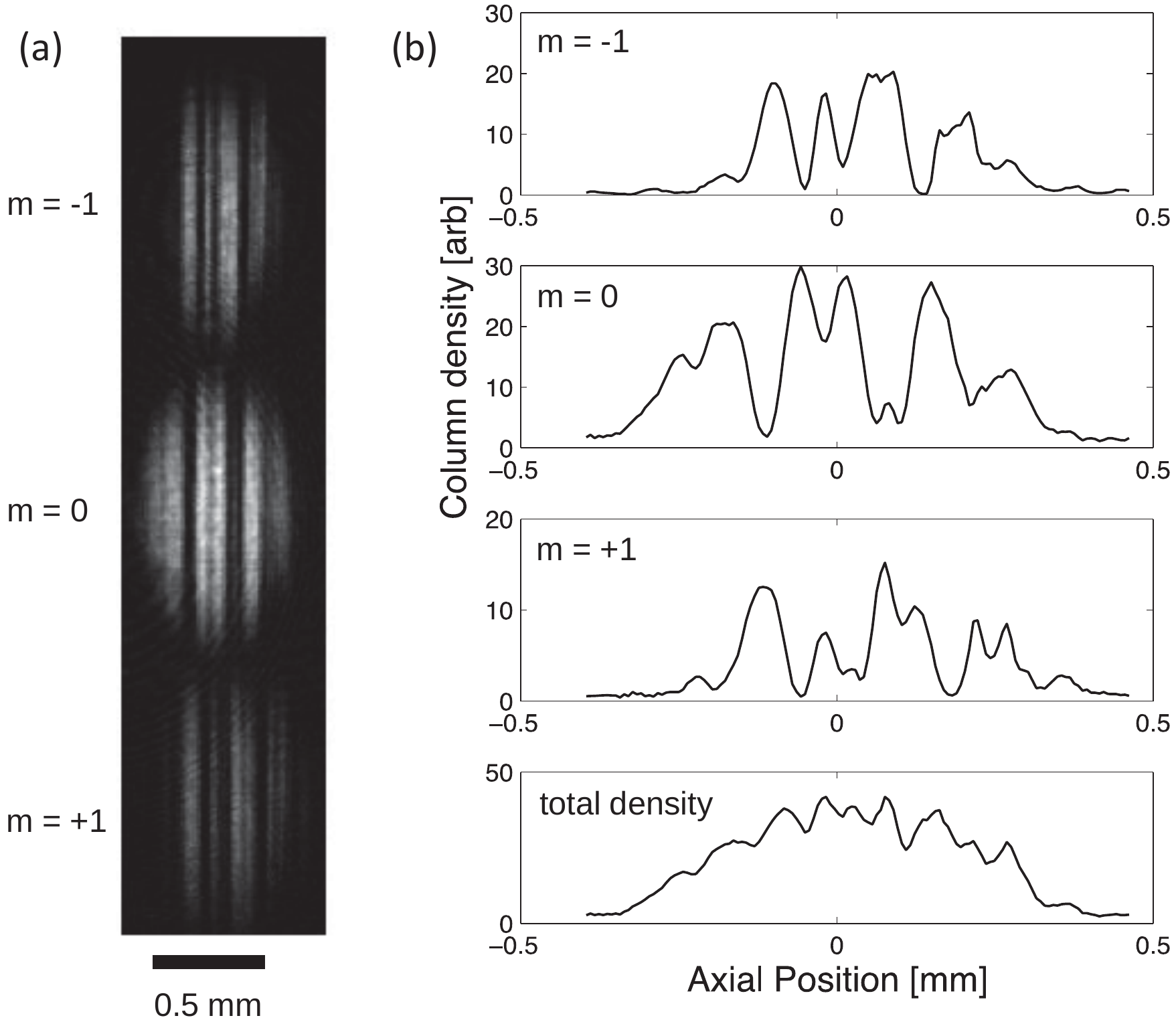}
\end{center}
\caption{Stern-Gerlach and time-of-flight analysis reveals the axial spin-state distribution of an elongated $^{23}$Na spinor Bose-Einstein condensate, 50 ms after a quench from positive to negative quadratic Zeeman shift.  The trapped condensate is prolate, with radial and axial dimensions of 4 and 270 $\mu$m (Thomas-Fermi radii).  The optical trap is extinguished, after which the condensate expands rapidly in the radial direction (vertical in image), so that the \emph{in situ} axial spin distribution (horizontal in image) is nearly preserved.  During the free expansion, a magnetic field gradient separates the Zeeman populations onto separate portions of the image plane (a), allowing the column density (b) of each population to be quantified.  Here, a condensate prepared initially in the $|m_z=0\rangle$ state undergoes a dynamical instability, evolving toward the transversely aligned polar state (as indicated by the correlation between the $|m_z=\pm 1\rangle$ distributions).  Figure prepared by Chandra Raman, based on \citet{book11}.}
\label{fig:sgtof}
\end{figure}

Characterizing the magnetic order in a spinor gas requires knowledge of both the populations in and the coherences among basis states of the atomic spin.  A straightforward method for measuring populations is releasing the gas from its optical confinement, using a magnetic field gradient to separate atoms in different magnetic sublevels into different spatial regions, and then imaging the separated portions of the gas; this is an amalgam of the Stern-Gerlach experiment \cite{gerl24} and time-of-flight analysis (Fig.\ \ref{fig:sgtof}).  In typical experiments, the magnetic field is varied adiabatically between the uniform field applied during optical trapping to the gradient field applied during the Stern-Gerlach separation; thus, one obtains a projective measurement of the distribution of atoms among the Zeeman sublevels in the trap.

Additionally, as in measurements of scalar quantum gases, the time-of-flight images carry information on the spatial, momentum, and energy distribution of atoms in the trap \cite{kett99var}.  The measurement resolution for the in-trap spatial distribution is limited by the minimum time of flight needed to separate the spin components, a minimum established not only by the maximum available magnetic field gradients, but also by the need to allow the gas to expand before applying field gradients so as to avoid interparticle scattering as the components separate.

By scattering many photons off each atom being imaged, the populations in the different sublevels can be determined with arbitrarily high accuracy. Sub-Poisson measurement uncertainty, i.e.\ with an imprecision $\Delta N < N^{1/2}$ for measurements on $N$ atoms, has been achieved by several groups \cite{book11correlations,haml12squeezing,luck11squeeze,gros11spinor} to probe correlations predicted from solutions for the  many-body spin Hamiltonian (Sec.\ \ref{sec:fragmentation}) and the spin squeezing expected to result from spin-mixing instabilities (Sec.\ \ref{sec:spinsqueezing}).

\subsubsection{Dispersive optical measurements}
\label{sec:dispersive}

To achieve higher spatial resolution while imaging the trapped and usually optically dense gas, it is advantageous to turn to dispersive imaging using off-resonant light, for which the refractive index of the cold gas is dominantly real \cite{andr96}.  The dispersion of light passing through the sample conveys information not only on the density of the atomic gas, but also on its spin polarization.  The real part of the linear optical susceptibility $\chi$ (which is unitless) for monochromatic illumination is a tensor quantity related to the reduced one-body, local density matrix $\rho(\mathbf{r})$ as
\begin{equation}
\chi_{ij}(\mathbf{r}) = \frac{1}{4 \pi \epsilon_0} \mbox{Tr}\left( \rho(\mathbf{r}) \sum_{g,e} \frac{1}{\hbar \Delta_{eg}} \hat{d}_i |e\rangle \langle e|\hat{d}_j | g\rangle \langle g| \right)
\end{equation}
where $i$,$j$ denote Cartesian directions, $\hat{\mathbf{d}}$ is the electric dipole operator, and $\epsilon_0$ is the vacuum permittivity.  Here we assume the population is entirely in the electronic ground state manifold.  The sum runs over the energy eigenstates of the ground-state manifold (labeled with $g$) and those of the electronic excited states (labeled with $e$), with $\Delta_{eg} = \omega - \omega_{eg}$ indicating the optical detuning for each transition.

To simplify this expression further, we assume the excited states are divided into several manifolds  of well defined angular momentum, and make the approximation that the detuning for transitions between the ground states (taken as a single spin manifold) and each of the excited-state manifolds is independent of the magnetic quantum numbers of each.  This approximation is well suited to spinor gases, studied under conditions of small Zeeman and Stark shifts compared to the optical detuning.  The susceptibility now becomes
\begin{equation}
\chi_{ij}(\mathbf{r}) =  \frac{1}{4 \pi \epsilon_0} \mbox{Tr}\left( \rho(\mathbf{r}) \hat{d}_i \left[ \sum_e \frac{|e\rangle \langle e|}{\hbar \Delta_{eg}} \right]\hat{d}_j \right)
\end{equation}
where the operator in square brackets is scalar.  As the expectation value of a dyadic tensor, the optical susceptibility is identified as a sum of spherical tensors of ranks 0, 1 and 2, i.e.\
\begin{equation}
\chi_{ij}(\mathbf{r}) = \chi^{(0)} \delta_{ij} n(\mathbf{r}) + \chi^{(1)} i \epsilon_{i j k} {\cal F}_k(\mathbf{r}) + \chi^{(2)} {\cal N}_{ij}(\mathbf{r}).
\end{equation}
Thus, the scalar, rank-1 and rank-2 components of the optical susceptibility tensor relate directly to the density $n(\mathbf{r})$, the vector magnetization $\mathbf{\cal F}(\mathbf{r})$ and nematicity ${\cal N}_{i,j}(\mathbf{r})$, respectively, of the spinor gas. The vector and tensor contributions give rise to circular and linear birefringence, respectively.  This optical sensitivity summarizes magneto-optical properties of atomic gases \cite{sute05book,budk02rmp}, and was expressed in relation to studies of spinor gases by \citet{caru04imag}.


One means of measuring the susceptibility tensor is to send collimated monochromatic probe light through the spinor gas.  In passing through a thin slice of the gas, the electric field of probe light propagating along the $\mathbf{z}$ direction is modified as
\begin{equation}
\mathbf{E}(x, y, z+dz) = \left[ 1 +  i k \left(1+ \frac{1}{2}\chi_\perp(x,y, z) \right) dz \right] \cdot \mathbf{E}(x,y,z)
\end{equation}
where $k$ is the wave number in vacuum.  The electric field lies in the $\mathbf{x}$-$\mathbf{y}$ plane, and only the $x,y$ elements of the susceptibility tensor are retained in $\chi_\perp$, owing to the transverse nature of the optical field.  When the dispersive phase shifts are small, and applying the thin-lens approximation, the field of the light after passing through the entire gas is approximated as
\begin{widetext}
\begin{equation}
\mathbf{E}_{\rm out}(x,y)  \simeq e^{i \bar{\phi}} \left[1 + \frac{i}{2} k_0 \left( \begin{array}{c c} \frac{\chi^{(2)}}{2}  \left(\tilde{\cal N}_{xx}(x,y) - \tilde{\cal N}_{yy}(x,y)\right) & i \chi^{(1)} \tilde{\cal F}_z(x,y) + \chi^{(2)}  \tilde{\cal N}_{xy}(x,y) \\
-i \chi^{(1)} \tilde{\cal F}_z(x,y) + \chi^{(2)}  \tilde{\cal N}_{xy}(x,y) & - \frac{\chi^{(2)}}{2} \left(\tilde{\cal N}_{xx}(x,y) -  \tilde{\cal N}_{yy}(x,y)\right)
 \end{array} \right) \right]\cdot \mathbf{E}_{\rm in},
\label{eq:disp}
\end{equation}
\end{widetext}
where the common phase shift is $\bar{\phi} = (k_0/2) [\chi^{(0)} \tilde{n} + \chi^{(2)} (\tilde{\cal N}_{xx} + \tilde{\cal N}_{yy})/2]$ and the tilde denotes the column averaged quantity along the optical axis.

\begin{figure}[tb]
\begin{center}
\includegraphics[width=\columnwidth]{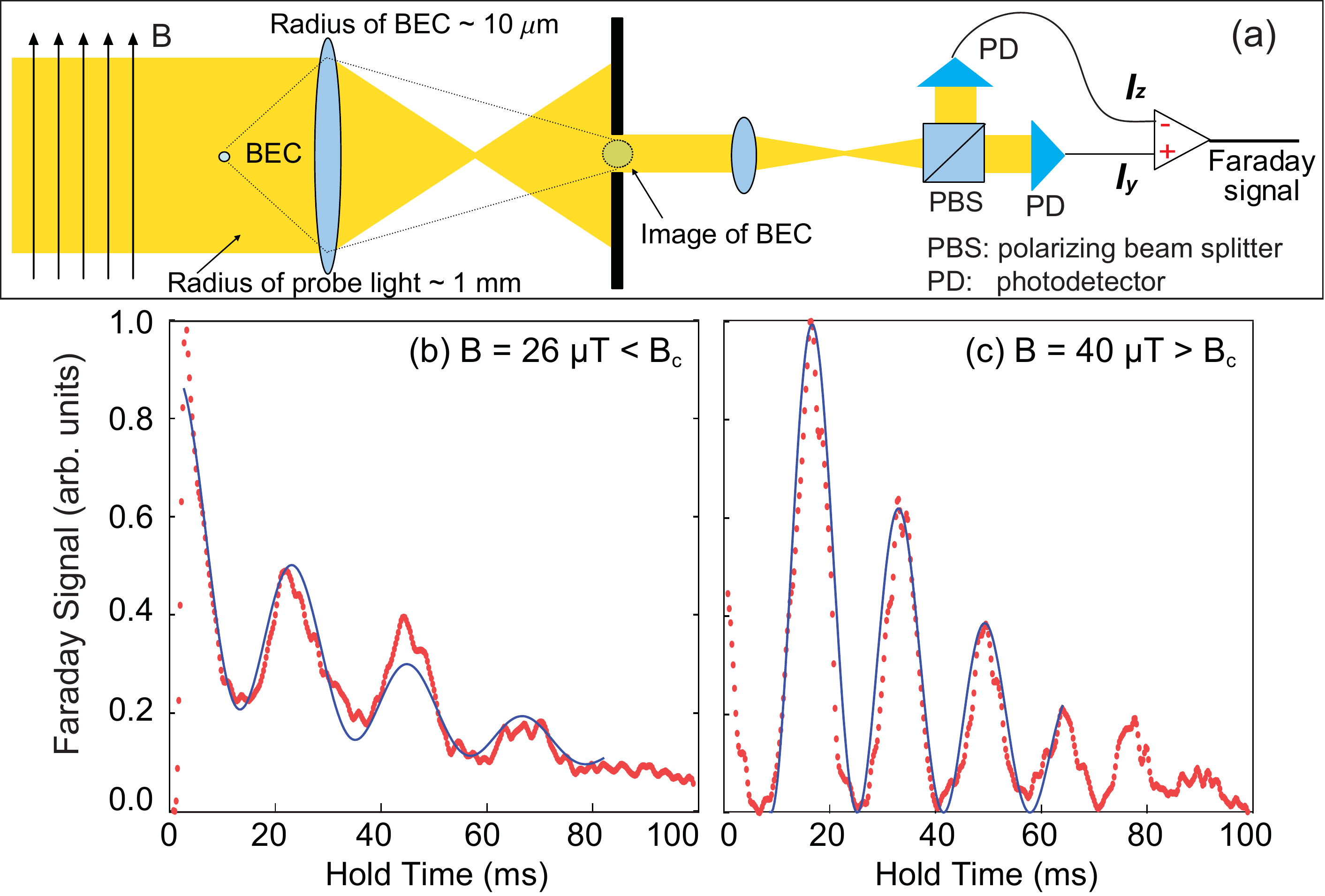}
\end{center}
\caption{(Color) Faraday rotation measurement, revealing spin-mixing dynamics in a $^{23}$Na spinor Bose condensate.  (a) Magnetization along the probe light axis causes the optical linear polarization to rotate, as measured by a photocurrent difference in a polarization analyzer.  With the condensate exposed to a magnetic field perpendicular to the imaging axis, the rotation signal measures the transverse magnetization, which oscillates at the Larmor precession frequency.  (b,c) The Larmor precession amplitude is recorded for a condensate that is prepared in the transversely magnetized state and allowed to evolve at constant $q$.  (b) For low $|q|$, attained at low magnetic field, the spin-dependent interactions cause the spin-mixing oscillations to follow an orbital trajectory around the transversely magnetized state; the condensate remains mostly magnetized, and, thus, the Larmor precession amplitude varies only slightly.  (c) In contrast, for high $|q|$, the condensate undergoes complete orientation-to-alignment conversion as expected in the absence of interactions.  This is indicated by the periodic oscillation between large precession amplitude, characteristic of the oriented (magnetized) state, and zero precession amplitude, characteristic of the aligned (unmagnetized) state.  Figure prepared by Yingmei Liu, adapted from \citet{liu09qpt}.}
\label{fig:lettfaraday}
\end{figure}

Dispersive optical detection has been used to measure the magnetization dynamics of $F=1$ spinor gases of $^{23}$Na \cite{liu09qpt} and $^{87}$Rb \cite{higb05larmor}.  The sodium experiments highlighted the capacity for performing weak continuous measurements on a single trapped spinor gas.  Strong optical confinement constrained the sodium spinor condensate to a single spatial mode.  Far detuned linear polarized light was passed once through the sample and then through a linear polarizer rotated $\pi/4$ with respect to the incident polarization in order to determine the circular birefringence.  Through Eq.\ (\ref{eq:disp}) and consistent with the stated assumptions, the measured signal is proportional to $k_0 \chi^{(1)} \tilde{\cal F}_z$, giving the magnetization component of the gas along the optical axis.  By imposing a magnetic field transverse to this axis, the rapid Larmor precession of the magnetization allowed both components of the transverse magnetization to be detected as a temporal oscillation of the optical rotation signal.  The amplitude of this oscillation (Fig.\ \ref{fig:lettfaraday}) gave a continuous measure of spin mixing dynamics in the polar spinor condensate \cite{liu09qpt}.  In experiments on nondegenerate gases, it has been demonstrated that a continuous Faraday rotation measurement, supplemented with imposed linear and quadratic Zeeman effects, can be used for complete characterization of the spin density matrix for arbitrary spin \cite{smit04cont}.  This powerful method may be extended to the study of spinor gases in future works.

\begin{figure*}[tb]
\begin{center}
\includegraphics[width=5 in]{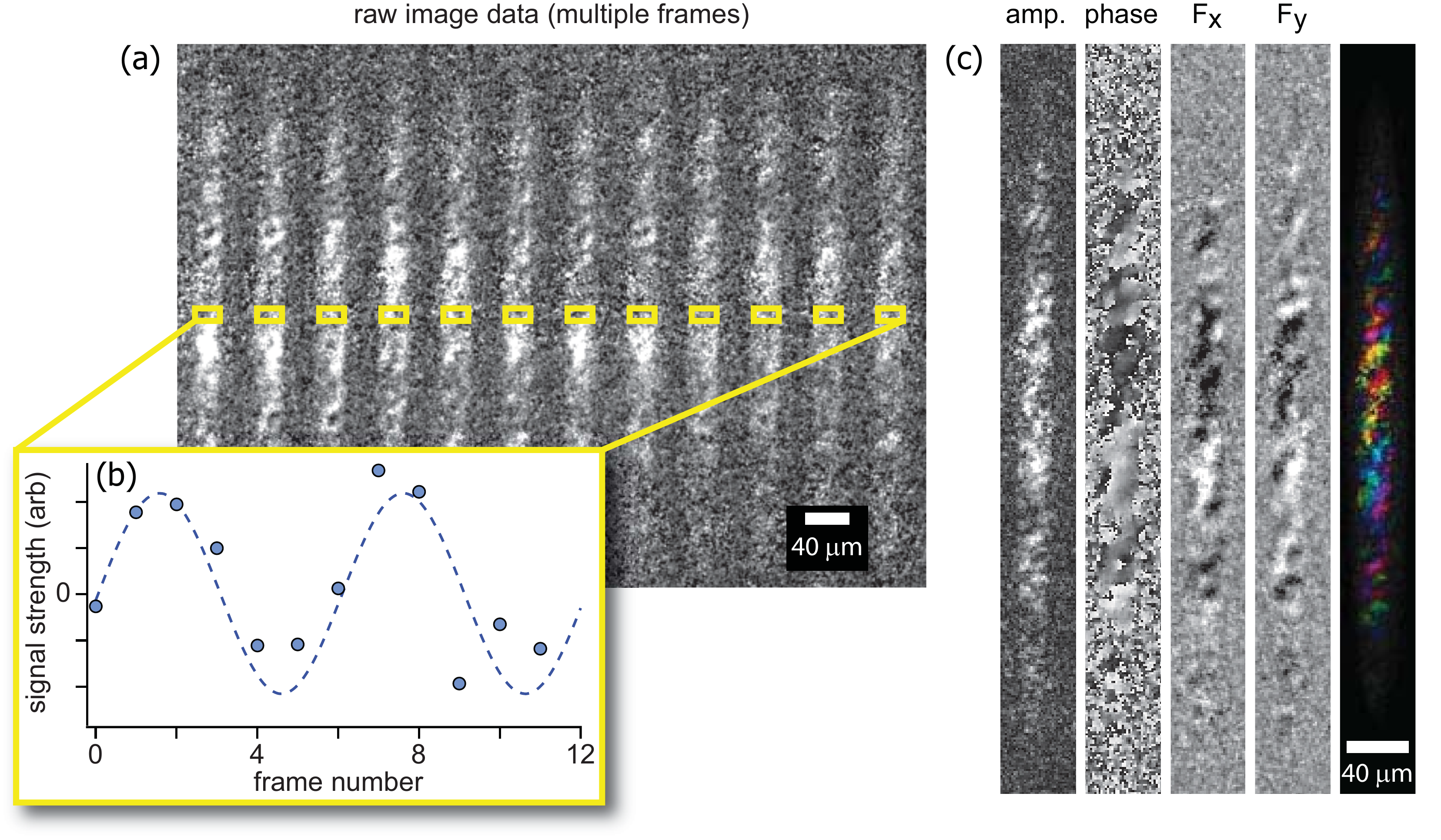}
\end{center}
\caption{(Color) Repeated polarization-contrast imaging is used to measure both components of the transverse magnetization of a spin-1 $^{87}$Rb gas. (a) Several brief polarization-rotation images are taken while the transverse magnetization undergoes Larmor precession between image frames.  (b) Such precession is seen in data selected from a common location in each image frame (yellow boxes in a).  (c) Analyzing data from each image pixel, one extracts the spatially varying Larmor precession amplitude and phase, or, equivalently, the two components of the transverse magnetization, $\tilde{F}_x$ and $\tilde{F}_y$.  These are represented either in gray scale, or in a color scale with brightness indicating amplitude and hue indicating orientation (color wheel shown in Fig.\ref{fig:equilibriumrb})).  This image analyzes a degenerate gas equilibrating to a ferromagnetic state \cite{guzm11}.}
\label{fig:larmor}
\end{figure*}

Applications of dispersive birefringent imaging to rubidium gases have highlighted the ability to measure the vector magnetization at high spatial resolution.  Circular polarized light was sent through the gas (along the $\mathbf{y}$ axis), and a polarization-independent phase-contrast method \cite{hech89} was used to convert the dispersive signal into an intensity image \cite{higb05larmor}.  Following Eq.\ (\ref{eq:disp}), the signal is proportional to $\chi^{(0)} \tilde{n} + \chi^{(1)} \tilde{\cal F}_y + \chi^{(2)} (\tilde{\cal N}_{xx} + \tilde{\cal N}_{zz})/2$.  To isolate the magnetization signal, a series of images was taken while the atomic spins underwent Larmor precession in a transverse applied magnetic field (along $\mathbf{z}$), with each probe pulse lasting a small fraction of the Larmor precession period.  The pixel-by-pixel amplitude and phase of the signal variation at the Larmor precession frequency (aliased by the slower frequency at which images were taken) gave a detailed spatial map of both the magnitude and orientation of the transverse magnetization (Fig.\ \ref{fig:larmor}).  In later work, a spatial map of the longitudinal magnetization was obtained also by applying a $\pi/2$ rf pulse in the middle of the imaging sequence \cite{veng10periodic}.  The nematicity signal, precessing at twice the Larmor frequency and with $\chi^{(2)}$ being small for the imaging settings, was ignored.

\subsubsection{Fundamental limits to dispersive measurements}

The back action of optical measurements of the atomic spin influences the gas being probed, necessitating a compromise between measurement precision and the fidelity of repeated measurements.  When measurements are made of several spin moments of the gas, the limitations to measurement fidelity are fundamental, arising from the non-commutativity of the measured observables.  These limitations have been studied extensively for measurements of the vector spin for gases of spin-1/2 atoms \cite{taka09} and also higher-spin atoms \cite{kuzm98,kuzm00}, and for pseudo-spin-1/2 systems \cite{schl10states,appe09,ried10entanglement,gros10nonlinear,este08squeeze, hamm10rmp}.  For measurements on a spinor gas that is fully magnetized along the $\mathbf{e}$ axis, standard quantum limits apply to measurements of the two magnetization projections transverse to $\mathbf{e}$, defining the condition when the requisite back action from measurements of one projection contributes significant fluctuations to the other.  This limit to the measurement variance is given as $(\Delta M_\perp)^2 = n V F/2$ where $n$ is the atomic density, $F$ is the (dimensionless) quantum number of the total angular momentum, and $V$ is the volume over which the magnetization is measured.

Dispersive measurements on optically thick gases are capable, in principle, of surpassing the standard quantum limits for spin detection non-destructively.  Dispersive measurements on spinor gases have yet to achieve this precision, being limited by collective optical scattering from the optically dense gas \cite{veng07mag}.  However, the prospect of doing so is compelling.  Analyzing the fluctuations in a low-noise dispersive optical measurement may distinguish between the many-body correlated spin states of spinor gases \cite{ecke07}.  Similar to the cases cited above for spin- and pseudo-spin-1/2 atoms, quantum non-demolition measurements of the atomic spin provide a means to spin squeezing. Achieving such measurements in high-resolution dispersive imaging would allow one to transform a spinor Bose-Einstein condensate into a non-classical, spatially extended quantum field \cite{sau10njp}.  It is interesting to consider how dispersive imaging, or other non-destructive detection methods, can achieve squeezing and entanglement that is specific to ensembles of high-spin particles, criteria for which have been specified by \citet{vita11arb} (see also Sec.\ \ref{sec:spinsqueezing}).

\subsubsection{Cavity aided detection and manipulation}

Placing a dispersive medium inside an optical resonator leads to an enhancement of the dispersive optical phase shift by the cavity finesse.  The cavity-enhanced detection of the magnetization of a cold atomic gas enabled a recent implementation of magnetic resonance imaging for optically trapped ultracold atomic gases.  The atomic magnetization was measured with an imprecision below the Poisson limit, although not as a quantum non-demolition measurement due to perturbations to the atomic motional states \cite{brah11mri}.  This sensitivity is sufficient to detect quantum-correlated ground states of spinor Bose condensates according to recent proposals \cite{cui08fluc,zhan09probing}.

Optical cavities have also been examined as a means to influence the dynamics of spinor Bose gases.  Variations in the atomic spin populations alter the optical susceptibility of the gas, changing the resonance frequency of differently polarized optical cavity modes.  In turn, for an externally driven cavity, these changes modulate the cavity fields and thereby act back on the atomic spin dynamics through the varying ac Stark shifts.  This coherent back action is predicted to lead to nonlinearities and bistability in spin-mixing dynamics \cite{zhou09bistab,zhou10cavity}.

Missing from these theoretical descriptions is a full accounting for the effects of quantum noise and quantum measurement. Aside from influencing the atomic spin dynamics, the cavity field also effects a continuous measurement of the atomic spin.  Such measurement requires a back action on the atomic spin, which comes typically from the quantum-optical (shot noise) fluctuations of the cavity field.  Some aspects of this back-action, in terms of disrupting the otherwise coherent spin-mixing dynamics, are discussed by \citet{zhan11backaction}.  Cavity quantum electrodynamics offers a clean setting in which to explore such measurement back action effects in future experiments.

%% file: rmp_ch_groundstates.tex
\section{Ground-state properties of spinor Bose gases}
\label{sec:groundstates}

\subsection{Mean-field ground states}
\label{sec:meanfieldground}

Gathering energy terms discussed in the previous sections, and continuing to ignore the MDDI (see later discussion in Sec.\ \ref{sec:mddi_sma}), the total Hamiltonian of the spinor Bose gas is given by
\begin{eqnarray}
\hat{H} & =& \int d\mathbf{r}\sum_m\hat{\psi}_m^\dagger(\mathbf{r})\left[-\frac{\hbar^2\nabla^2}{2 M}+U(\mathbf{r})
\right]\hat{\psi}_m(\mathbf{r}) \nonumber \\
& &
+\hat{H}_{\rm Z}+
\hat{V},
\label{3.1}
\end{eqnarray}
where $U$ is the trapping potential, assumed to be scalar, and $\hat{H}_Z$ describes the linear and quadratic Zeeman effects given by
\begin{equation}
\hat{H}_{\rm Z}\!=\!\int d\mathbf{r}\sum_{m_1 m_2}\hat{\psi}_{m_1}^\dagger(\mathbf{r})
\left[
p \hat{F}_{z} + q \hat{F}^2_z \right]_{m_1, m_2}
\hat{\psi}_{m_2}(\mathbf{r}),
\label{3.2}
\end{equation}
where we have assumed that the magnetic field is oriented uniformly along $\mathbf{z}$, and that the quadratic Zeeman shift also selects the $\mathbf{z}$ axis.

To focus only on the spin-dependent energies, let us make the single-mode approximation (SMA)~\cite{law98spin2} whereby we express field operators as $\hat{\psi}_m(\mathbf{r})=\Phi(\mathbf{r})\hat{\psi}_m$, with $\Phi(\mathbf{r})$ defining the spatial mode of the Bose-Einstein condensate.  This approximation is valid so long as the energetic cost of spatial variations in the spin state of the gas exceed any consequent reductions in the spin-dependent energies, or equivalently, if the system's size is much smaller than the spin healing length $\xi_{\rm sp}$. By the discussion of Sec.\ \ref{sec:opticaltraps}, such a condensate is zero-dimensional for spin degrees of freedom. When the system's size is larger than $\xi_{\rm sp}$, the SMA may also be applied locally within the local density approximation.  The ground states obtained under the SMA can thus clarify how spin-dependent interactions are locally minimized, even in spatially extended samples.

\subsubsection{Spin-1 case}

Within the SMA, the spin-dependent part of the Hamiltonian for a spin-1 Bose gas reads as
\begin{eqnarray}
\hat{H}^{(1)}&=& \frac{c_1^{(1)} n}{2} \sum_{\substack{m_1, m_2 \\ m_3, m_4}} \hat{\psi}^\dagger_{m_1} \hat{\psi}^\dagger_{m_2} {\mathbf{F}}_{m_1 m_3} \cdot {\mathbf{F}}_{m_2 m_4} \hat{\psi}_{m_3} \hat{\psi}_{m_4} \nonumber \\
& +& \sum_{m_1,m_2} \hat{\psi}^\dagger_{m_1} \left[ p F_z+ q F_z^2\right]_{m_1, m_2}
\hat{\psi}_{m_2},
\end{eqnarray}
where $n = \int d^3 \mathbf{r} \, |\Phi^4| / \int d^3 \mathbf{r} \, |\Phi^2| \propto N$ is the average density, proportional to the number of atoms in the gas $N$.

If we ignore quantum fluctuations, and take expectation values with respect to the mean values of the mode operators, we obtain the mean-field energy functional
\begin{equation}
E^{(1)} = \frac{c_1^{(1)} n}{2} \langle \hat{\mathbf{F}} \rangle^2 + p \langle \hat{F}_z \rangle + q \langle \hat{F}_z^2\rangle.
\label{eq:spin1energyfunctional}
\end{equation}

%

The spin-dependent interaction term is rotationally symmetric.  For $c_1^{(1)}>0$, as for the $F=1$ spinor gas of $^{23}$Na, this term favors the manifold of polar states, for which the $\left|\langle \hat{\mathbf{F}} \rangle \right| = 0$.  Such interactions are often denoted as ``antiferromagnetic,'' though unlike solid-state antiferromagnets, there is no N\'{e}el order. For $c_1^{(1)} < 0$, pertinent to the $F=1$ spinor gas of $^{87}$Rb, this term favors the manifold of ferromagnetic states, for which $\left|\langle \hat{\mathbf{F}} \rangle \right| = 1$.  Both manifolds of states are inert because they depend only on the sign and not on the magnitude of $c_1^{(1)}$.

\begin{figure*}[tb]
\begin{center}
\includegraphics[width=4.5 in]{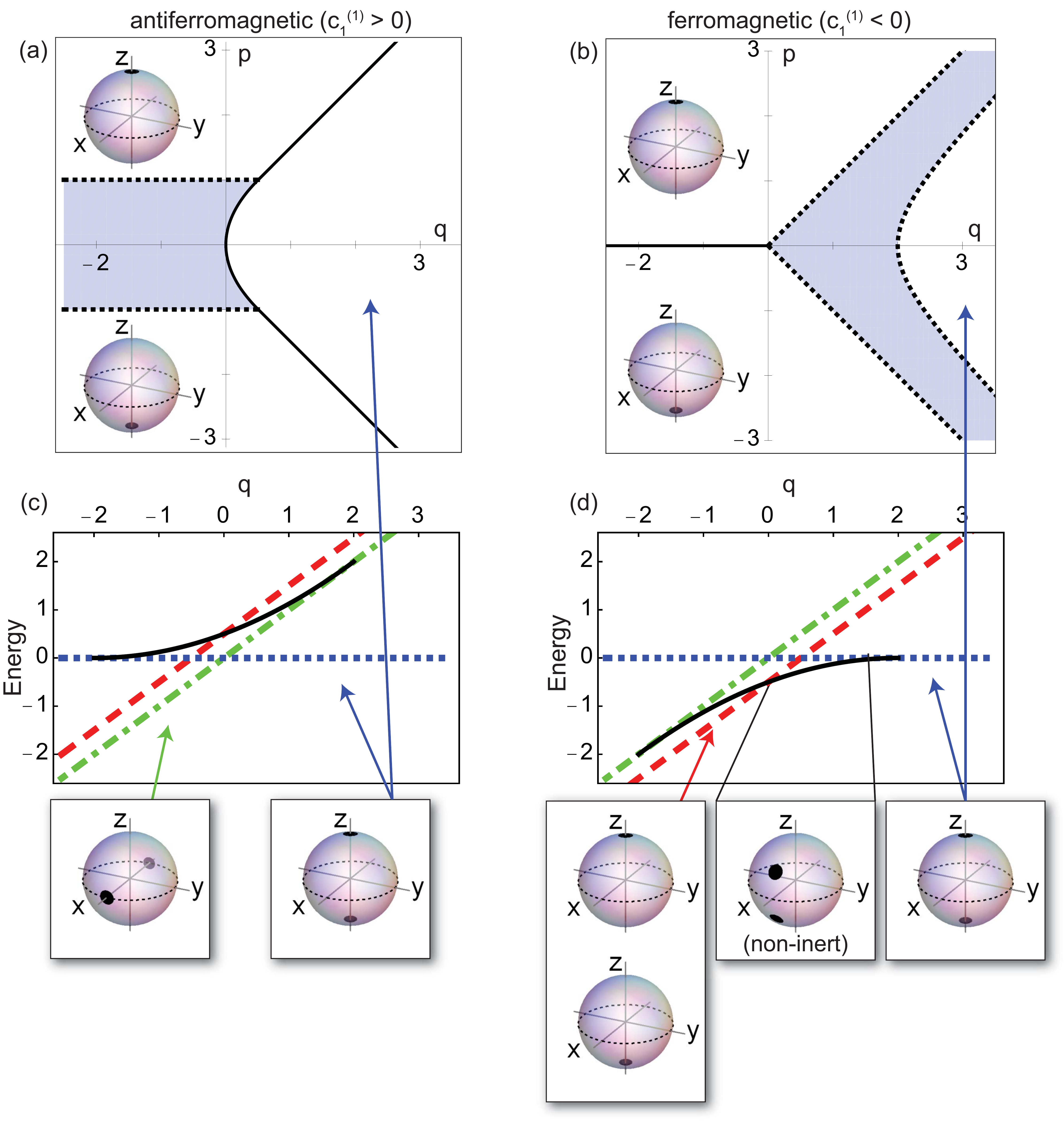}
\end{center}
\caption{(Color online) Mean-field phase diagram for $F=1$ spinor Bose condensates with (a,c) antiferromagnetic or (b,d) ferromagnetic s-wave interactions.  (a,c) Ground-states for variable linear ($p$) and quadratic Zeeman shifts ($q$), as derived by \citet{sten98spin}.  For the antiferromagnetic case, the solid curve parameterized by $p = \pm \sqrt{2 q}$ ($0<q<1/2$) and $p=\pm (q+1/2)$ ($q>1/2$) marks the sudden (first-order) transition from the longitudinal polar state ($|m_z=0\rangle$) to either one ($|p|\geq1$) or a superposition of both ($|p|<1$, shaded region) longitudinal ferromagnetic states.  The dashed lines indicate smooth (second-order) transitions in state composition.  For the ferromagnetic case, the dashed lines $p= \pm q$ ($q>0$) and $p = \pm \sqrt{q(q-2)}$ ($q>2$) bound the regions where the ground state transforms smoothly between the longitudinal polar and the longitudinal ferromagnetic states.  (c,d) Extrema of the mean-field energy functional for $p=0$,  plotted for three inert states -- the longitudinal ferromagnetic (red dashed line), the longitudinal polar (blue dotted), and the transverse polar states (green dot-dashed) -- and the non-inert extremal state for $|q|<2$ (black).  The antiferromagnetic ground state is either a longitudinal or transverse polar state for positive or negative $q$, respectively.  The ferromagnetic ground state is either the longitudinal polar state ($q>2$), the partially magnetized state ($0<q<2$) or the longitudinal ferromagnetic state ($q<0$).  Majorana representations are shown.  In quench experiments, a condensate is prepared in the longitudinal polar state at low $q$, an extremal high energy state that is subject to dynamical instabilities.  All energies scaled by $|c_{1}^{(1)} n|$.}
\label{fig:spin1phasediagrams}
\label{fig:spin1pzero}
\end{figure*}

The application of external fields, described by the effective linear and quadratic Zeeman energies $p$ and $q$, respectively, breaks rotational symmetry.  The minimum-energy states of $E^{(1)}$ are represented in Fig.\ \ref{fig:spin1phasediagrams}.

Let us consider the case of $p=0$, on which many experiments have focused.  Three classes of inert states attain extremal values of the quadratic Zeeman term, and are thus identified as extrema of $E^{(1)}$: the longitudinal ferromagnetic states $\psi = (1,0,0)^T$ and $(0, 0, 1)^T$; the longitudinal polar state $\psi = (0,1,0)^T$, and the transverse polar states $\psi = (1, 0, e^{i \phi})^T/\sqrt{2}$ for arbitrary $\phi$.  In the regime $|q|<q_0 = |2 c_1^{(1)} n|$, we identify an additional, non-inert energy extremum of partial transverse magnetization that arises from the competition between the spin-dependent interaction and quadratic Zeeman effect.  The energies of these extremal states are evaluated in Fig.\ \ref{fig:spin1pzero}.

For ferromagnetic coupling, the non-inert transversely magnetized state emerges as the lowest energy state at $q=q_0$, marking the point of a second-order phase transition that breaks axial symmetry.  At $q=0$, we observe a transition between easy-plane ($q>0$) and easy-axis ($q<0$) ferromagnetism.

Both transitions have been observed experimentally.  The onset of transverse ferromagnetism at $q = q_0$ was observed by the Chapman group \cite{chan04,chan05nphys}.  In a first experiment, the Zeeman populations of an $F=1$ $^{87}$Rb condensate were measured after the gas was prepared with zero longitudinal magnetization and allowed to equilibrate for several seconds at variable $q$.  At high $q$, the condensate equilibrated with its entire population in the $|m_z=0\rangle$ state, while in the range $0>q>q_0$, this fractional population diminished to about one half, as predicted by the mean-field treatment \cite{chan04}.  A second experiment provided evidence that this state of mixed Zeeman populations was indeed transversely magnetized.  For this, the equilibrium state at low $q$ was coherently manipulated by briefly pulsing a higher value of $q$, launching spin mixing dynamics, as described in Sec.\ \ref{sec:spinmixing}.  Such dynamics depend on the coherence between Zeeman populations, and confirmed the magnetized nature of the initial state \cite{chan05nphys}.

\begin{figure*}[tb]
\begin{center}
\includegraphics[width=7 in]{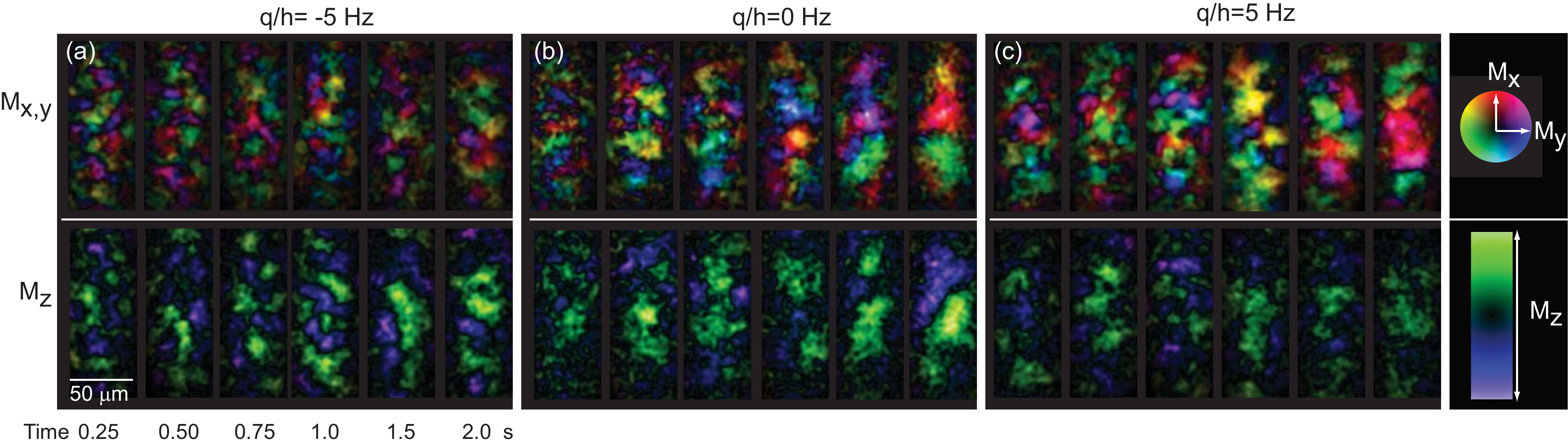}
\end{center}
\caption{(Color) Transverse (top) and longitudinal (bottom) magnetization of an $F=1$ $^{87}$Rb spinor Bose gas that was prepared initially in a nondegenerate incoherent $(1/3, 1/3, 1/3)$ population mixture of the Zeeman states and then cooled to quantum degeneracy.  The degenerate spin texture evolves for variable time at a quadratic shift of $q/h=$ -5 Hz (a), 0 Hz (b), and 5 Hz (c) before the column density of the vector magnetization is measured. The transverse magnetization is represented using the color wheel shown, where the hue indicates the magnetization orientation and the brightness its magnitude.  The longitudinal magnetization is represented by the color bar. The maximum brightness is set by a fully polarized condensate at 2 s evolution time.  Textures form small domains of commonly oriented magnetization and then coarsen to where only a few domains span the condensate. The late-time images show a spin-space anisotropy;  for positive $q$ the transverse magnetization appears brighter while for negative $q$ the opposite trend is observed.  The condensates contain $\sim 10^6$ atoms and are two-dimensional with respect to spin dynamics.  Figure reproduced from \citet{guzm11}.} \label{fig:PCvstime}
\label{fig:equilibriumrb}
\end{figure*}

The transition between easy-axis and easy-plane ferromagnetism at $q=0$ was measured by the Berkeley group, who applied \emph{in-situ} magnetization-sensitive imaging to probe the spin structures formed in $^{87}$Rb spinor condensates after several seconds of equilibration (Fig.\ \ref{fig:equilibriumrb}).  The observed textures contained large commonly magnetized spin domains with the predicted spin-space anisotropy \cite{guzm11}.

The antiferromagnetic interactions of $F=1$ $^{23}$Na condensates were identified by allowing such condensates to equilibrate in the presence of a magnetic field gradient.   \citet{sten98spin} observed that, for sufficiently large $q>0$, the spatial extent of the $|m_z = 0\rangle$ population of the condensate was larger than predicted for spin-independent interactions, demonstrating that spin-dependent interactions make the longitudinal polar state lower in energy than the longitudinal ferromagnetic states.

A second signature of antiferromagnetic interactions is the immiscibility of the $|m_z = 0\rangle$ and $|m_z=-1\rangle$ (or $|m_z = +1\rangle$) states \cite{sten98spin,mies99meta}.  This immiscibility (Fig.\ \ref{fig:miesner}) results from the relation $a_{i j} > \sqrt{a_{ii} a_{jj}}$ between the self- and cross-scattering s-wave scattering lengths for two equal-mass distinguishable atomic states labeled $i$ and $j$.  For the $F=1$ spinor system, $a_{-1 -1} = a_{0,-1} = a_2$ and $a_{0 0} = (2 a_2 + a_0)/3$.  For antiferromagnetic interactions, $a_0 > a_2$ and the immiscibility condition holds; in the diagrams of Fig.\ \ref{fig:spin1phasediagrams}, this immiscibility is reflected in the sharp transitions between the $|m_z = \pm1\rangle$ and $|m_z = 0\rangle$ phases (for $q>0$).

Along the $p=0$ axis, the antiferromagnetic $F=1$ spinor condensate should always favor a polar ground state.  Its alignment is predicted to vary from longitudinal, for $q>0$, to transverse, for $q<0$, as has been observed experimentally \cite{book11}.

\subsubsection{Spin-2 case}

As in the spin-1 case, we assume that the system is spatially uniform and ignore the kinetic energy.
Within the SMA and neglecting the MDDI, we can write the spin-dependent part of the mean-field energy functional as~\cite{koas00,ciob00,ueda02spin2}
\begin{equation}
E^{(2)} = \frac{c_1^{(2)} n}{2} \langle \hat{\mathbf{F}} \rangle^2 +\frac{c_2^{(2)} n}{2} |\langle \hat{A} \rangle|^2
+ p \langle \hat{F}_z \rangle + q \langle \hat{F}_z^2\rangle.
\label{eq:spin2energyfunctional}
\end{equation}
Compared with the spin-1 case, the energy functional involves an additional term $\hat{A}_2\equiv\sum_{m=-2}^2(-1)^m\hat{\psi}_m\hat{\psi}_{-m}/\sqrt{5}$ which is the annihilation operator of a spin-singlet pair of $F=2$ atoms.

\begin{figure}[tb]
\begin{center}
\includegraphics[width=\columnwidth]{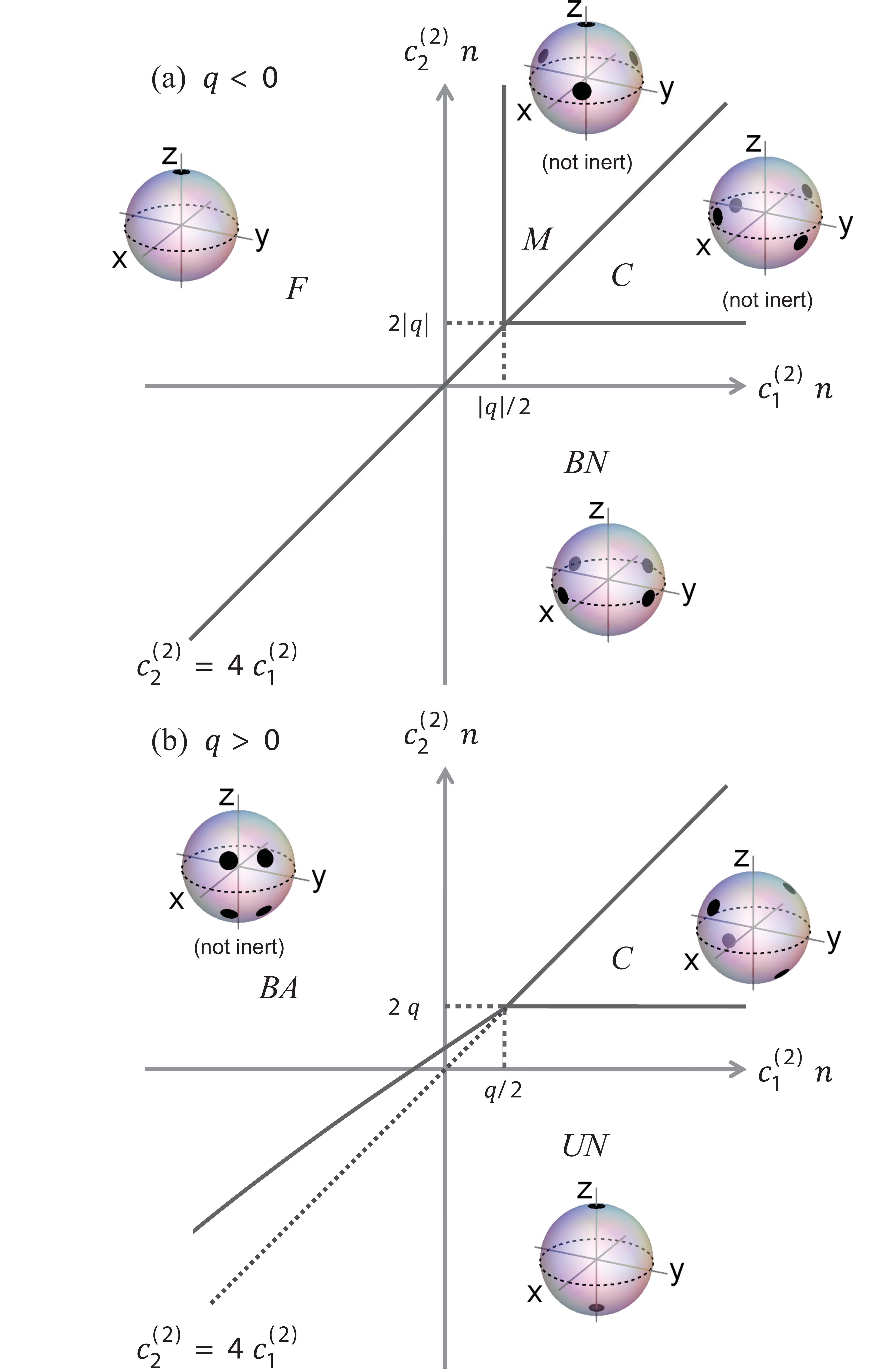}
\end{center}
\caption{(Color online) Phase diagrams of spin-2 Bose-Einstein condensates in the presence of the quadratic Zeemen
energy $q$. (a) Case of $q < 0$, where BN, C, F, and M stand for the
biaxial nematic, cyclic, ferromagnetic, and mixed phases, respectively.
In the absence of an external magnetic field ($q = 0$), the C and M
states can be transformed to each other by a rotation in space. (b) Case
of $q > 0$, where the UN and BA stand for the uniaxial nematic and
broken-axisymmetric phases, respectively. The broken-axisymmetric phase
has a transverse magnetization and a spin-singlet pair amplitude, both
of which depend on $q$, as in the case of the spin-1 broken-axisymmetric
phase.  The thick lines show the phase boundaries. The Majorana
representation is inserted in each phase.
Figure adapted from \citet{uchi10lhy}.}
\label{fig:spin2phasediagram}
\end{figure}

The mean-field phase diagram of a spin-2 Bose-Einstein condensate is summarized in Table \ref{tab:spin2} and Fig.\ \ref{fig:spin2phasediagram}.  Consider first the ground state phases in the absence of the quadratic Zeeman energy.  When $c_1^{(2)}<0$ and $c_2^{(2)}>4c_1^{(2)}$, the interaction energy is minimized by maximizing the magnetization and minimizing the spin-singlet amplitude, so that the ground state is ferromagnetic.  When $c_2^{(2)}<0$ and $c_2^{(2)}<4c_1^{(2)}$, the interaction energy is minimized by maximizing the spin-singlet amplitude and minimizing the magnetization, so that the ground state is antiferromagnetic with the spinor given by $(0,0,1,0,0)^T$ (and the family of states generated therefrom by rotations). It is not appropriate to call this state polar because it has no polarity.  In fact, the order parameter of this phase is proportional to the $m=0$ component of the spherical harmonic function of rank 2: $Y_2^0(\theta,\phi)\propto 3\cos^2\theta-1$ which is even under $\theta\rightarrow\theta+\pi$. In contrast, the spin-1 polar state $(0,1,0)^T$ does feature polarity, changing sign upon a rotation by $\pi$ about a transverse axis.

\begin{table*}[tb]
\begin{center}
\begin{tabular}{| l|c|c|}
\hline \hline
Phase & Order Parameter & Parameter Region \\
\hline  \hline
F ($q<0$) & $(1,0,0,0,0)$ & $c_2^{(2)}>4c_1^{(2)}, \
 \ c_1^{(2)} < \frac{|q|}{2n}$  \\
\hline
BN ($q<0$) & $\left(\frac{1}{\sqrt{2}},0,0,0,\frac{1}{\sqrt{2}}\right)$  & $c_2^{(2)}<4c_1^{(2)},
\   c_2^{(2)} < \frac{2|q|}{n}$  \\
\hline
M  ($q<0$) & $\left(\cos\theta_M,
0,0,\sin\theta_M,0\right)$  & $c_2^{(2)}>4c_1^{(2)}, \
c_1^{(2)}>\frac{|q|}{2n},$  \\
\hline
C   & $\left(\frac{\sin\theta_C}{\sqrt{2}},
0,-i\cos\theta_C,0,\frac{\sin\theta_C}{\sqrt{2}}\right)$  & $c_2^{(2)}<4c_1^{(2)},
\ c_2^{(2)} >\frac{2|q|}{n}$  \\
\hline
BA ($q>0$) & $(\pm a,b,c,b,\pm a)$ &  + (-) \
\text{sign for}\  $c_1^{(2)}<0\ (>0)$   \\
\hline
UN ($q>0$) &$(0,0,1,0,0)$ & \text{numerically determined}  \\
\hline  \hline
\end{tabular}
\caption{Definitions of the various states and those of
the parameter regions in which they are expected to be the ground state, where
F, BN, M, C, BA, and UN stand for ferromagnetic, biaxial nematic, mixed, cyclic, broken-axisymmetric, and uniaxial nematic phases, respectively, $q$ is the quadratic Zeeman coefficient, and $\cos\theta_M=\sqrt{1/3-q/(3c_1^{(2)}n)}$and $\cos\theta_C=\sqrt{1/2+5q/(c_2^{(2)}n)}$~\cite{sait05diag}.
The phase boundary between the BA and UN phases and the components $a,b,c$ are determined numerically~\cite{uchi10lhy}.}
\label{tab:spin2}
\end{center}
\end{table*}

The order parameter for the nematic phase is given by~\cite{ueda02spin2}
\begin{eqnarray}
\zeta^{\rm N}=\left(\frac{\sin\eta}{\sqrt{2}},0,\cos\eta,0,\frac{\sin\eta}{\sqrt{2}}\right)^T,
\label{nematic}
\end{eqnarray}
Depending on the value of $\eta$, this order parameter describes the uniaxial nematic (UN) phase ($\eta=0$), the biaxial nematic (BN) phase ($\eta=\pi/6$), and the dihedral-2 phase (otherwise)~\cite{uchi10lhy}. Although these states are degenerate at the mean-field level, zero-point fluctuations in the Bogoliubov modes lift the degeneracy, causing a first-order (order-by-disorder) quantum phase transition between the UN phase with $c_1^{(2)}>0$ and the BN phase with $c_1^{(2)}<0$~\cite{song07,turn07order}.  It is interesting to note that the parameter $\eta$ characterizes the geometric shape of the order parameter, as seen, for example, in the Majorana representations of the various states in Fig.\ \ref{fig:spin2phasediagram}. However, it is not related to a symmetry of the Hamiltonian, and thus does not lead to a conserved current, even though the eigenspectrum is degenerate with respect to $\eta$.

When $c_1^{(2)}>0$ and $c_2^{(2)}>0$, neither ferromagnetic nor antiferromagnetic order is energetically favorable. Rather, the zero-field ground state is the state $(1/2,0,-i/\sqrt{2},0,1/2)^T$ and the set of states generated therefrom by rotations.  This state, called either the cyclic or the tetrahedral state, has the symmetry of a tetrahedron under rotations and inversions.  As in the $F=1$ polar state, these discrete geometric operations must be accompanied by gauge transformations to leave the state unchanged; a fact that we discuss further in Sec.\ \ref{sec:textures}.  The cyclic/tetrahedral phase has neither spin-singlet amplitude nor magnetization, but it breaks time-reversal symmetry.

%

In the presence of an external magnetic field, the quadratic Zeeman term lifts the degeneracy between the UN and BN phases in favor of the UN (BN) phase for $q>0$ ($q<0$). The quadratic Zeeman effect also lifts the degeneracy of the two cyclic/tetrahedral phase described above, distorting these states into either $\zeta^{\rm C}=(\sin\theta/\sqrt{2},0,-i\cos\theta,0,\sin\theta/\sqrt{2})$ with $\cos^2\theta=1/2+5q/(c_2^{(2)}n)$ or $\zeta^{\rm M}=(\cos\theta,0,0,\sin\theta,0)$ with $\cos^2\theta=1/3-q/(3c_1^{(2)}n)$, in different portions of the phase diagram~\cite{sait05diag}. These phases retain either the original three-fold ($\zeta^{\rm M}$) or four-fold ($\zeta^{\rm C}$) symmetries, but they are distorted along the $\mathbf{z}$ direction by the quadratic Zeeman effect (see Fig.~\ref{fig:spin2phasediagram}). The phase $\zeta^{\rm C}$ remains nonmagnetic, whereas $\zeta^{\rm M}$ acquires magnetization proportional to $2\cos^2\theta-\sin^2\theta=-q/(3c_1^{(2)}n)$; the ``mixed" phase is named after such magnetic hybridization.


At present, the only experimentally realized spin-2 spinor Bose-Einstein condensate is a ${}^{87}$Rb condensate~\cite{kuwa04spin2,schm04,chan04}. Although this state is in the upper hyperfine manifold, the lifetime is as long as a few hundreds of milliseconds.  A theoretical calculation of the interaction parameters predicts that this condensate lies slightly on the antiferromagnetic side of the phase boundary between antiferromagnetic and cyclic/tetrahedral phases, i.e.\ that $c_1^{(2)}>0$ and $c_2^{(2)}$ is only slightly positive \cite{klau01rbspin}.

The coupling constant $c_1^{(2)}$ characterizes the spin-mixing processes with $\Delta m_F=\pm1$, while $c_2^{(2)}$ characterizes the processes with $\Delta m_F=\pm2$. \citet{schm04} observed the time evolution of the $F=2$ condensate by starting with the condensate with all atoms in the $|m_z=0\rangle$ state. The subsequent evolution shows that the $m_z=\pm1$ components first appear, followed by the appearance of the $m_z=\pm2$ components, indicating that $c_1^{(2)}$ is significantly larger than $c_2^{(2)}$. They also investigated the stability of various initial states against spin-mixing collisions, and found that the antiferromagnetic configuration $|m_z = 2\rangle+|m_z = -2\rangle$ is stable and that the cyclic/tetrahedral configuration shows nearly no spin dynamics, suggesting that the ground state is slightly on the antiferromagnetic side of the phase boundary.  However, it should be noted that their experiments were performed in the presence of the quadratic Zeeman effect of the order of a few nK which may well stabilize the antiferromagnetic configuration.

By placing a pair of atoms in each site of an optical lattice and letting them undergo spin-mixing oscillations, \citet{wide06precision} were able to determine the values of scattering length difference within the $^{87}$Rb $F=1$ and $F=2$ manifolds.  They suggested that ``the data points toward the antiferromagnetic ground state'' for the $F=2$ state with the reservation that the cyclic/tetrahedral ground state cannot be excluded because of the error bars.

\subsection{Exact ground states and fragmentation}
\label{sec:fragmentation}

In the previous discussion, we assume \emph{a priori} that the spinor Bose gas forms a Bose-Einstein condensate with a single macroscopically occupied single-particle ground state, described by the mean-field condensate spinor wavefunction.  However, in the presence of spin-dependent interactions, is it necessarily the case that such a mean-field, coherent spin-state condensate correctly describes the gas?

In the theory of \emph{scalar} Bose-Einstein condensates, the validity of the mean-field assumption was quantified by \citet{bogo47}.  Due to interparticle scattering, the many-body ground state of the Bose gas includes a non-zero population in the excited single-particle states.  However, this excited state population is kept small because of the additional single-particle energy required to occupy such states.  The fraction of atoms found outside the single-particle ground state, called the quantum depletion, remains small as long as the typical excited-state energy (scaling with density as $n^{2/3}$) is much larger than the typical interaction energy (scaling as $n a^2$).

However, for spinor Bose-Einstein condensates, the energetic barrier against the interaction-induced mixing of different many-body \emph{spin} states is absent.  For example, the spin-1 many-body Hamiltonian contains the spin mixing term given in Eq.\ (\ref{eq:spinmixingterm}).  The mean-field state of an $N$-atom, polar $F=1$ condensate aligned with the $\mathbf{z}$ axis, written as $|0, N, 0\rangle$ in the Fock-state $\hat{F}_z$ eigenbasis, couples by spin mixing with the state $|1, N-2, 1\rangle$.  The latter state overlaps with the polar mean-field state canted slightly from the $\mathbf{z}$ axis, a state that, for $q=0$, is equal in energy to the initial state.  Thus, spin mixing is seen to couple degenerate coherent spin states.  Following the arguments of Bogoliubov regarding the scalar Bose gas, one might therefore expect the quantum depletion due to such spin mixing to be massive.

Without making the \emph{a priori} mean-field approximation, \citet{law98spin2} showed that, within the SMA, the many-body Hamiltonian of a spin-1 condensate can be diagonalized exactly.  The spin-dependent interactions are symmetric both under rotation and under particle exchange; thus, naturally, the many-body s-wave spin-dependent scattering Hamiltonian can be written in terms of the total spin operator, $\hat{\mathbf{F}}_{\rm tot} = \sum_i \hat{\mathbf{F}}_i$, as
\begin{equation}
\hat{\hami} = \frac{c_1^{(1)} n}{2} \left( \frac{\hat{\mathbf{F}}_{\rm tot}^2}{N} - 2\right)
\label{eq:bigelowhami}
\end{equation}

The ground states of this Hamiltonian can be read off immediately.  An $N$-atom ferromagnetic $F=1$ condensate has a ground-state manifold described by the states of maximum $F_{\rm tot} = N$.  A constraint on the total longitudinal spin, e.g.\ $F_{z, \rm tot} = M_F$, is accommodated by selecting the corresponding magnetic sublevel of the $F_{\rm tot} = N$ manifold.  For example, in the case of strictly zero longitudinal magnetization, the exact ferromagnetic ground state is $|F_{\rm tot} = N, M_F=0\rangle$.  This state differs from the transversely oriented mean-field state by the exact equality between the $|m_z = +1\rangle$ and $|m_z = -1 \rangle$ single-particle Zeeman state populations.

The ground state of an antiferromagnetic $F=1$ condensate is more subtle.  At zero magnetization, the ground state, in the case that $N$ is even, is the unique state with $F_{\rm tot} = 0$.  Unlike the mean-field ground states, this many-body state retains rotational symmetry.  The character of this state is clarified by constructing the many-body spin state using the two-particle spin-singlet creation operator (for $F=1$ atoms) $\hat{A}_1^\dagger = \left( \left.\hat{\psi}_0^\dagger\right.^2 - 2 \hat{\psi}^\dagger_{+1} \hat{\psi}^\dagger_{-1}\right)/\sqrt{3}$.  Generally, one finds the many-body spin state with atom number $N = F_{\rm tot} + 2 N_2$, spin $F_{\rm tot}$, and longitudinal spin $M_F$ is written as \cite{koas00,ho00frag}
\begin{equation}
|N, F_{\rm tot}, M_F\rangle = Z^{-1/2} \left. \hat{A}_1^\dagger\right.^{N_2} \hat{F}_{-}^{F_{\rm tot}-M_F} \left. \hat{\psi}^\dagger\right.^{F_{\rm tot}}|\mathrm{vac}\rangle
\label{eq:koashistate}
\end{equation}
where here $\hat{F}_{-}$ is the collective spin lowering operator, $Z$ is a normalization constant, and $|\mathrm{vac}\rangle$ is the zero particle state.  The state with $|F_{\rm tot} = 0, M_F=0\rangle \propto \left. \hat{A}_1^\dagger\right.^{N/2} |\mathrm{vac}\rangle$ is thus seen as one in which all particles form spin-singlet pairs.  From the form of $\hat{A}_1^\dagger$, one sees that all single-atom magnetic sublevels in the many-body spin state are equally populated, a result also understood from the isotropy of the spin-singlet state.  This distribution is distinctly different from that expected for a mean-field polar state.  The sublevel populations of the many-body ground state also exhibit large fluctuations, with the standard deviation of the order of $N$ \cite{ho00frag}.

According to the definition of \citet{penr56}, a system is Bose-Einstein condensed if one of the eigenvalues of the reduced single-particle density matrix is extensive, i.e., of the order of the number of particles. \citet{nozi82} pointed out the possibility that there is more than one such macroscopic eigenvalue and called that state a fragmented Bose-Einstein condensate. The $F=1$ spinor condensate with $c_1^{(1)}>0$ at zero magnetic field is thus fragmented.  Indeed, this state can be characterized as a superposition of mean-field Bose-Einstein condensate states \cite{muel06frag},
\begin{equation}
|F_{\rm tot}=0, M_F=0\rangle \propto\int d\Omega \, |\theta,\phi\rangle,
\label{spinsinglet}
\end{equation}
where $|\theta,\phi\rangle$ is the polar coherent spin state aligned with the axis specified by the polar and azimuthal angles $\theta$ and $\phi$.

More generally, it has been shown that the spin state of an $F=1$ condensate within the SMA can be expressed in the form \cite{barn10rotor,barn11rotor}
\begin{equation}
\int d\Omega \, \psi(\theta, \phi)\, |\theta,\phi\rangle,
\label{rotorstate}
\end{equation}
where $\psi(\theta, \phi)$ is a properly normalized probability amplitude function on the unit sphere.  The state of the condensate is thus mapped onto that of a quantum rotor. When the quadratic Zeeman term $q$ is absent, $\psi(\theta, \phi)$ distributes uniformly over the entire unit sphere and the state (\ref{rotorstate}) reduces to Eq.~(\ref{spinsinglet}). When $q>0$ and the number of atoms is large, $\psi(\theta, \phi)$ tends to be localized, and the mean-field polar solution is favored.
It has been shown that the complete spectrum of the many-body Hamiltonian of Eq.\ (\ref{eq:bigelowhami}) with $c_1^{(1)}>0$ agrees with the lowest set of eigenvalues of a quantum rotor Hamiltonian. A similar representation has been developed for spin-2 condensates \cite{barn11rotor}.

It has been argued that fragmented Bose-Einstein condensates are fragile against symmetry-breaking perturbations~\cite{muel06frag}.  In the present case, it can be shown that when the longitudinal spin of the system is $M_F$, the population of the $|m_z = 0\rangle$ Zeeman state is calculated to be~\cite{koas00,ho00frag}
\begin{equation}
\langle N_0 \rangle
= \frac{N-M_F}{2 M_F+3}
\label{eq:mzerofragile}
\end{equation}
This result suggests that if the isotropy of the system is broken and the system acquires magnetization $M_F$, the $m_z=0$ population decreases rapidly; it decreases to $\sqrt{N}$ as $M_F$ increases to $\sqrt{N}$ and the distribution of populations rapidly approaches a mean-field result $(N/2,0,N/2)$. Thus the mean-field theory breaks down when the system is exactly isotropic, but it is rapidly restored in the presence of symmetry-breaking perturbations.

%% file: rmp_ch_textures.tex
\section{Spin textures of degenerate spinor gases}
\label{sec:textures}

A single coherent (i.e., unfragmented) Bose-Einstein condensate can be described by the order parameter which is given by the eigenfunction corresponding to the largest and extensive eigenvalue of the reduced single-particle density matrix. Here by extensive, we mean that the eigenvalue, which gives the number of condensate particles, is proportional to the total number of particles. Apart from the particle-number density distribution, a scalar Bose gas has only the phase degree of freedom whose spatial variation gives rise to mass current. A degenerate spinor Bose gas has the additional degrees of freedom of spin whose spatial variation produces not only mass currents but also spin currents.

Spatial variations of the spin configuration are called spin textures. Spin textures can vary over space and time in response to an external magnetic field and boundary conditions as well as due to spin-dependent contact interactions and the MDDI. Their rich structure allows spinor gases to accommodate a very rich variety of topological excitations such as vortices, monopoles, skyrmions, and knots.

\subsection{Order parameter manifolds}

Within the SMA phase diagrams developed above (Sec.\ \ref{sec:meanfieldground}), each condensate phase is represented by a standard spinor $\psi_0$. For example, the ferromagnetic and polar phases of a spin-1 BEC are represented by $(1,0,0)^T$ and $(0,1,0)^T$, respectively. In the absence of an external magnetic field and neglecting the MDDI, the free energy of the system is invariant under the $U(1)$ gauge transformations and the $SO(3)$ rotations in spin space. The entire group of symmetry transformations that leave the free energy invariant is given by $G=U(1)\times SO(3)$. Therefore, a general order parameter is given by performing a gauge transformation by the phase $\phi$ and rotating $\psi_0$ to an arbitrary direction specified by Euler angles $\alpha$, $\beta$, and $\gamma$: $\psi=e^{i\phi}U(\alpha,\beta,\gamma)\psi_0$, where $U(\alpha,\beta,\gamma)=e^{-i F_z \alpha}e^{-i F_y \beta}e^{-i F_z \gamma}$; we will use the subscript $\mathbf{S}$ to denote such geometric (spin) rotations. The dependencies of $\phi,\alpha,\beta$ and $\gamma$ on space and time characterize spin textures.

Each phase of the condensate has distinct symmetry characterized by the isotropy group $H$, whose elements $h$ leave $\psi$ unaltered: $h\psi=\psi$. Then, the order parameter space is given by the coset $M=G/H$. Acting every element $m$ of $M$ on the standard spinor generates a complete set of spinors, called an orbit, which represent the order parameter manifold~\cite{merm79}.

Let us consider the order parameter manifolds for several spinor condensate phases.  The order parameter of the $F=1$ ferromagnetic condensate transforms under arbitrary gauge transformations and spin rotations as
\begin{equation}
\mathbf{\zeta}^F = e^{i\phi}U(\alpha,\beta,\gamma)\left( \begin{array}{c} 1 \\ 0 \\ 0 \end{array}\right)
= e^{i(\phi-\gamma)}
\left( \begin{array}{c}
e^{-i\alpha}  \cos^2 \frac{\beta}{2}\\
\frac{1}{\sqrt{2}} \sin \beta\\
e^{i\alpha} \sin^2 \frac{\beta}{2} \end{array}
\right).
\label{eq:rotateferro}
\end{equation}
The crucial observation here is that the gauge angle $\phi$ and the rotation angle $\gamma$ about the spin direction appear as a linear combination $\phi-\gamma$, so that the order parameter is invariant under the simultaneous gauge transformation and the spin rotation about its orientation by the same amount. Therefore, $H=U(1)_{\phi+S_\gamma}$, where the subscripts $\phi+S_\gamma$ is displayed to emphasize the combined nature of the group operations.  The order parameter manifold is given by~\cite{ho98}
\begin{eqnarray}
M^{\rm F}
= \frac{U(1)_\phi \times SO(3)_{\bf S}}{U(1)_{\phi + S_\gamma}}
= SO(3)_{\phi, {\bf S}}.
\label{OPM_Ferromagnetic}
\end{eqnarray}

The polar phase of the $F=1$ spinor condensate transforms as follows:
\begin{equation}
\mathbf{\zeta}^P= e^{i\phi} U (\alpha,\beta,\gamma) \left( \begin{array}{c} 0 \\ 1 \\ 0 \end{array}\right)
=
e^{i\phi}
\left( \begin{array}{c}
-\frac{e^{-i\alpha}}{\sqrt{2}} \sin\beta \\
\cos\beta \\
\frac{e^{i\alpha}}{\sqrt{2}} \sin\beta \end{array}
\right).
\label{eq:rotatepolar}
\end{equation}
This expression illustrates that the polar state has a $U(1)$ symmetry with respect to rotation about the quantization axis because $\mathbf{\zeta}^P$ does not depend on $\gamma$.  Also, we see that rotating the $|m_z = 0\rangle$ polar state by $\pi$ about the $\mathbf{y}$ axis generates a minus sign, as can be seen by making the change $\beta\rightarrow\beta+\pi$ in Eq.\ (\ref{eq:rotatepolar}).  Combining this rotation with a gauge transformation by $\pi$ returns the original state.  Thus the polar state manifold has a discrete $\mathbf{Z}_2$ symmetry.  Altogether, the isotropy group is given as $H=U(1)_{S_\gamma}\times(\mathbf{Z}_2)_{\phi,S_\beta}$ and the order parameter manifold is given by~\cite{zhou01}
\begin{eqnarray}
M^{\rm P}
=\frac{U(1)_\phi \times SO(3)_{\bf S}}{U(1)_{S_\gamma} \times (\mathbf{Z}_2)_{\phi, S_\beta}}
=\frac{U(1)_\phi \times S^2_{\bf S}}{(\mathbf{Z}_2)_{\phi, S_\beta}}.
\label{OPM_Polar}
\end{eqnarray}
We note here that the spin and gauge remain coupled in a discrete manner.  As discussed in Sec.\ \ref{sec:topological}, this discrete spin-gauge symmetry causes the polar-state manifold to support half-quantum vortices~\cite{leon00alice}.

Spin-2 condensates have even richer order parameter manifolds. The ferromagnetic phase with $\psi_0=(1,0,0,0,0)^T$ has the same symmetry as that of the spin-1 case except that the order parameter manifold (Eq.\ (\ref{OPM_Ferromagnetic})) is divided by $(\mathbf{Z}_2)_{\phi,{\bf S}}$~\cite{make03defects} because the phase winding is doubled compared with the spin-1 case.
The uniaxial nematic phase with $\psi_0=(0,0,1,0,0)^T$ looks similar in form to the polar phase of a spin-1 condensate, but has one important difference. Both of these systems have the spin inversion symmetry about an arbitrary axis perpendicular to the quantization axis, but only the spin-1 order parameter changes sign upon inversion. Thus, only the spin-1 condensate can couple the spin inversion with the gauge and generate a half-quantum vortex accompanied by mass circulation. The biaxial nematic phase with $\psi_0=(1,0,0,0,1)^T$ has four-fold symmetry about the quantization axis and  $\pi$ rotation symmetry about four perpendicular axes. Therefore the isotropy group is the dihedral-four group $\left(D_4\right)_{\phi, \mathbf{S}}$, and the order parameter manifold is given by~\cite{song07}
\begin{eqnarray}
M^{\rm biaxial}
=\frac{U(1)_\phi \times SO(3)_{\bf S}}{(D_4)_{\phi,{\bf S}}}.
\label{OPM_BN}
\end{eqnarray}
The cyclic/tetrahedral phase has the tetrahedral symmetry with the isotropy group $H=\mathbf{T}$.  The order parameter manifold is thus given by~\cite{make03defects,seme07onethird}
\begin{eqnarray}
M^{\rm cyclic}
=\frac{U(1)_\phi \times SO(3)_{\bf S}}{({\mathbf T})_{\phi, {\bf S}}}.
\label{OPM_C}
\end{eqnarray}
In both these cases, as in the polar case, geometric rotations and inversions are accompanied by gauge transformations to return to the original quantum state. Because two elements in dihedral-four and tetrahedral groups do not commute in general, some vortices created in these phases are non-Abelian~\cite{koba09nonabelian}.
A list of symmetry groups of spin-1 and spin-2 condensates as well as liquid crystals and superfluid $^3$He systems is given by \citet{koba12abe}.

\subsection{Superflow and spin-gauge symmetry}
\label{sec:superflow}

In a spatially extended spinor Bose-Einstein condensate, the spin order parameter can everywhere minimize the local spin-dependent interactions that arise from s-wave scattering, but still be characterized by spatial variation in the order parameter within its degenerate manifold.  In the absence of magnetic dipole-dipole interactions, the resulting spatial pattern, a spin texture, is a low-energy phenomenon, with excess energy only due to the kinetic energy.  As exhibited in Eq.\ (\ref{eq:rotateferro}) for the $F=1$ ferromagnetic case, the combined spin-gauge symmetry or the order-parameter manifold implies that local spin rotations (spatially varying Euler angles) can be regarded equivalently as local variations in the condensate phase $\phi$, to which the condensate responds by supercurrent flow \cite{ho96}. Conversely, a supercurrent can decay by forming a spin texture.

For the $F=1$ ferromagnetic spinor gas, this connection is exhibited in the expression for the superfluid velocity:
\begin{eqnarray}
\mathbf{v}_s^F=\frac{\hbar}{M}[\nabla(\phi-\gamma)-\cos\beta\nabla\alpha],
\label{6.2}
\end{eqnarray}
where $M$ is the atomic mass.
Unlike for the scalar Bose-Einstein condensate, in which the superflow is irrotational ($\nabla \times \mathbf{v}_s=0$), here, $\mathbf{v}_s^F$ satisfies the Mermin-Ho relation \cite{merm76}:
\begin{eqnarray}
\nabla \times \mathbf{v}_s^F=\frac{\hbar}{M}\sin\beta\nabla\beta\times\nabla\alpha.
\label{6.3}
\end{eqnarray}
This result implies that the superflow in the ferromagnetic phase is not a potential flow, and consequently the circulation alone is not quantized.  Rather, the difference between the circulation along an arbitrary contour $C$ and the Berry phase enclosed by it is quantized:
\begin{equation}
\oint_C {\bf v}_s^F d{\bf r} - \frac{\hbar}{M}
\oint_C (1-\cos\beta)\nabla\alpha\,  d{\bf r}
=\frac{\hbar}{M} \times
\  \text{integer}.
\label{6.4}
\end{equation}

For the $F=1$ polar phase, the superfluid velocity is calculated to give
\begin{eqnarray}
{\bf v}_{\rm s}^{\rm P}
=\frac{\hbar}{M}\nabla\phi
\label{3.14}
\end{eqnarray}
which is irrotational, and hence the circulation alone is quantized.
It should be noted that although Eq.~(\ref{3.14}) assumes the same form as that of scalar condensates, the quantum of circulation is $h/2M$ which is half of the latter. This is because If the phase $\phi$ changes only $\pi$ around a loop, the single-valuedness condition can be met if the Euler angle $\beta$ changes by $\pi$ simultaneously, as can be seen from Eq.~(\ref{eq:rotatepolar}).

\subsection{Spin current and spin nematicity}

A spin current can flow in the absence of mass current by carrying the same number of particles with different spin states in the opposite directions. In the absence of an external magnetic field and the MDDI, the magnetization satisfies the continuity equation:
\begin{eqnarray}
\frac{\partial {\cal F}_\mu}{\partial t}+\nabla\cdot\mathbf{j}_\mu^{\rm spin}=0,
\label{spin_conteq}
\end{eqnarray}
where
\begin{eqnarray}
\mathbf{j}_\mu^{\rm spin}=\frac{\hbar}{2Mi}\sum_{m,n=-F}^F(F_\mu)_{mn}
\left[\psi_m^*\nabla\psi_n-(\nabla\psi_m^*)\psi_n
\right]
\label{spinvelocit}
\end{eqnarray}
is the spin current. However, additional terms that break spin conservation appear on the right-hand side of Eq.~(\ref{spin_conteq}) in the presence of an external magnetic field or the MDDI \cite{kudo10hydro}:
\begin{equation}
\frac{\partial {\cal F}_\mu}{\partial t}+\nabla\cdot\mathbf{j}_\mu^{\rm spin} =
\frac{c_{\rm dd}}{\hbar}(\mathbf{b}\times\mathbf{\cal F})_\mu+\frac{p}{\hbar}(\mathbf{z}\times\mathbf{\cal F})_\mu+\frac{2q}{\hbar}\epsilon_{\mu z\nu}{N}_{z\nu},
\label{spin_conteq2}
\end{equation}
where $c_{\rm dd}$ is the strength of the MDDI, $\mathbf{b}$ is the dipole field, $p$ and $q$ are the strengths of the linear and quadratic Zeeman effects, respectively, and $\epsilon_{\mu z\nu}$ is the Levi-Civita tensor, and $N_{\mu \nu}$ is the spin nematic tensor.  This equation shows that the dipole field and the linear Zeeman term generate spin torques and cause spin precessions, while the quadratic Zeeman term combined with spin nematicity produces a spin-nonconserving term.  This last term implies that even if the initial state of the system is nonmagnetic, a system having spin nematicity can dynamically develop transverse magnetization. Once the transverse magnetization grows, the MDDI leads to longitudinal magnetization due to the first term on the right-hand side of Eq.~(\ref{spin_conteq2}).

\subsection{Topological excitations}
\label{sec:topological}

Topological excitations are the defects that are stable against weak perturbations. Such stability is ensured by discrete quantum numbers that characterize the topology of the order parameter manifold. Naturally, the types of topological excitations depend crucially on the properties of the order parameter manifold which, in turn, is determined by the phase of the condensate. A rich variety of order-parameter manifolds of spinor condensates allow many distinct types of topological excitations. Examples include integer and fractional vortices~\cite{yip99,leon00alice,zhou01,seme07onethird,isos01qvort,make03defects}, non-Abelian vortices~\cite{koba09nonabelian}, 't Hooft-Polyakov~\cite{stoo01} and Dirac monopoles~\cite{blah76,ruos03monopole,volo03uni}, skyrmions (particle-like solitons)~\cite{volo77soliton,shan77,khaw01skyrmion}, and knots~\cite{kawa08knots}.

\begin{figure}[tb]
\centering
\includegraphics[width=0.8 \columnwidth]{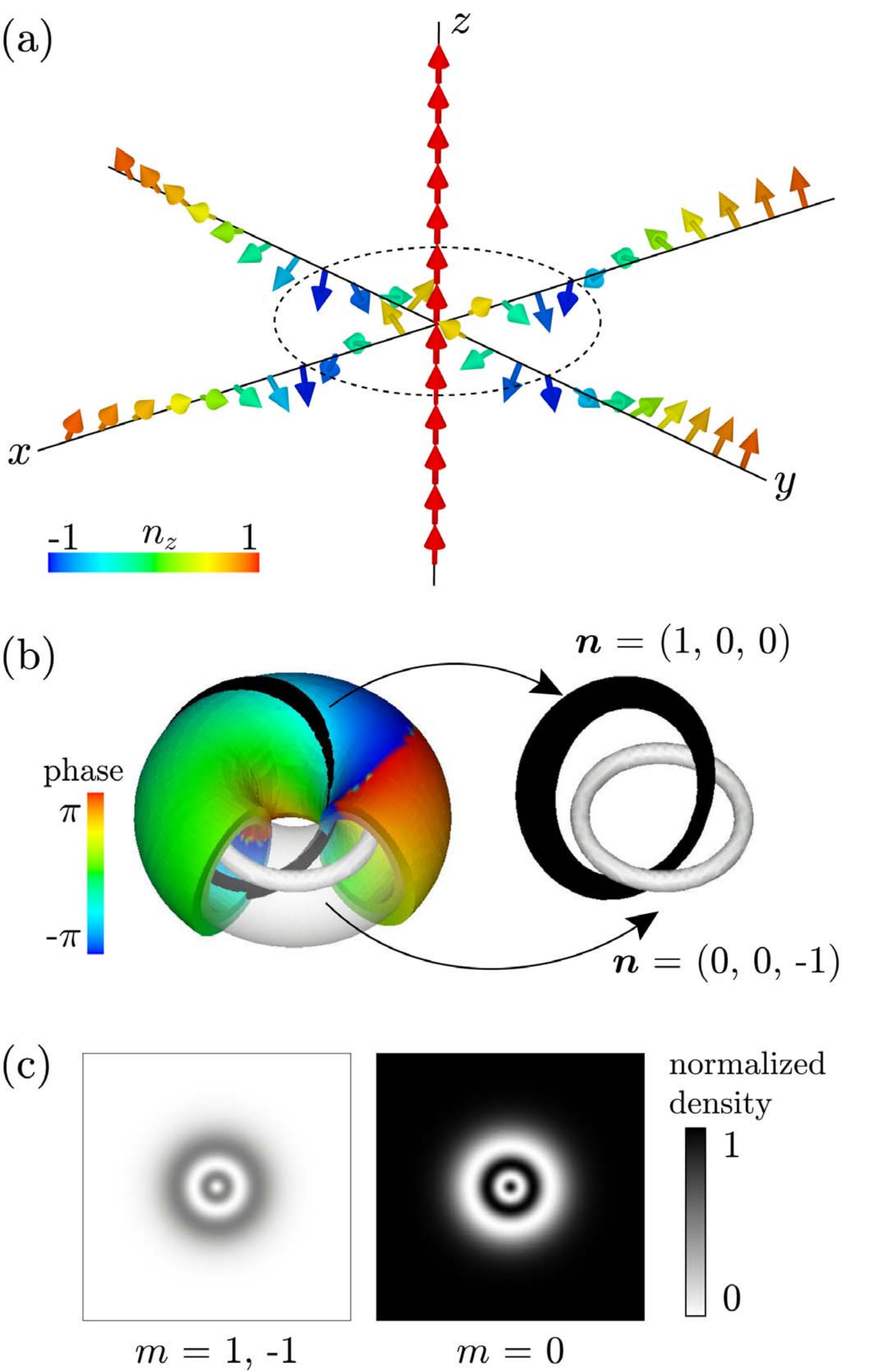}
\caption{(Color) (a) Three-dimensional configuration of the $\mathbf{n}$ field (shown as arrows) of a knot with the unit linking number, where $\mathbf{n}=(n_x,n_y,n_z)^T$ gives the normalized spinor order parameter as $e^{i\theta}((-n_x+in_y)/\sqrt{2},n_z,(n_x+in_y)/\sqrt{2})$. In the figure, only $\mathbf{n}$ on the $\mathbf{x}$, $\mathbf{y}$, and $\mathbf{z}$ axes are shown. The dashed circle follows the direction $\mathbf{n}=(0,0,-1)^T$. The color represents the value of $n_z$ according to the scale shown.  (b) The torus on the figure shows the isodensity surface of $|\psi_{-1}|^2=0.47$ with the color showing the phase of $\psi_{-1}$ according to the color gauge on the left.  The white and black tubes on the right are the cutouts on which $\mathbf{n}$ points in the directions $(0,0,-1)^T$ $(n_z<-0.95)$ and $(1,0,0)^T$ ($n_x>0.95)$, respectively, The two tubes cross once (linking number 1).  (c) Cross sections of the density of the $m=\pm1$ (left) and $m=0$ (right) components in the $\mathbf{x}$-$\mathbf{y}$ plane. The torus shapes of the $m=\pm1$ components manifest themselves as double rings in the cross section, which can serve as the signature of a knot.
Figure adapted from \citet{kawa08knots}.}
\label{fig:topological}
\end{figure}

The underlying physics that supports fractional vortices is discrete spin-gauge symmetries. Consider, for example, the polar state of a spin-1 condensate whose order parameter changes its sign upon inversion in spin space. Now, we consider a closed loop in real space and assume that the order parameter undergoes inversion in spin space as one circumnavigates the loop. The single-valuedness of the order parameter is met if the order parameter has an additional phase change of $\pi$, which is realized if the winding number of the state is a half integer. Therefore, under such circumstances, the system possesses half-quantum or Alice vortices~\cite{leon00alice}. Let us next consider the cyclic/tetrahedral phase of a spin-2 condensate.  As discussed earlier, the order parameter has a three-fold symmetry with respect to $2 \pi/3$ and $4 \pi/3$ geometric rotations about the symmetry axes of the tetrahedron.  This geometric rotation adds a phase of $2 \pi/3$ or $4 \pi/3$ to the spinor wavefunction.  Consider a loop in real space and assume that the order parameter rotates geometrically by $2\pi/3$ as one makes a complete circuit of the loop.  The single-valuedness of the order parameter is only met if the gauge phase changes by an integer multiple of $2\pi/3$. Thus, the cyclic phase can accommodate vortices with circulations of one-third or two-thirds of $h/M$~\cite{seme07onethird}.

The biaxial nematic and cyclic/tetrahedral phases of a spin-2 condensate can host non-Abelian vortices because their isotropy groups are non-Abelian~\cite{koba09nonabelian}. The unique feature of non-Abelian vortices lies in the fact that the generators of vortices do not commute with each other. Consequently, when two vortices collide, they cannot reconnect or pass through each other. Rather, they form a rung vortex that sticks the two vortices together. This implies that a collection of vortices will eventually form an interwoven network in sharp contrast to the case of Abelian vortices which usually reconnect upon collisions and tend to produce vortices with smaller scales. Such a distinction is expected to lead to a major difference in statistical properties of quantum turbulence.

A spin-1 polar condensate is predicted to accommodate knot excitations~\cite{kawa08knots}. Here the objects that form knots are field lines along which local spins point in a given direction. Suppose that a polar condensate is held in an optical trap and then a quadrupole magnetic field is applied on the condensate. Then, the magnetic moment of each atom undergoes Larmor precession according to the local magnetic field. Numerical simulations show that one after another knot, which consists of two loops linking together with unit linking number, enters the condensate from its periphery. Experimentally, such a knot formation can be probed by performing a Stern-Gerlach experiment on a sliced condensate; if there is a knot, one can observe a double-ring density distribution on every spin component.

Experimental efforts to create and study these myriad topological objects are ongoing.  Investigations thus far have revealed two such objects: the polar-core spin vortex \cite{sadl06symm}, skyrmions \cite{choi12skyrmion}, and Mermin-Ho textures \cite{lesl09skyrmion,choi12njp}.

The polar-core vortex was observed in $F=1$ $^{87}$Rb condensates following a quench of an $|m_z=0\rangle$ condensate to a setting of the quadratic Zeeman shift $q$ that favored ferromagnetic ordering; the relevant phase diagram is shown in Fig.\ \ref{fig:spin1pzero} and the dynamics of such a quantum quench are detailed in Sec.\ \ref{sec:spinmixing}.  Summarizing, the quench leads to a dynamical instability that introduces spontaneous spin currents within the initially uniform condensate.  The spontaneous symmetry breaking occurs independently in different regions of the condensate.  Altogether, this spontaneously generated spin-current field is then expected to contain circulating spin currents.  The polar-core spin vortex represents one quantum of spin-current circulation.  An isolated polar-core spin vortex may be described by the following condensate wavefunction (in cylindrical coordinates $(r, \theta)$):
\begin{equation}
\mathbf{\psi}(r, \theta) = \left( \begin{array}{c c c} f(r) & \times & e^{i \theta} \\
g(r) & \times & 1 \\
h(r) & \times & e^{-i \theta}
\end{array}\right)
\label{eq:pcv}
\end{equation}
The radial functions $f(r)$, $g(r)$ and $h(r)$ have the limiting values $\lim_{r \rightarrow \infty}(f,g,h)= (1/2, 1/\sqrt{2}, 1/2)$ and  $\lim_{r \rightarrow 0}(f, g, h) =  (0,1,0)$.  The former limit yields a transversely magnetized texture away from the vortex core, a configuration that is locally in the ground state for the ferromagnetic $F=1$ condensate with $q \simeq 0$.  The latter limit applies because the $|m_z = \pm 1\rangle$ components have non-zero circulation, and thus they must vanish at the origin.  However, the non-rotating $|m_z=0\rangle$ component remains within the vortex core, where the spinor condensate remains in the higher-energy polar state.  That is, such a topological defect captures the ``false vacuum'' represented by the pre-quench state of the condensate.  A polar-core spin vortex of opposite circulation would be described by replacing $\theta \rightarrow -\theta$ in Eq.\ (\ref{eq:pcv}).

\begin{figure}[tb]
\centering
\includegraphics[width=0.8\columnwidth]{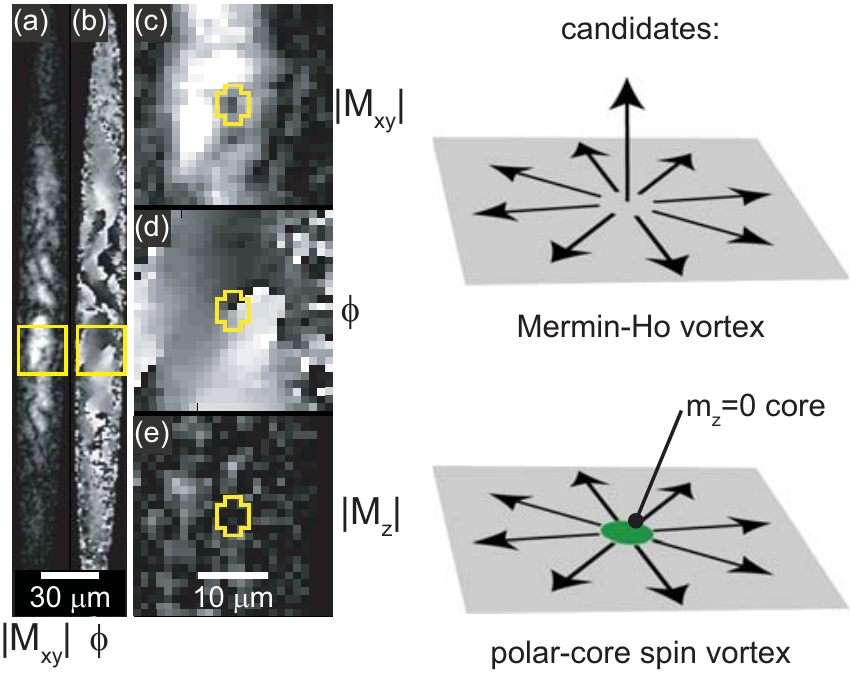}
\caption{(Color online) \emph{In-situ }detection of a polar-core spin vortex.  Spatial maps of the transverse magnetization (a) magnitude ($|M_{xy}|$) and (b) orientation ($\phi$) are shown for a gas imaged 156 ms after being quenched from the polar to the ferromagnetic state.  Data from a portion of the image are magnified, showing the transverse magnetization (c) magnitude and (d) orientation and also (e) the magnitude of the longitudinal magnetization ($|M_z|$).  The transverse magnetization orientation along a closed path (indicated in yellow) shows a net winding of $2 \pi$,
revealing the presence of a spin-vortex defect.  The core (within the closed path) shows no significant longitudinal magnetization, identifying the defect as a polar-core spin vortex.  Gray scale for images (a,c,e) is the same.  Figure adapted from \citet{sadl06symm}.}
\label{fig:pcv}
\end{figure}

The polar-core vortex was identified using \emph{in situ} magnetization-sensitive imaging (Fig.\ \ref{fig:pcv}).  A closed path within the condensate texture was identified as having non-zero transverse magnetization along the entire path, with the magnetization orientation winding by $\pm 2 \pi$.  Within that closed path, the magnetization was found to be consistent with zero, allowing this magnetization winding to be identified as a polar-core spin vortex, which has a non-magnetic core, rather than as a Mermin-Ho texture, which would be longitudinally magnetized at its core.  The polar-core spin vortex represents the spontaneous breaking of chiral symmetry and is reproduced in numerical simulations of such quench experiments \cite{sait06chiral}.

\begin{figure}[tb]
\centering
\includegraphics[width=\columnwidth]{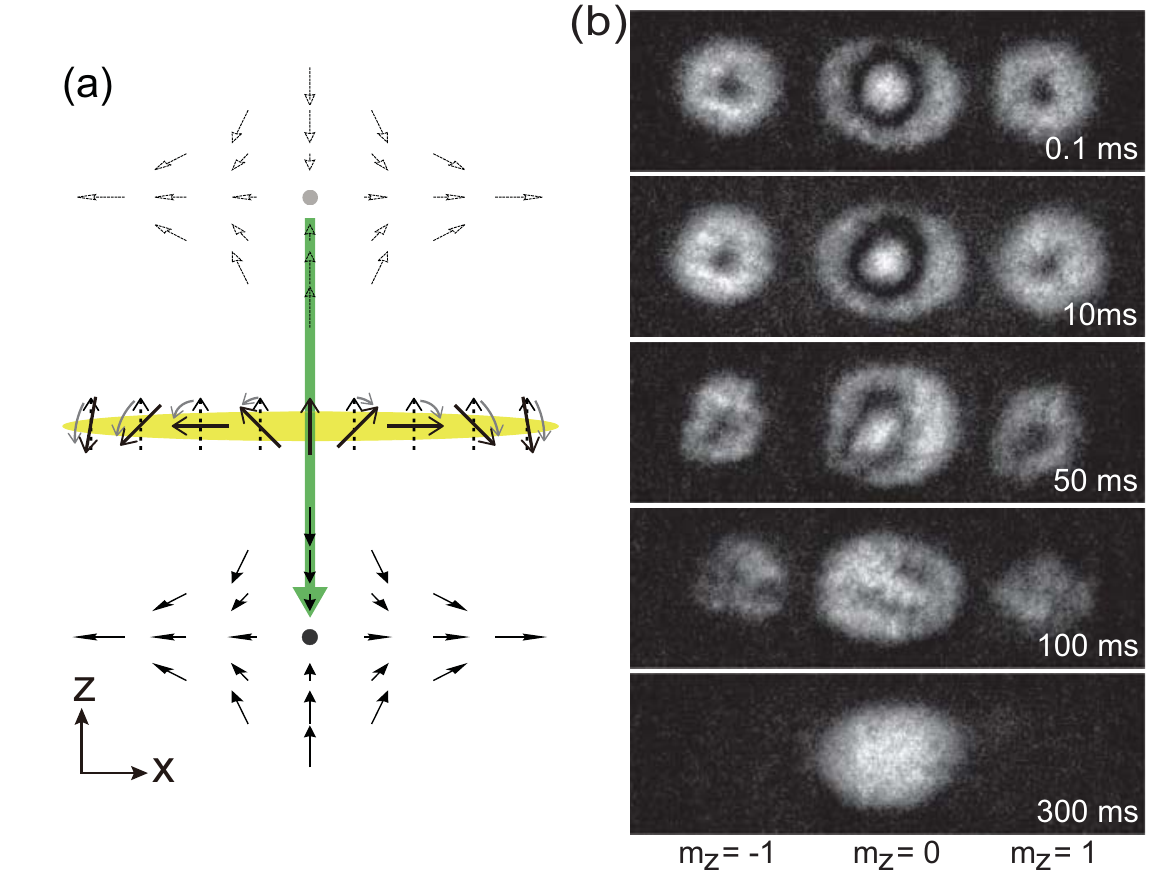}
\caption{(Color online) Creation of a skyrmion in a two-dimensional antiferromagnetic $F=1$ spinor condensate.  (a) A spherical quadrupole magnetic field, with field orientations shown by arrows, is translated from being centered above to below the Bose condensate (yellow horizontal region at the center).  The nematic director $\mathbf{n}$ follows the rotating field at large distances from where the magnetic field zero crosses the condensate, and remains fixed near the crossing, generating the skyrmion texture.  (b) Stern-Gerlach absorption imaging shows the $|m_z = 0\rangle$ component is divided between an inner region (where $\mathbf{n} = + \mathbf{z}$) and an outer ring (where $\mathbf{n} = - \mathbf{z}$).  Between these regions, the nematic director lies in the transverse plane; the state is thus a superposition of the $|m_z = \pm 1\rangle$ sublevels as seen in the data.  The skyrmion decays after 100 ms.  At long times, the entire gas returns to the $|m_z = 0\rangle$ state.  Figure reproduced from \citet{choi12skyrmion}.}
\label{fig:skyrmion}
\end{figure}

Two-dimensional skyrmion spin textures were generated in an antiferromagnetic $F=1$ spinor condensate in the polar state.  Such a skyrmion is defined as a defect in which the entire order parameter manifold is mapped continuously onto the plane, in such a manner that the order parameter is identical at the periphery of the two-dimensional system -- or at infinity in an infinitely extended sample.  Such a skyrmion was created by preparing the condensate in a spatially uniform polar state with the nematic director $\mathbf{n}$ aligned with the $\mathbf{z}$ axis, and then rotating the condensate spin in a coherent, spatially dependent manner to generate a texture with $\mathbf{n} = +\mathbf{z}$ at the center and $\mathbf{n} = -\mathbf{z}$ at the outer edge of the gas.  To effect such a rotation, following earlier realizations of a similar technique with ferromagnetic textures \cite{naka00,lean02imprinting}, \citet{choi12skyrmion} produced a spherical-quadrupole magnetic field, of the form $\mathbf{B} = B^\prime \left[ x \mathbf{x} + y \mathbf{y} - 2 z \mathbf{z}\right]$ where the Cartesian coordinates are defined with respect to the location of zero field.  The location of the field zero was gradually translated perpendicular to and across an optically trapped $^{23}$Na condensate (Fig.\ \ref{fig:skyrmion}).  Everywhere within the condensate, the magnetic field reverses its orientation about an axis that varies with the azimuthal angle $\theta$ in the plane of the condensate.  Far from the trajectory of the field zero, the field is always strong and the atomic spin follows the rotating field adiabatically, rotating the director $\mathbf{n}$ by $\pi$ radians.  At the location of the field zero, the magnetic field changes its direction diabatically; the atoms cannot follow this field reversal, and thus their nematic director remains unchanged.  At intermediate distances from the field zero, the spin texture smoothly interpolates between these settings, generating a continuous skyrmion texture.

The texture was characterized by time-of-flight measurements of the Zeeman-state distributions, preceded by rf-pulses which allowed for a characterization of the phase winding in each of the spin components.  The skyrmion was stable for about 100 ms.  The mechanism for its decay has yet to be identified.


Applying a similar procedure to an $F=1$ condensate prepared initially in a uniform ferromagnetic state produces a Mermin-Ho spin texture \cite{naka00,lean02imprinting,choi12njp}.  This texture differs from a skyrmion because the edge of the condensate, while magnetized in a common direction, carries a $4 \pi$ phase winding.  This difference between the polar and ferromagnetic states is reflective of the fact that the $SO(3)$ group does not support two-dimensional skyrmions.

A texture similar to the Mermin-Ho ferromagnetic texture was produced in an untrapped $F=2$ $^{87}$Rb spinor gas using optical Raman transitions \cite{lesl09skyrmion}.  Starting with a gas spin-polarized in the $|m_z = 2\rangle$ state, circular polarized light was used to drive $\Delta m_F = 2$ transitions, populating the $|m_z = 0\rangle$ and $|m_z=-2\rangle$ states.  One of the beams used for this transition was in the first-order Laguerre-Gaussian mode, so that the Raman transition also imprinted a $+ 2 \pi$ phase winding onto the product Zeeman state.  The spin configuration produced with such Raman transitions resembled the Mermin-Ho textures of ferromagnetic $F=1$ spinor condensates, which are composed of $m_z = (+1, 0, -1)$ states with circulations of $(0, 1, 2)$ quanta.

Topological excitations are also expected to appear upon rotating spinor condensates.  Ferromagnetic spin-1 condensates are expected to accommodate rotation by forming Mermin-Ho textures, which are shown to be stabilized by such rotation  \cite{mizu02,muel04text}.  Under certain circumstances, such spinor condensates may also adopt non-axisymmetric vortices, ones in which all Zeeman components are circulating, but in which the vortex cores of the components do not overlap.  Such spin configurations are no longer fully ferromagnetic \cite{mizu02nonaxis}.  Antiferromagnetic spin-1 condensates are expected to store angular momentum in the form of half-vortices \cite{isos02axis,muel04text}.  The landscape of spin condensates under rapid rotation is expected to be very complex \cite{kita02many,mizu04coreless}.

\subsection{Hydrodynamic description}
\label{sec:Hydrodynamics}

The Gross-Pitaevskii equation deals with the order parameter, from which physical quantities such as magnetization, spin nematicity, mass current and spin current can be calculated.  Alternately, one may develop descriptions that are equivalent to the Gross-Pitaevskii equation and deal directly with the quantities of experimental interest. Such hydrodynamic theory has been instrumental in describing scalar condensates, and it is of interest to ask if it is possible to describe spinor condensates in terms of observable quantities only.

To produce manageable equations of motion, one may confine the description to low-energy variations within the order-parameter manifold.
Lamacraft derived such a set of low-energy hydrodynamic equations for the case of a spin-1 ferromagnetic condensate~\cite{lama08helix}. When the particle density is constant and the system is locally fully magnetized, the normalized magnetization vector $\mathbf{F}$ obeys a modified Landau-Lifshitz equation that describes an advection of the magnetization by a superfluid velocity $\mathbf{v}$:
\begin{eqnarray}
\left(\partial_t+\mathbf{v}\cdot\nabla\right)\mathbf{F}=\frac{1}{2}\mathbf{F}\times\nabla^2\mathbf{F},
\label{LLE}
\end{eqnarray}
where the superfluid velocity satisfies the incompressibility condition $\nabla\cdot\mathbf{v}=0$. A self-contained set of hydrodynamic equations that are equivalent to the spin-1 Gross-Pitaevskii equation involve the spin quadrupole tensor in addition to the particle density, spin density and mass current~\cite{yuka12hydro}

The spin-gauge coupling of the spinor wavefunction permits some arbitrariness as to how the overall phase of the wavefunction is apportioned between variations of the spin texture orientation and of the superfluid phase.  The choice of such apportionment defines a vector potential, $\mathbf{a}$, which enters into the definition of the superfluid velocity: $\mathbf{v}=\nabla\phi-\mathbf{a}$, where $\phi$ is the superfluid phase. As in the vector potential in electromagnetism, this vector potential has an arbitrariness of the gradient of a scalar potential, but its curl is fixed by the Mermin-Ho relation:
\begin{eqnarray}
\nabla\times\mathbf{a}=-\frac{F}{2}\epsilon_{ijk}F_i\nabla F_j\times\nabla F_k,
\label{Mermin-Ho}
\end{eqnarray}
In analogy of electromagnetism, the right-hand side may be interpreted as an effective magnetic field which is generated by spin textures.

In addition to the vector potential, we may also introduce a scalar potential $\eta$ via the Berry connection $\eta=i\langle\psi|\partial_t|\psi\rangle$. The set $(\mathbf{a},\eta)$ then forms a four-vector $a_\alpha$, and by analogy with electromagnetism, we may introduce the electromagnetic tensor $f_{\alpha\beta}=\partial_\alpha a_\beta-\partial_\beta a_\alpha$ whose time and space components give the electric and magnetic fields, respectively, and the right-hand side of the Landau-Lifshitz equation (\ref{LLE}) now includes an analogue of the Lorentz force \cite{volo87,barn09geometrical}.

Including the effects of the MDDI changes the hydrodynamic description because it permits non-local spin-spin interactions.  \citet{taka07classical} included the MDDI in a hydrodynamic description of classical spins (ignoring spin-gauge coupling).   A hydrodynamic treatment for the ferromagnetic condensate including the MDDI and the quadratic Zeeman energy is discussed by \citet{kudo10hydro}.



%% file: rmp_ch_magnetization.tex
\section{Examining the connection between magnetic order and Bose condensation}
\label{sec:magneticorder}

Magnetism in solid-state systems may be ascribed broadly to two different origins: the interaction among local magnetic moments, or the interaction among itinerant fermions.  In spinor Bose gases, we encounter magnetic ordering that occurs in a system of itinerant bosons.  We argue that this magnetization occurs due to Bose-Einstein condensation, by which the influence of weak interactions is amplified through bosonic enhancement.

Here, we explore the question of how this statistically enhanced magnetic order responds when the effects of bosonic enhancement are weakened.  We consider three scenarios: increasing the temperature toward or above the Bose-Einstein condensation temperature, restricting the dimensionality of the system to one or two dimensions so as to enhance the role of quantum and thermal fluctuations, and placing the spinor Bose gas within a periodic potential so as to increase the low-energy density of states and enhance the role of interactions.

\subsection{Spinor Bose gases at non-zero temperature}
\label{sec:nonzeroT}

\begin{figure}[tb]
\centering
\includegraphics[width=0.8\columnwidth]{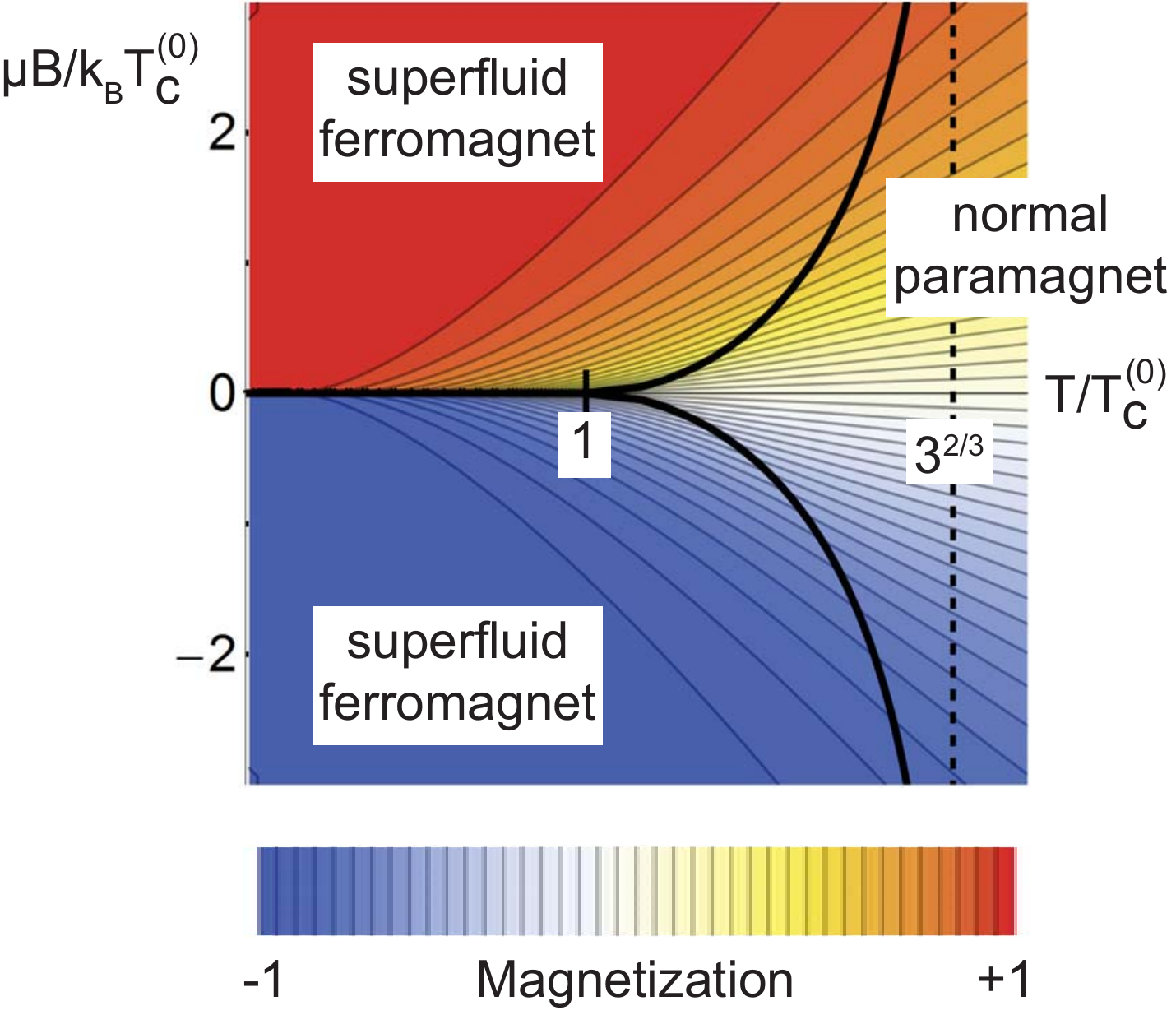}
\caption{(Color) Phase diagram for a uniform non-interacting $F=1$ spinor Bose gas as a function of applied magnetic field $B$ and temperature $T$.  The normal density for the $m_z$ Zeeman sublevel is determined by its chemical potential $\mu_0 - \mu B m_z / k_B T$, where $\mu$ is the magnetic moment and $\mu_0$ is determined by number conservation.  The unmagnetized gas at $B=0$ condenses at the critical temperature $T_c^{(0)} = 2 \pi \hbar^2/m k_B \times (n\, \zeta(3/2) \, /3)^{2/3}$ determined by the density $n/3$ of each Zeeman sublevel.  For $|\mu B/k_B T_c^{(0)}| \gg 1$, the critical temperature (solid black line) increases by a factor $(2F+1)^{2/3}$ (indicated by the dashed line).  Contours show the longitudinal magnetization, with color bar shown.  The superfluid state is ferromagnetic, as indicated by the discontinuous magnetization at $B = 0$ and $T<T_c^{(0)}$.}
\label{fig:yamafigure}
\end{figure}

The magnetic properties of a gas of non-interacting bosons can be derived from basic statistical mechanics \cite{yama82}.  A phase diagram evaluated as a function of the applied magnetic field $B$ and the temperature $T$, given in Fig.\ \ref{fig:yamafigure} specifically for a spin-1 gas, shows two notable features.  First, the quantum degeneracy temperature varies with the magnetic field.  This occurs because the Zeeman energy differentiates the chemical potential of the different Zeeman sublevels, increasing the fractional population, and hence the phase-space density, in the lowest-energy Zeeman state.  Thus, the Bose-Einstein condensation temperature $T_c$ increases by a maximum factor of $(2 F+1)^{2/3}$ in the limit that the Zeeman splitting greatly exceeds the thermal energy.  Second, ferromagnetism accompanies Bose-Einstein condensation, as evident in the diverging zero-field susceptibility of the gas below the Bose condensation temperature.  It may be said that ferromagnetism is a ``parasitic'' phenomenon, taking advantage of the zero entropy of the condensed fraction to establish magnetic order.  Refined treatments of this system, e.g.\ utilizing renormalization group theory and evaluating critical exponents, support the picture that the two phenomena of condensation and magnetization occur through the same transition \cite{simk99,cara96,frot84,gu03phenom}.

In the case considered here, the ferromagnetic phase -- as opposed to nematic, cyclic, or other phases -- is selected due to the application of a symmetry-breaking magnetic field and the absence of spin-dependent interactions.  This ferromagnetic tendency persists even under arbitrary \emph{spin-independent} interactions, as can be shown both for pseudo-spin $1/2$ \cite{sigg80,yang03rigorous} and higher-spin systems \cite{eise02}.

\begin{figure}[tb]
\centering
\includegraphics[width=0.8\columnwidth]{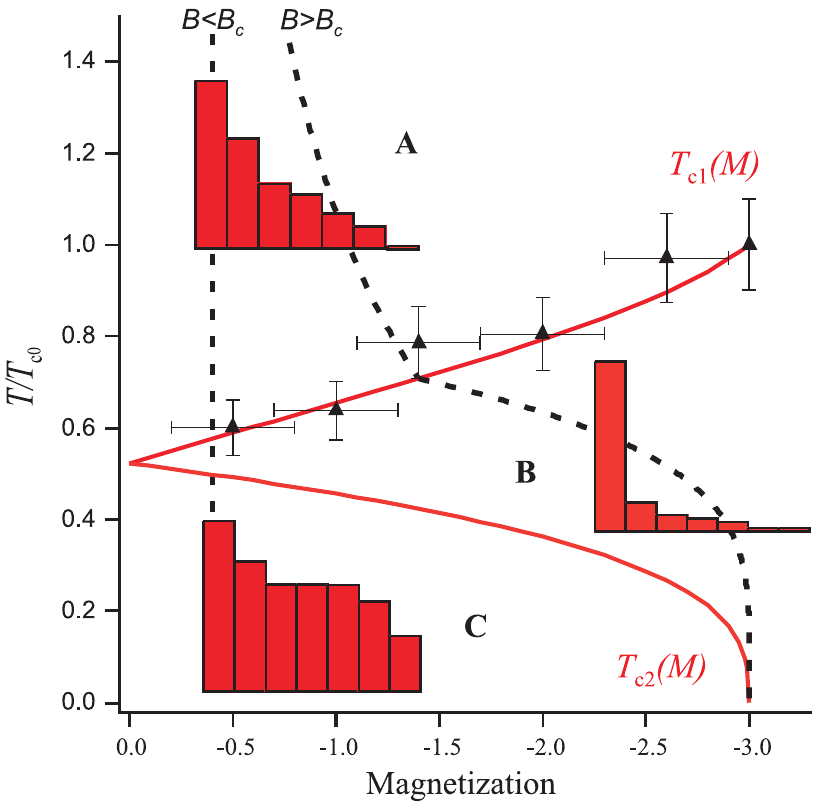}
\caption{(Color) Thermodynamic phase diagram of a chromium spinor Bose gas.  The large dipolar relaxation rate of chromium allows the longitudinal magnetization (abscissa, defined as the per-atom spin divided by $\hbar$) to achieve its thermodynamic equilibrium value as a function of the applied magnetic field and the gas temperature (ordinate, scaled by the transition temperature $T_{c0}$ of a fully magnetized gas).  The Bose-Einstein condensate transition temperature increases with magnetization, as predicted (see Fig.\ \ref{fig:yamafigure}).  Below the condensation temperature, the Bose-Einstein condensate is either fully magnetized (region B) or partially magnetized (region C), according to whether the Zeeman energy is larger or smaller than the non-ferromagnetic spin-dependent interaction energy.  Figure from \citet{pasq12thermo}.}
\label{fig:laburthe}
\end{figure}

This basic thermodynamic picture was confirmed in experiments with chromium gases \cite{pasq12thermo}.  Due to the large dipolar relaxation rate in this gas, the longitudinal magnetization is a free parameter, determined by thermodynamics, and also by spin-dependent interactions.  The Bose-Einstein condensation temperature was observed to vary with magnetic field, and hence with the magnetization of the gas, as predicted (Fig.\ \ref{fig:laburthe}).  Below the Bose-Einstein condensation temperature, the gas adopts a magnetization that is determined by both the linear Zeeman effect and also the spin-dependent interactions in the gas \cite{pasq11crspinor}: at higher magnetic field, where the Zeeman energy dominates, the condensate is fully magnetized, while at lower magnetic field, where the non-ferromagnetic spin-dependent interaction dominates, the condensate occupies multiple Zeeman states.

In the absence of dipolar relaxation, such as for the alkali spinor Bose gases, the equilibrium states of spinor gases need be evaluated under the constraint of constant longitudinal magnetization, rather than as occurring at constant magnetic field.  To the extent that the magnetization in Fig.\ \ref{fig:yamafigure} spans all values for $T>T_c^{(0)}$, the constrained equilibrium state can be taken as that which occurs at the value of $B$ (now a Lagrange multiplier) that gives the correct longitudinal magnetization.

However, for $T<T_c^{(0)}$, the equilibrium diagram of Fig.\ \ref{fig:yamafigure} fails to describe spinor Bose gases within a range of longitudinal magnetization around zero.  In this case, a gas of non-zero longitudinal magnetization is predicted to undergo a sequence of Bose-Einstein condensation transitions \cite{isos00double,zhan04trapped,kiss07polar}.  A first condensation, at $T>T_c^{(0)}$, produces a purely longitudinally polarized condensate.  Below this temperature, the non-condensed populations in all Zeeman states approach a common value, whereupon, at $T <T_c^{(0)}$, Bose-Einstein condensation occurs into multiple Zeeman sublevels.  The magnetic order of that low-temperature condensate depends on the character of the spin-dependent interactions.

Experimental evidence of this two-step condensation phenomenon has been obtained by the Paris group working with sodium condensates of non-zero magnetization\footnote{F.\ Gerbier, personal communication, 2012}.  Two-step condensation can also occur due the quadratic Zeeman shift, which, similarly, can imbalance the distribution of atoms within the non-condensed gas so as to initiate Bose-Einstein condensation into a single Zeeman sublevel (the $|m_z = 0\rangle$ state if $q>0$) before condensation can occur into other states as well \cite{phuc11}.

The role of spin-dependent interactions on the Bose-Einstein condensation transition has been debated.  In a mean-field description, magnetic order in the non-condensed or the condensed gas produces a Weiss field through the spin-dependent s-wave interaction.  Calculating thermodynamic functions in the presence of this molecular field indicates that, in the case of a ferromagnetic $F=1$ spinor gas ($c_1^{(1)}<0$), a ferromagnetic state has a lower free energy than the paramagnetic state at the Bose-Einstein condensation temperature, i.e.\ ferromagnetism occurs separately and before Bose-Einstein condensation, even for a weak ferromagnetic s-wave interaction \cite{gu03,gu04spinwaves,tao08thermo,kiss05}.

However, \citet{natu11pairing} point out that such a calculation neglects the dominant effect of exchange interactions.  As discussed in Secs.\ \ref{sec:collisions} and \ref{sec:spinwaves}, zero-range interactions take place between identical bosons or fermions if their total spin $F_{\rm pair}$ is even.  For example, for two electrons in an atom, the strong Coulomb repulsion at short range is thereby minimized for spin-triplet states.  For two bosons, the repulsion due to a strong spin-independent s-wave interaction is minimized by pairing the bosons in an odd spin state.  This direct bosonic exchange effect \emph{disfavors} ferromagnetism in the nondegenerate gas, and thus a weak spin-dependent interaction cannot yield ferromagnetism before condensation.  For condensed atoms, the direct exchange is absent, and weak spin-dependent interactions can indeed dictate the magnetic order of the gas.

The magnetic properties of spinor Bose gases close to the Bose-Einstein condensation transition were examined experimentally by \citet{veng10periodic}.  A non-magnetized gas was prepared well above the Bose-Einstein condensation transition, and then gradually cooled by evaporation from an optical trap.  A time-of-flight analysis indicated the formation of a low-momentum, quantum degenerate gas at the temperature $T_c^{(0)}$ expected given the atom number per Zeeman state and the presence of a harmonic confining potential.  \emph{In-situ} imaging revealed local magnetic ordering at all temperatures below $T_c^{(0)}$.  However, the experimental sensitivity to magnetization and the level of control over the gas temperature were insufficient to settle the question of whether magnetization appeared exactly at, or else slightly above, $T_c^{(0)}$.

\subsection{Two- and one-dimensional systems}
\label{sec:lowdimensions}

Another interesting arena in which to examine the connection between magnetic order and superfluidity is in low-dimensional systems.  For example, in a scalar Bose fluid in two dimensions, the link between superfluidity and condensation is severed.  Below the Berezinskii-Kosterlitz-Thouless (BKT) phase transition, the fluid becomes suddenly superfluid, via a jump in the superfluid stiffness, but the gas eschews off-diagonal long-range order and is thus not Bose-Einstein condensed.  The transition can be understood from the behavior of topological defects in the scalar fluid -- vortices of single positive or negative quantum of circulation -- that are bound below the BKT transition and unbound above it.  Signatures of the BKT transition have been observed in atomic gases, where the presence of a trap transforms the transition into a gradual crossover, and where the low-temperature phase is indeed Bose-Einstein condensed in that the coherence length becomes as large as the finite size of the gas \cite{hadz11review}.

Topological structures in spinor Bose-Einstein condensates differ from those of scalar condensates, and thus one expects different behavior in two dimensions.  Identifying the correct symmetry group of the $F=1$ polar spinor Bose-Einstein condensate, \citet{muke06topo} discussed the role of half-quantum vortices in the behavior of such a gas in two dimensions.  From analytical arguments and supported by numerical studies, they identified a BKT-like transition from a normal to a superfluid state.  Given the lower kinetic energy of the half-quantum vortex, they predict such a transition is characterized by a larger jump in the superfluid stiffness than is observed in the scalar case, a tell-tale sign to be sought in future experiments.  It is important to note that this larger jump was predicted much earlier in considerations of classical two-dimensional nematic films \cite{dege71} and of p-wave superfluids \cite{kors85}.

Remarkably, this topological phase transition leads to a gas that is superfluid, but still lacking in long-range spin order.  This phenomenon can be understood by considering the Hamiltonian of the gas in the form \cite{jame11twoD}
\begin{equation}
H = \int d^2 \mathbf{r} \, \left[ t_n \left( \nabla \mathbf{n} \right)^2 + t_\phi \left(\nabla \phi\right)^2 - q n_z^2 \right].
\end{equation}
Here, the gas is restricted to the ground-state manifold, and thus defined by the locally varying director $\mathbf{n}(\mathbf{r})$ of the polar state and the superfluid phase $\phi(\mathbf{r})$.  The quantities $t_n$ and $t_\phi$ denote the magnetic and phase stiffness (kinetic energy) of the gas, derived for example from a lattice model of the fluid.  In the $q=0$ case considered by \citet{muke06topo}, the dynamics of $\mathbf{n}$ and $\phi$ are decoupled, owing to the lack of spin-gauge coupling for the polar state, except for the topological linking of nematic disclinations and  vorticity of the two fields, respectively, by the half-quantum vortices.  Below the BKT transition, the half-quantum vortices of opposite charge become bound to each other, and the two fields become unlinked from one another.  Algebraic order develops in the quantity $e^{2 i  \phi}$, yielding superfluidity without condensation.  However, the director $\mathbf{n}$, which explores the larger space of rotations, cannot order according to the Mermin-Wagner-Hohenberg theorem, and thus no nematic spin order is obtained.

\begin{figure}[tb]
\begin{center}
\includegraphics[width=0.8\columnwidth]{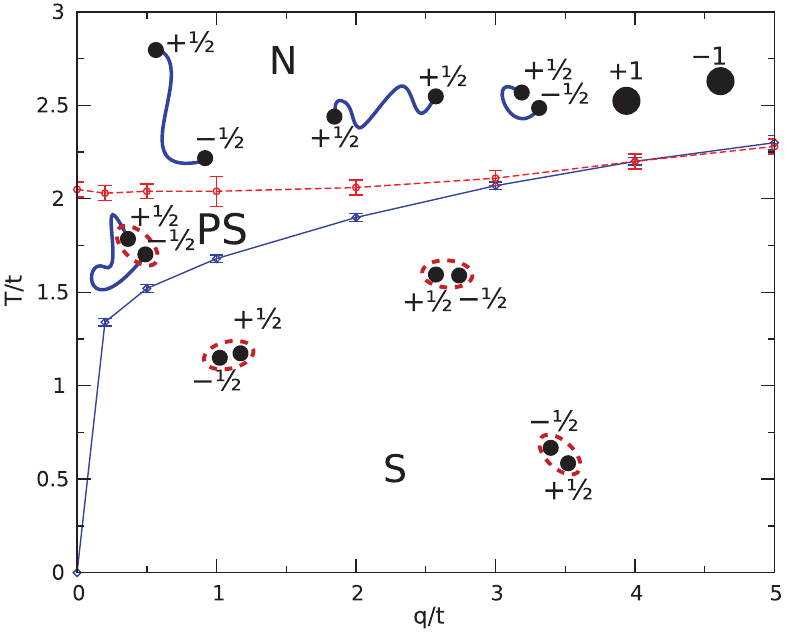}
\end{center}
\caption{(Color online) Phase diagram for a two-dimensional $F=1$ polar spinor Bose gas vs.\ temperature and quadratic Zeeman shift.  The dashed red line indicates the BKT transition from the normal to the spin-disordered superfluid state (PS), marked by the binding of half-quantum vortices \cite{muke06topo}.  A non-zero quadratic Zeeman shift favors the longitudinal polar state.  A spin texture containing a bound half-quantum-vortex pair necessitates regions where the polar-state director deviates from the $\mathbf{z}$ axis; these regions acquire a domain-wall width and tension defined by $q$.  At high temperature, in the PS state, these domain walls proliferate, causing the spin texture of the superfluid to be disordered.  Below an Ising-type phase transition (blue line), the domain walls are confined, allowing for long-range spin ordering and a conventional 2D superfluid (S).  Here, $t$ is the hopping energy in a discrete lattice model of the 2D gas.  Figure reproduced from  \citet{jame11twoD}.} \label{fig:lamapolar}
\end{figure}

The application of a quadratic Zeeman energy breaks the rotational symmetry of the polar-state manifold.  For $q>0$, one expects the spins to order at sufficiently low temperature at an Ising-type phase transition (Fig.\ \ref{fig:lamapolar}).  A polar superfluid is expected also at $q<0$, though such a scenario has been considered only under the additional influence of tuning $t_n$ and $t_\phi$ separately \cite{podo09novels1}.  Numerical simulations have been performed taking into account the finite size of a trapped polar gas, revealing similar spatial variations between coherent and incoherent portions of the gas as observed for trapped scalar gases, though the precision of such calculations is difficult to assess without a reliable non-zero temperature theory for the spinor Bose gas \cite{piet10finite2d}.

In one dimension, the scalar Bose gas is described by Luttinger liquid theory, and, for delta-function interactions, can also be treated exactly by the coordinate and thermodynamic Bethe Ans\"{a}tze.  At high density, the low temperature phase is a quasi-condensate, essentially described by mean-field theory but lacking true long-range order.  At low density, in the Tonks-Girardeau regime, the gas properties resemble those of non-interacting fermions as the repulsive interactions suppress the short-range two-body correlation function \cite{caza11rmp}.

A one-dimensional Bose gas with spin still admits an exact solution in the case of fully symmetric delta-function interactions.  For example, for the $F=1$ spinor gas with $SU(3)$ symmetric s-wave interactions (corresponding to $c_1^{(1)} = 0$), the ground state is a ``color ferromagnetic'' liquid: the phonon excitations and phase coherence properties are those of the scalar Luttinger liquid, while the spin sector shows local magnetic order and free-particle-like ($\propto k^2$) excitations \cite{li01su3}.

Significantly different behavior is seen for antiferromagnetic spin-dependent interactions.  As discussed for the low-temperature 2D polar gas, the dynamics of the gas separate into the charge-phase dynamics and the nematicity-director dynamics.  The zero-temperature state shows a power-law decay in the spin-singlet amplitude $A_1$, a state which can be regarded as a spin liquid that is gapped against spin-triplet excitations \cite{cao07epl,essl09polar}.

The separated dynamics of the charge and spin degrees of freedom hark to the phenomenon of spin-charge separation expected for one-dimensional Fermi systems.  Here, the mass and spin degrees of freedom of the spinor Bose gas evolve separately, showing, for example, differing sound velocities \cite{essl09polar}.  However, in the Bose system, this separation is not unique to one dimension, appearing also in three-dimensional gases.

Adding a non-zero magnetization to the gas transforms the polar spin-liquid state into a ferromagnetic liquid state.  \citet{lee09oneD} suggest that, between these two limits, the two liquids coexist as a two-component Luttinger liquid.  In this regime, the spin excitations of the liquid are no longer gapped.

\subsection{Spinor Bose gases in optical lattices}
\label{sec:lattices}

An optical lattice makes it possible to control the filling fraction -- the number of atoms per lattice site --  realizing an artificial carrier-doped many-body quantum system that features, among other things, the superfluid-Mott-insulator (SF-MI) transition~\cite{jaks98lattice,grei02mott}. The fluctuations in the number of atoms per lattice site are controlled by changing the ratio of the tunneling amplitude $t$ to the on-site Hartree interaction (charge gap) $U_0$.  At zero temperature, if the ratio $t/U_0$ is below a critical value, the system is in the MI phase; otherwise, the system is in the superfluid phase. With a spinor gas of spin-$F$ atoms loaded into the lattice, new energy scales concerning spin-dependent interactions, $U_i$ $(i=1,\cdots,F)$, come into play, making the system an ideal playground to study quantum magnetism. The dynamics of spinor gases in optical lattices also offers unique possibilities because spin-mixing dynamics can be coupled with density modulations such as Bloch oscillations and self-trapping~\cite{li10dyna}. An optical lattice loaded with spinor gases can also be utilized to determine various scattering lengths of spinor gases precisely~\cite{wide06precision} and for quantum information processing~\cite{wide05dyn}.

A spinor gas in a spin-independent optical lattice can be described with the Bose-Hubbard Hamiltonian~\cite{jaks98lattice} supplemented by the spinor interaction at each lattice site. For the spin-1 case, the Hamiltonian is given by
\begin{eqnarray}
\hat{H}&=&-t\sum_{\langle ij\rangle,m}(\hat{\psi}_{im}^\dagger \hat{\psi}_{jm}+\hat{\psi}_{jm}^\dagger \hat{\psi}_{im})
+\frac{U_0}{2}\sum_i \hat{n}_i(\hat{n}_i-1) \nonumber \\
& & +\frac{U_1}{2}\sum_i(\hat{\mathbf{F}}_i^2-2n_i)-\mu\sum_i\hat{n}_i,
\label{spinorBH}
\end{eqnarray}
where $\hat{\psi}_{im}$ is the annihilation operator of a boson in the magnetic sublevel $m_F=1,0,-1$ at site $i$, $\hat{n}_i=\sum_m\hat{\psi}_{im}^\dagger\hat{\psi}_{im}$, and $\mu$ is the chemical potential that controls the filling fraction of the system.  The spin operator $\hat{\mathbf{F}}_i$ has three components $\hat{F}_{i\alpha}=\hat{\psi}_{im}^\dagger (F_\alpha)_{mn}\hat{\psi}_{in}$ ($\alpha=x,y,z$), where $F_\alpha$ is the $\alpha$ component of the spin-1 operator. The ratio $U_1/U_0$ is about 0.04 for $^{23}$Na and -0.005 for $^{87}$Rb (see Table \ref{tab:scatlengths}). The control parameters that determine fundamental properties of the system are $t/U_i$, filling fraction, dimensionality, and size of the atomic spin.

When the number of atoms per lattice site is large, a ferromagnetic mini-condensate on each lattice has a large magnetic moment and interacts with other condensates through the MDDI. Such an array of large spins is shown to yield ferromagnetism~\cite{pu01ferroprl} and magnetic solitons~\cite{xie04pra,li05pra} in one dimension, and domain structures in higher dimensions~\cite{gros02lattice}.

When the number of atoms per lattice site is small, the MDDI can be ignored. Then, ferromagnetic condensates are expected to show similar behavior as scalar condensates with respect to the MI-SF transition.  For the antiferromagnetic case, the state in each Mott lobe is expected to show properties such as a nematic insulator and a spin-singlet  insulator~\cite{deml02break,imam03spinexchange}, where the nematic state corresponds to the polar state of a bulk condensate.  The nematic state breaks spin-rotation symmetry but preserves time-reversal symmetry and has gapless spin-wave excitations, while the spin-singlet state has spin-rotation symmetry and gapped excitations (spin gap).  A remarkable feature of the antiferromagnetic MI is an even-odd parity effect in which even fillings stabilize the MI state because all particles can then form spin-singlet pairs with the spin gap in addition to the charge gap~\cite{deml02break,tsuc04pra}.

In one dimension, antiferromagnetic spin-1 systems with odd fillings and small $t/U_0$ are predicted to be dimerized in such a manner that adjacent atoms tend to form spin-singlet dimers with an excitation gap~\cite{deml02break,yip03dimer,zhou03singlet}. The underlying physics is that virtual hopping to neighboring sites gives rise to an effective antiferromagnetic coupling between adjacent sites. Since each atom has two nearest neighbors to be dimerized, the dimer phase has two-fold degeneracy and breaks lattice translational symmetry.  A renormalization-group method confirms that for odd fillings the insulating phase is always dimerized~\cite{rizz05}, although the obtained dimer amplitude is very small and the state might be quantum nematic near the SO(3)-symmetric point~\cite{lauc06nematics}.  On the other hand, quantum Monte Carlo calculations show that the entire first Mott lobe has the dimerized ground state up to a relatively large $t/U_0$ which may be probed by the Bragg spectroscopy with peaks locating at half-integer multiples of the inverse lattice vector~\cite{apaj06}. A Monte Carlo study also shows that the SF-MI transition is first order for even fillings due to the spin gap of spin-singlet pairs, whereas it is second order for odd fillings as in the ferromagnetic case~\cite{batr09}.

The spin-1 system with small $t/U_i$ can be described with an effective Hamiltonian called the bilinear-biquadratic Hamiltonian~\cite{yip03dimer,imam03spinexchange,rodr11,dech11bilinear}
\begin{eqnarray}
\hat{H}&=&J\sum_{\langle ij\rangle}\left[
\cos\theta(\hat{\mathbf{F}}_i\cdot\hat{\mathbf{F}}_j)+\sin\theta(\hat{\mathbf{F}}_i\cdot\hat{\mathbf{F}}_j)^2\right]\nonumber \\
& & +D\sum_i(\hat{F}_{iz})^2,
\label{biquad}
\end{eqnarray}
where the $\cos\theta$ term arises from the virtual second-order hopping to neighboring sites, and $D$ characterizes the strength of the quadratic Zeeman effect. In one dimension, this model describes many different phases such as the Haldane phase including the Affleck-Kennedy-Lieb-Tasaki point~\cite{dech11bilinear}.

In higher dimensions, insulating states with an odd number of atoms per site are predicted to be nematic, whereas those with an even number are either nematic or singlet with a first-order transition between them~\cite{imam03spinexchange}.  Based on the mean-field or Gutzwiller approximation, it has been shown~\cite{kimu05first,krut05mi} that the SF-MI transition may be first order around the tip of each Mott lobe.  This is in sharp contrast with the ferromagnetic case in which the phase boundary is always second order, as in the scalar case~\cite{fish89}. The physical origin of the first-order transition is that in the MI phase the spin state is fixed, whereas in the superfluid phase it is not; thus the spin symmetry can have a discontinuity at the phase boundary. A possible experimental signature of the first-order transition is a discontinuous jump in the transverse magnetization from zero in the MI phase to a nonzero value in the SF phase~\cite{kimu05first}.  The spin gap in the MI phase for even fillings leads to quantized magnetization plateaus, while the second-order virtual hopping of spinor atoms induces an effective quadrupolar interaction between spins~\cite{imam03spinexchange,snoe09spin2} (see the second term in Eq.~(\ref{biquad})) which leads to a canted nematic order in which magnetization is not quantized but changes continuously as a function of an external magnetic field~\cite{imam04plateau}.  As the ratio $U_1/U_0$ increases, even Mott lobes expand and odd Mott lobes shrink until the latter disappear at a critical value around $U_1/U_0=0.5$~\cite{lack11disordered}, presumably because the spin gap due to spin-singlet pair formation works in favor of the even Mott lobes.
At unit filling, phase diagrams in the presence of the quadratic Zeeman effect were investigated based on the density-matrix renormalization method and Lanczos diagonalization~\cite{rodr11}, showing the field-induced transition between the transverse and longitudinal magnetization (or nematic) for the ferromagnetic (or antiferromagnetic) case.

For the case of spin-2 atoms, we may expect even richer phase diagrams~\cite{zhou06insulating}. The phase diagram was studied by treating the intersite hopping perturbatively~\cite{hou03,ecke07njp} and by a variational approach~\cite{snoe09spin2}. In one dimension, a trimerized state was predicted~\cite{chen12dimerized}.  The SF-MI transition in spin-3 condensates with dipolar interactions was also discussed~\cite{bern07spin3}.

We note that these theoretical predictions for quantum magnetism in optical lattices are extremely appealing, but also very challenging to achieve.  The spin-dependent energies between atoms trapped at low filling within optical lattices are typically much smaller than the on-site interaction energies that have produced the Mott insulating states studied to date.   The task of lowering the temperature of lattice-trapped gases to the point where such energy scales dominate, without the aid of the Bose enhancement which makes spinor Bose-Einstein condensates accessible, is a major hurdle.  Moreover, opening the spin degree of freedom for lattice-trapped gases may add a large amount of entropy per particle that may be difficult to reduce thereafter by evaporation or other techniques.  In this regard, it is worthwhile to mention two major experimental advances: the demagnetization gradient cooling method demonstrated by the MIT group~\cite{medl11demag} and the Pomerantchuk-type cooling observed with spinful fermions of Yb by the Kyoto group~\cite{suga11filling}.

%% file: rmp_ch_dipolar.tex
\section{Dipolar spinor gases}
\label{sec:dipolar}


The magnetic dipole-dipole interaction (MDDI) couples the spin degrees of freedom with the orbital degrees of freedom.  For magnetic dipoles $\bmu_1$ and $\bmu_2$ separated by the relative coordinate ${\bf r}$, the interaction energy is
\begin{equation}
V_{\rm dd}(\mathbf{r})
= c_{\rm dd} \frac{{\bmu}_1 \cdot {\bmu}_2 - 3({{\bmu}_1} \cdot {\bf r}/r)({\bmu}_2\cdot {\bf r}/r)}{r^3}.
\label{eq:vdd_twodipoles}
\end{equation}

The MDDI is invariant under simultaneous rotation of spins and coordinates, but not under the separate rotation of each.  Therefore, in the presence of the MDDI, the constraint of global spin conservation, assumed frequently above, no longer strictly holds.  The MDDI conserves the total angular momentum, and, thus, a dipole-induced relaxation of the spin angular momentum should lead to a change in the orbital angular momentum, causing a previously stationary spinor gas to begin rotating \cite{kawa06dehaas}. This phenomenon is analogous to the Einstein-de Haas effect in a ferromagnet \cite{eins16dehaas}, where the spin angular momentum is converted to a mechanical rotation. Conversely, a rotation of the system can lead to its magnetization -- a phenomenon known as the Barnett effect~\cite{barn15}.

In many experiments on cold atomic gases, the MDDI is negligible.  To assess its importance, one may compare the MDDI to the spin-independent contact interaction, which determines the compressibility and excitation spectrum of a spin-polarized Bose-Einstein condensate.  To make such a comparison, recall that the magnetic field inside a sphere with uniform magnetization $\mathbf{M}$ is $\mathbf{B} = (2/3) \mu_0 \mathbf{M}$, where $\mu_0$ is the vacuum permeability.  Setting $\mathbf{M} = \bmu n$ for a gas of atoms with magnetic moment $\bmu$ and density $n$, the magnetic-dipole energy per atom, $E_{\rm dd} = (1/3) \mu_0 \mu^2 n$, shows the same scaling with density as the mean-field interaction energy per atom $E = 4 \pi \hbar^2 a n / M$ with $a$ being the scattering length and $M$ the atomic mass.  The ratio of their energies is thus independent of density:
\begin{equation}
\epsilon_{{\rm dd}, a} = \frac{\mu_0 \mu^2 M}{12 \pi \hbar^2 a}
\label{eq:epsilondd}
\end{equation}

For alkali atoms, this ratio is quite small away from Feshbach resonances where the interaction energy can be tuned near zero \cite{fatt08dipolar,poll09extreme}; for example, for $^{87}$Rb in the $F=1$ state, $\epsilon_{{\rm dd}, a} = 0.002$.  Thus, the long-range dipole interaction can typically be neglected in experiments on scalar alkali gases.  In contrast, for Cr ($\epsilon_{{\rm dd}, a} = 0.15$), and even more so for Dy ($\epsilon_{{\rm dd}, a} \simeq 1$ for a typical scattering length), the dipolar interaction is an important factor in the structure and stability of a spin-polarized gas (see reviews by \citet{bara02dipolarreview,bara08,laha09review}).

However, for spinor Bose gases, the MDDI can have significant effects, both because it is spin-dependent and because it is long-ranged.  As a spin-dependent interaction, the MDDI may compete with the spin-dependent contact interaction to determine the \emph{local spin ordering} of the degenerate spinor gas.  Here, the importance of the MDDI may be assessed by examining the ratio of the spin-dependent contact to the dipolar interaction energy, given as $\epsilon_{{\rm dd}, \Delta a}$ in which we now use the scattering length difference $\Delta a$ that governs the spin-dependent interactions.  For example, in the $F=1$ manifold of $^{87}$Rb, considering $\Delta a = (a_2 - a_0)/3 = 0.25$ \AA, we find $\epsilon_{{\rm dd}, \Delta a} = 0.4$, so that we may expect dipolar interactions to play a significant role, as first pointed out by \citet{yi04dipolar} and as discussed in Sec.\ \ref{sec:mddi_sma}.

Even if the MDDI does not disrupt the local spin order selected by the spin-dependent contact interaction, it may nevertheless dictate the \emph{long-range spin ordering} of the gas.  Here, the proper comparison is between the MDDI and the quantum confinement energy, $\hbar^2 / (2 m l^2)$.  Equivalently, one may define a dipolar healing length, $\xi_{dd} = \sqrt{\hbar^2 / (2 M E_{\rm dd})}$ and compare it to the length $l$ of the condensed gas.  As a result, for spatially extended samples, the ground state spin configurations are predicted to be inhomogeneous spin textures that spontaneously rotate \cite{yi06texture,kawa06dipolar}, as discussed in Sec.\ \ref{sec:Flux-closure}.

\subsection{Dipolar interactions in the cold-collision regime}


The s-wave approximation cannot be used to treat the MDDI, owing to its peculiar scattering properties in three dimensions. The scattering phase shift $\delta_\ell(k)$ for the $\ell$th partial wave is proportional to the wave vector $k$ in the low-energy limit $k\rightarrow0$, regardless of $\ell$. This implies that all partial waves contribute to the low-energy scattering~\cite{land77qm}. Moreover, the anisotropy of the MDDI induces coupling between different partial waves.


Rather, the MDDI is retained explicitly in the interaction Hamiltonian.  At low magnetic fields, where the hyperfine spin $F$ is a good quantum number, the matrix elements of the magnetic dipole operator are given by $\langle F,m|\bmu|F,m'\rangle=g_F\mu_B\mathbf{F}_{mm'}$, where $g_F$ is the Land\'{e} $g$-factor.  The MDDI is then written in second quantization as
\begin{equation}
\hat{V}_{\rm dd} = \frac{c_{\rm dd}}{2}\iint d\mathbf{r}_1 \, d\mathbf{r}_2 : \left(\frac{\mathbf{F}\cdot\mathbf{F}-3(\mathbf{F}\cdot \mathbf{r}/r)(\mathbf{F}\cdot\mathbf{r}/r)}{r^3}\right) :
\label{eq:vdd}
\end{equation}
where in the integral $\mathbf{r}\! =\! \mathbf{r}_1\! -\! \mathbf{r}_2$, and $c_{\rm dd} = \mu_0 (g_F \mu_B)^2 / 4 \pi$.  The coupling between different hyperfine manifolds can be ignored because the hyperfine splitting $\sim 100$ mK is much larger than the $\sim 1$ nK energy scale of the MDDI.

\subsection{Interactions between rapidly precessing dipoles}

The rapid precession of the gas magnetization under a strong applied magnetic field alters the effective form of the MDDI.  For conditions where the strength of the dipolar field, on the order of $B \sim g_F \mu_0 \mu_B n$, is much smaller than the applied field, we account for the rapid precession in the interaction between two spatially fixed dipoles by identifying the dipole moments $\bmu^{\rm (rot)}$ in the Larmor precessing frame as $\mu_z^{\rm (rot)} = \mu_z$ and $\mu_\pm^{\rm (rot)} \equiv \mu_x^{\rm (rot)} \pm i \mu_y^{\rm (rot)} = e^{\pm i \omega_L t} \mu_\pm$, where we have specified $\mathbf{z}$ as the magnetic field axis $\mathbf{b}$.  Taking the time average, we obtain the effective interaction as
\begin{equation}
V_{\rm dd}^{\rm eff}(\mathbf{r})
= \frac{\mu_0}{4 \pi} \frac{\bmu_1^{\rm(rot)} \cdot \bar{\mathbf{Q}}(\bf{r}) \cdot \bmu_2^{\rm (rot)}}{r^3},
\label{time-averagedDDI}
\end{equation}
where
\begin{equation}
\bar{Q}_{\mu\nu}(\mathbf{r}) =\frac{1-3((\mathbf{r}/r)\cdot \mathbf{b})^2}{r^3} \frac{3 b_\mu b_\nu - \delta_{\mu \nu}}{2}.
\label{eq:dipole_vdd_ave_general}
\end{equation}

The effective MDDI, written now for arbitrary orientation of the magnetic field, distinguishes the longitudinal from the transverse magnetization, and expresses a non-local XXZ-type interaction between atomic spins.  This interaction can effect variations in the magnetization even when the dipolar field is small compared to the applied field, for the reason that the dipolar field produced by a transversely magnetized region of the gas is modulated at precisely the Larmor precession frequency.  This time-varying field can cause resonant Rabi nutation of the magnetization in another portion of the gas.

\subsection{Dipolar interactions in the single spatial-mode regime}
\label{sec:mddi_sma}

Let us revisit the mean-field description of spinor Bose-Einstein condensates in the single-mode approximation, presented above in Sec.\ \ref{sec:meanfieldground}, wherein the condensate spin character is described simply by a spatially independent spinor wavefunction.

Accounting for the MDDI introduces a new spin-dependent energy, potentially changing the local order preferred by the gas.  Within the SMA, this additional energy is determined by taking the spatial integral of ${V}_{\rm dd}$ -- or, in the presence of rapid Larmor precession, of ${V}_{\rm dd}^{\rm eff}$ -- over the condensate density distribution.  Considering a density distribution that is cylindrically symmetric about the $\mathbf{z}$ axis, the spatial integral of ${V}_{\rm dd}$ yields the following addition to the energy functional \cite{yi04dipolar}:
\begin{eqnarray}
E_{\rm dd} &=& c_{\rm dd}^\prime n \left( 2 \langle F_z \rangle^2 - \langle F_x \rangle^2 - \langle F_y \rangle^2\right) \nonumber \\
c_{dd}^\prime &=& \frac{c_{dd}}{n} \iint d\mathbf{r}_1 \, d\mathbf{r}_2 \frac{n(\mathbf{r}_1) n(\mathbf{r}_2)^2}{|\mathbf{r}_1 - \mathbf{r}_2|} \left(1- 3 \cos^2 \theta_{12}\right),  \nonumber
\end{eqnarray}
where $n(\mathbf{r})$ is the condensate density, and $\theta_{12}$ is the polar angle of the vector $\mathbf{r}_1 - \mathbf{r}_2$.

We see that the dipolar interaction adds an energy similar to that of the spin-dependent contact interactions, so that local effects of the MDDI can be regarded, roughly, as changing the value of $c_1$.  However, the MDDI energy term is anisotropic, i.e.\ the coupling between spin and orbital degrees of freedom causes the spatial anisotropy in the density of the condensate, determined by the trapping geometry, to translate into spin-space anisotropy.  By varying the trapping geometry, one can change the sign of $c_{\rm dd}^\prime$ and thus tune the effective sign of the spin-dependent dipolar energy.

\citet{yi04dipolar} examined the impact of dipolar interactions on the exact many-body ground-state predicted for the $F=1$ case by \citet{law98spin2} (see Sec.\ \ref{sec:fragmentation}).  For ferromagnetic interactions, the dipolar interaction breaks spin-rotational symmetry, so that the ground state becomes the state of maximum spin, $|F_{\rm tot} = N, F_{z, \rm tot} = M_F\rangle$, where now $M_F=N$ for $c_{dd}^\prime <0$ (prolate) and $M_F=0$ for $c_{dd}^\prime >0$ (oblate).  The dipolar interaction also stabilizes these maximal-spin states as the ground state even in the presence of weak antiferromagnetic interactions.  A similar stabilization of the ferromagnetic ground states was noted by \citet{kjal09}, who considered also the impact of the quadratic Zeeman shift on the $F=1$ mean-field phase diagram.  Dipolar interactions, treated within the single-mode approximation, are also predicted to affect the dynamics of the condensate magnetization under applied magnetic fields \cite{yasu11resonance}, an effect which will ultimately affect the accuracy of spinor-gas based magnetometers \cite{veng07mag} (discussed in Sec.\ \ref{sec:magnetometry}).

The dipolar interaction introduces additional spin-mixing terms to the Hamiltonian, e.g., terms of the form $(\hat{\psi}^\dagger_0  \hat{\psi}^\dagger_0  \hat{\psi}_{+1}  \hat{\psi}_{-1} + \hat{\psi}^\dagger_{+1}  \hat{\psi}^\dagger_{-1}  \hat{\psi}_{0}  \hat{\psi}_{0})$ for the case of the $F=1$ spinor gas.  Thus, the MDDI modifies the spin-mixing dynamics, such as coherent oscillations and dynamical instabilities, discussed in Sec.\ \ref{sec:spinmixing}.  For example, dipolar interactions can induce a transition across the separatrix that divides two types of spin-mixing trajectories, so that the spin-mixing dynamics from a given initial spin state will depend strongly on the trapping geometry \cite{rong05dipolar}.  Dipolar interactions are also predicted to break the rotational symmetry of dynamically unstable modes in single-mode and spatially extended $|m_z = 0\rangle$ condensates quenched to low values of the quadratic Zeeman shift.  This dipole-induced anisotropy makes the critical quadratic Zeeman shift $q_0$ governing the instability depend on the orientation of the $\mathbf{z}$ axis with respect to the trap shape, and should also produce anisotropic patterns of spin domains after the quench \cite{sau09}.

\subsection{Flux-closure relation}
\label{sec:Flux-closure}

The MDDI favors spin waves and magnetic domain structures. This can be seen by expressing it in momentum space:
\begin{eqnarray}
\langle \hat{V}_{\rm dd} \rangle
\propto \int d^3\mathbf{k} \,
\left[3|{\bf F}({\bf k}) \cdot {\bf k}/k|^2-|{\bf F}({\bf k})|^2\right],
\label{dipole_energy_in_k}
\end{eqnarray}
where ${\bf F}({\bf k})$ is the Fourier transform of the magnetization vector ${\bf F}({\bf r})$.
The dipole interaction energy is minimized by minimizing the first term and maximizing the second term on the right-hand side. The latter is achieved by placing a large weight on a particular spin-wave component ${\mathbf{F}}(\mathbf{k})$, while the former can be achieved by requiring that $\mathbf{k}$ be perpendicular to ${\bf F}({\bf k})$. In real space, this condition is written as ${\bf \nabla}\cdot{\bf F}({\bf r})=0$, which implies that no net magnetic flux emanates from any point in real space, that is, the magnetic flux closes upon itself. This is known as the flux-closure relation~\cite{land84embook}, which favors magnetic domain structures in solid-state ferromagnets to minimize the magnetic-field energy.

\begin{figure}[tb]
\centering
\includegraphics[width=\columnwidth]{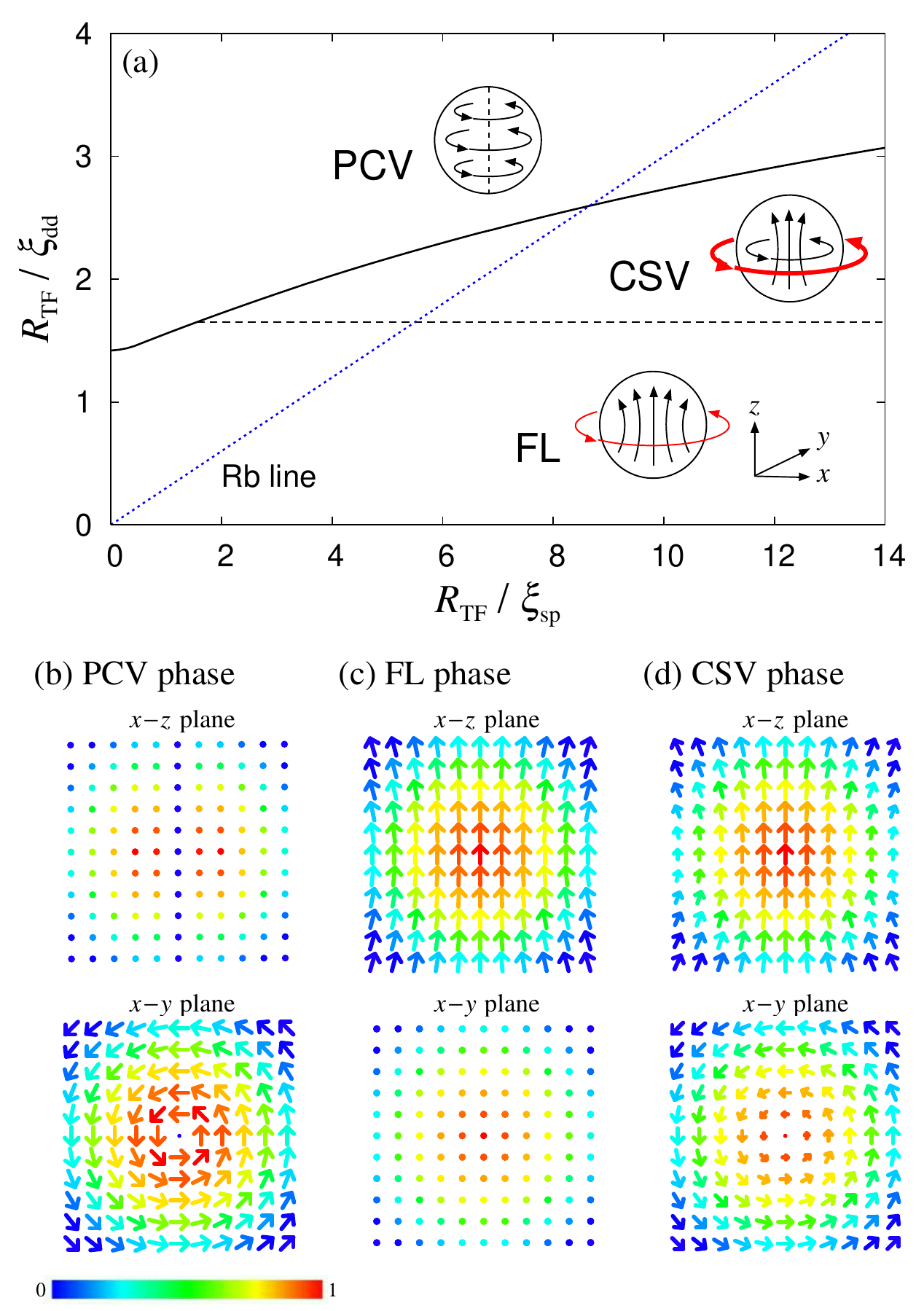}
\caption{(Color) (a) Phase diagram of a ferromagnetic spinor-dipolar condensate, where PCV, CSV, and FL stand for polar-core vortex, chiral spin-vortex, and flower phases, respectively, and $\xi_{\rm sp}$, $\xi_{\rm dd}$, and $R_{\rm TF}$ denote the spin healing length, the dipole healing length, and the Thomas-Fermi radius, respectively. The solid curve and dashed line show the first-order and second-order phase boundaries, respectively. The total angular momentum $J$ per particle is 0 for PCV, 1 for CSV and FL. The red and black circles with arrows show mass and spin circulations, respectively.
The $F=1$ $^{87}$Rb spinor gas traces the dotted line indicated as ``Rb line."  (b)-(d) Typical spin texture in each phase. The top and bottom
panels show the unit spin vector projected onto the $\mathbf{x}$-$\mathbf{z}$ and $\mathbf{x}$-$\mathbf{y}$ planes, respectively. The color of the arrows represents the magnitude of the normalized spin density according to the bottom scale.  Figure adapted from \citet{kawa06dipolar}.}
\label{fig:dipoletexture}
\end{figure}

In addition to the flux-closure relation, a ferromagnetic Bose-Einstein condensate features spin-gauge symmetry (see Sec.~\ref{sec:superflow}) that induces supercurrent when the spin texture of the system is nonuniform. As discussed in Sec.~\ref{sec:Hydrodynamics}, the spin textures also give rise to vorticity via the Mermin-Ho relation (Eq.\ (\ref{Mermin-Ho})). Since trapped systems in general produce nonuniform spin textures via the MDDI, we may expect a supercurrent to flow even in the ground state. In fact, the $F=1$ ferromagnetic Bose-Einstein condensates may undergo symmetry breaking into at least three different phases that feature inhomogeneous spin textures and circulating currents (Fig.\ \ref{fig:dipoletexture}).  These phases are selected by changing the atom number and trap frequencies \cite{kawa06dipolar,huht10}.

The phase diagram of dipolar spinor condensates at zero magnetic field has also been examined as a function of the dipole-dipole interaction and trap aspect ratio for both ferromagnetic and antiferromagnetic cases by \citet{yi06texture}, and spin textures are predicted to emerge in a pancake trap and relatively strong dipole-dipole interactions.
Spin textures in the MDDI-dominated regime are discussed by \citet{taka07classical} under the assumption that the spin behaves as a classical vector.

\subsection{Experimental signature}
\label{sec:SpinHelix}

Effects of the MDDI may have been observed in the dynamics of spin textures studied by the Berkeley group \cite{veng08helix}.  A ferromagnetic $F=1$ $^{87}$Rb condensate in a quasi-two-dimensional trap was prepared in a uniform transversely magnetized state.  Thereupon, a gradient of the magnitude of the magnetic field was briefly applied.  Larmor precession in the presence of this gradient gave rise to a helical spin texture, in which the orientation angle $\phi$ of the magnetization in the transverse spin plane increased linearly with distance along the direction of the field gradient.  The pitch of the helix was determined by the strength and duration of the pulsed gradient.  Thereafter, the texture was allowed to evolve under a uniform applied magnetic field before the gas was imaged.

\begin{figure*}[tb]
\centering
\includegraphics[width=5 in]{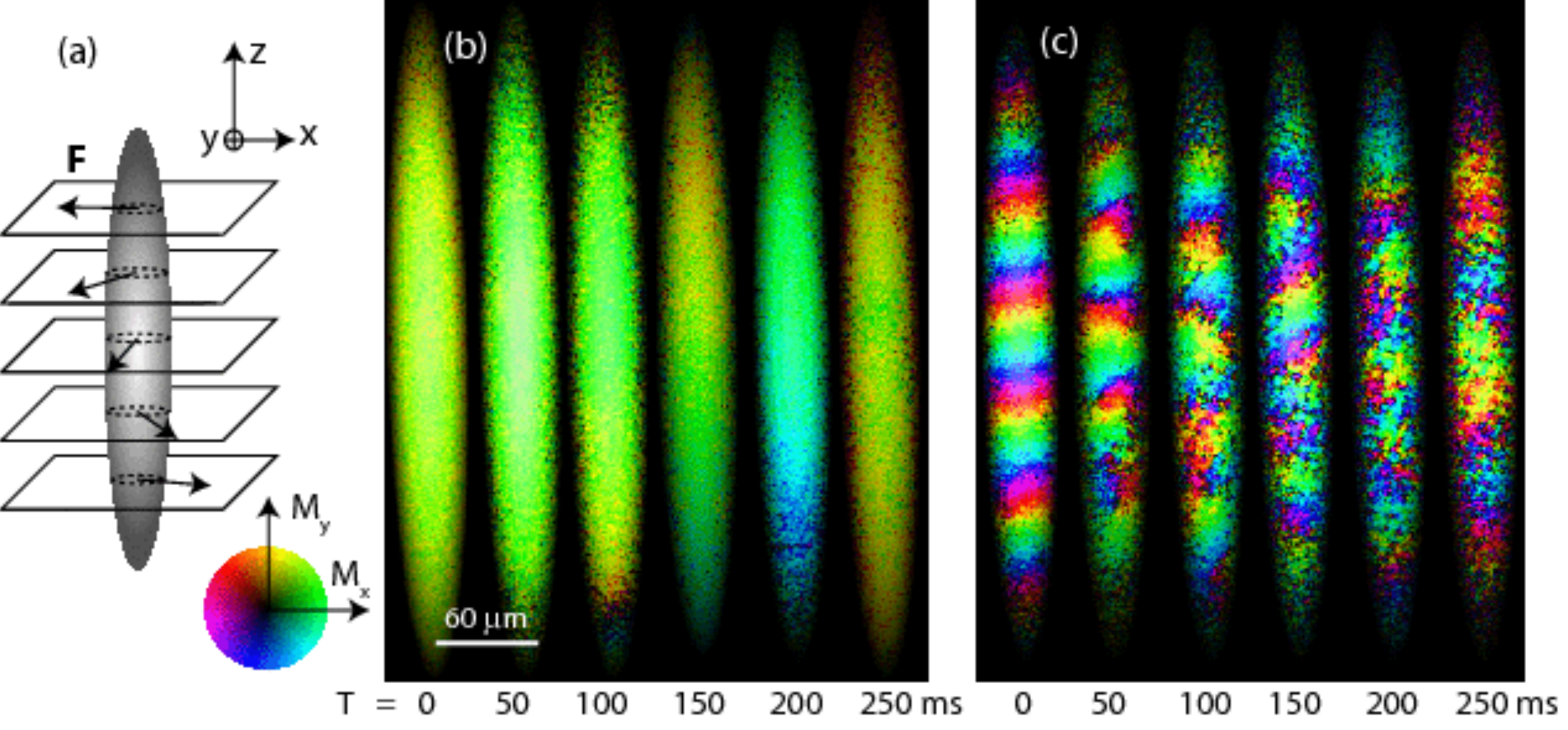}
\caption{(Color) Dissolution of helical spin textures in a ferromagnetic $F=1$ $^{87}$Rb spinor Bose-Einstein condensate.  A transient magnetic field gradient is used to prepare transversely magnetized (b) uniform or (a, c) helical magnetization textures. The transverse magnetization column density after a variable time $T$ of free evolution, in a homogeneous magnetic field, is shown in the imaged plane, with orientation indicated by hue and amplitude by brightness (color wheel shown). (b) A uniform texture remains homogeneous for long evolution times, while (c) a helical texture with pitch $\lambda = 60 \, \mu\mbox{m}$ dissolves over $\sim$200 ms, evolving into a sharply spatially modulated texture.  Figure from \citet{veng08helix}.}
\label{fig:spinhelix}
\end{figure*}

Considering just the s-wave interactions, a spin helix with a long pitch $\lambda$ is a low-energy texture, with excess energy per particle given just by the kinetic energy $\hbar^2 / 2 m \lambda^2$.  It was therefore surprising that such a helix underwent a rapid evolution toward a finely modulated spin texture (Fig.\ \ref{fig:spinhelix}) in which, given the short typical length scale of spin domains, the kinetic energy is much larger than its initial value.

The MDDI provides an additional source of energy to the spinor Bose-Einstein condensate, and may account for the observed dynamics.  To test this possibility, a rapid sequence of rf pulses with random spacing was applied to randomize the orientation of the atomic spins with respect to spatial coordinates.
Such rf pulses had the effect of diminishing the strength of the short length-scale modulation, supporting the picture that the MDDI causes such modulation to emerge.  We note that rf-pulse techniques have been suggested also as a way to isolate and emphasize, rather than to suppress, the effects of the MDDI in spinor gases \cite{yasu08spinecho}.

The stability of the helical texture was explored theoretically, both from the perspective of a low-energy hydrodynamic theory of the ferromagnetic state \cite{lama08helix}, and also using a Bogoliubov treatment \cite{cher08helix}.  The spectrum of modulational instability was found to reflect the anisotropy of dipolar interactions, suggesting a mode-softening toward a spatially periodic ground-state texture \cite{cher09roton}.  The possible emergence of modulated ground-state spin textures due to dipolar interactions was explored, focusing on the balance between the quadratic Zeeman effect and the MDDI~\cite{kawa10spont,kjal09,zhan10jltp}. When the quadratic Zeeman energy is positive and stronger than the MDDI, the longitudinal magnetization is suppressed, and the transverse magnetization forms a helix. In the opposite case in which the MDDI dominates the quadratic Zeeman energy, the longitudinal magnetization forms a domain structure to minimize the MDDI.  However, these works have not yet provided a consistent explanation for the experimental findings \cite{kawa10spont}.

%% file: rmp_ch_spinmixing.tex
\section{Non-equilibrium quantum dynamics}
\label{sec:spinmixing}

Cold atomic gases, and specifically spinor Bose gases, allow one to examine non-equilibrium dynamics in a manner that would be difficult with solids or liquids.  In contrast with condensed-matter systems, which are studied typically near equilibrium, cold-atom materials can be prepared far from equilibrium by applying state preparation techniques or rapidly changing the system Hamiltonian.  Atomic gases equilibrate slowly.  The kinetic degrees of freedom relax on millisecond timescales defined by the rate of elastic collisions.  The internal degrees of freedom in spinor gases equilibrate even more slowly.  The decay of the total longitudinal magnetization is either slow (as for $^{52}$Cr) or nearly imperceptible (as for many alkali gases).  The weak spin-dependent contact and dipolar interactions also produce dynamics over time scales --  many milliseconds -- that are long compared to the time needed to vary the trapping potential, the applied fields, or to probe the system.

In addition, unlike condensed-matter materials, cold-atom systems generally lack true reservoirs with which to exchange energy, angular momentum, or magnetization.  This raises the question of whether cold-atom systems equilibrate at all, whether equilibration is achieved within a given time frame for certain properties and not for others, and, if an equilibrium state is achieved, whether that state differs from a straightforward thermal equilibrium state.

Here, we highlight studies on spin mixing dynamics in degenerate spinor Bose gases.  Such dynamics are associated with redistribution of atomic populations among spin states, the formation of spin domains and spin textures, symmetry breaking and phase transitions, and nonlinear matter-wave optics.

\subsection{Coherent spin mixing in single mode experiments}

\subsubsection{Microscopic spin dynamics}
\label{sec:twoatommixing}

The spin-dependent contact interactions in a spinor gas can cause temporal oscillations of spin-state populations.  To exhibit this effect, consider a system with two $F=1$ atoms in the ground state of a tightly confining trap for which we make the single-mode approximation.  Consider an initial state $\psi(0)= |0, 2, 0\rangle$ written as a Fock state in the eigenbasis of a projection of the single-particle angular momentum, with magnetic quantum numbers $m_F=\{1, 0, -1\}$, respectively.  Owing to its being a superposition of $F_{\rm tot}=0$ and $F_{\rm tot}=2$ states, the initial state evolves as
\begin{equation}
\psi(t) = e^{-i \omega_0 t} \left[ \frac{2 + e^{-i \omega_1 t }}{3} |0, 2, 0\rangle - \frac{\sqrt{2}(1 - e^{-i \omega_1 t})}{3} |1, 0, 1\rangle \right]
\end{equation}
where $\omega_0 = (c_0^{(1)} + c_1^{(1)}) n / 2 \hbar$ and $\omega_1 = 3 c_1^{(1)} n / 2 \hbar$ with $n$ being the average density.  In a spin-mixing collision, a pair of atoms in the $|m_z=0\rangle$ spin state scatter inelastically into the $|m_z=+1\rangle$ and $|m_z=-1\rangle$ states.  This interaction is coherent, leading to periodic variation of the spin composition of the system.

\begin{figure}[tb]
\centering
\includegraphics[width=\columnwidth]{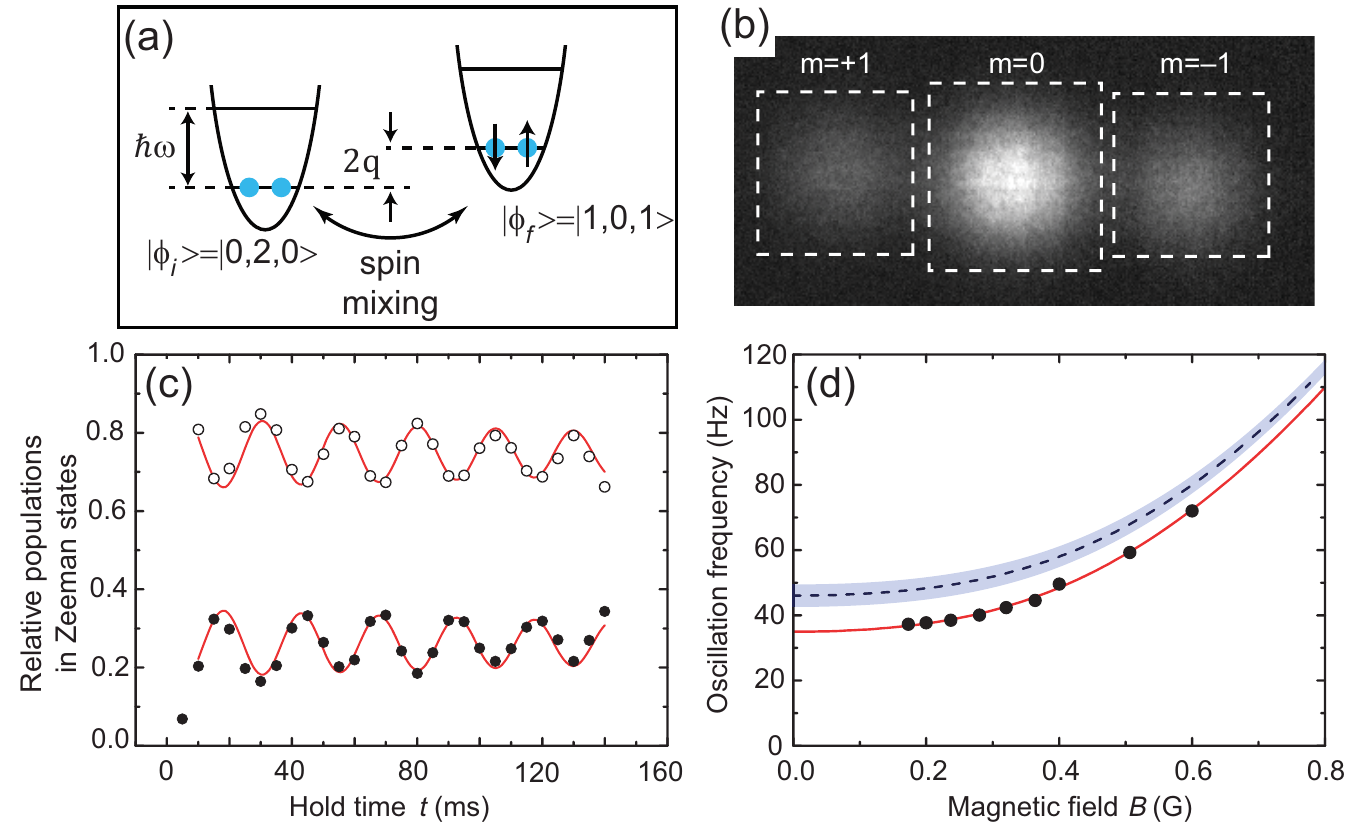}
\caption{(Color online) Spin-mixing oscillations of two trapped $F=1$ $^{87}$Rb atoms.  (a) An atom pair is prepared in the lowest motional state of an optical lattice site, in the initial state $|0, 2, 0\rangle$.  Spin mixing collisions convert the atom pair to the state $|1, 0, 1\rangle$, which is offset in energy from the initial state by the quadratic Zeeman energy $2 q \propto B^2$.  Spin mixing occurs in a large ensemble of such two-atom systems.  (b) Population oscillations are measured by Stern-Gerlach/time-of-flight imaging.  (c) Several coherent spin-mixing oscillations are observed, (d) the frequency of which varies with the quadratic Zeeman shift.  The frequency differs from that expected based on the scattering lengths calculated by \citet{vankemp02} (theory curved (dashed), with systematic uncertainty indicated by shading); indeed, 20\% discrepancies remain among reported values of $a_2 - a_0$ for the $F=1$ $^{87}$Rb system.  Figures adapted from \citet{wide06precision}.}
\label{fig:mainzosc}
\end{figure}

Such coherent microscopic spin mixing oscillations were observed by \citet{wide05dyn} (Fig.\ \ref{fig:mainzosc}).  Pairs of atoms were held in sites of an optical lattice, prepared with rf and microwave pulses, allowed to interact for a fixed time, and then probed by a Stern-Gerlach analysis.  Each atom pair serves as a clean microscopic laboratory to measure the strength of the spin-mixing terms \cite{wide06precision}, potentially also sensitive to the magnetic dipole interaction between individual cold atoms \cite{sun06twoatoms}.  Spin mixing oscillations were observed within both the $F=1$ and $F=2$ manifolds of states of $^{87}$Rb.  The experiments explored also the impact of quadratic Zeeman shifts \cite{gerb06rescontrol}, adding a controlled energy difference between the $|0,2,0\rangle$ and $|1,0,1\rangle$ two-atom spin states.  When this difference exceeds the spin-mixing interaction strength, the oscillations increase in frequency and decrease in amplitude, as expected for off-resonant Rabi oscillations.

\subsubsection{Single-mode mean-field dynamics}
\label{sec:meanfieldmixing}

Remarkably, the coherence of microscopic collisional dynamics of atom pairs is retained also in the spin-mixing dynamics of macroscopic collections of atoms.  Such dynamics in tightly confined Bose-Einstein condensates can be treated in a single-spatial-mode mean-field theory, i.e.\ specifying the condensate wavefunction, $\Psi = \Phi(\mathbf{r}) \psi(t)$, as a product of a common, fixed spatial wavefunction and a time-varying spin state.  We consider the ensuing spin-mixing dynamics in the presence of the contact interaction and a quadratic Zeeman shift $q$ along $\mathbf{z}$.  For an $F=1$ spinor gas, while the spinor wavefunction $\psi$ is specified by six real parameters, the dynamics is constrained by the overall $U(1)$ gauge symmetry of the spinor wavefunction (associated with conservation of the norm of $\psi$) and rotational symmetry about a given direction, say, $\mathbf{z}$ (associated with conservation of the $z$-projection of the spin, defined as $m = \left| \psi_{+1} \right|^2 - \left| \psi_{-1} \right|^2$ per atom), leaving just two dynamical variables \cite{pu99dyn,roma04,zhan05cohspin,kron05evol,kron08njp}.  Following Zhang \emph{et al.}, we define $\rho_0 = \psi^\dagger_0 \psi_0$ as the fractional population of the $|m_z = 0\rangle$ Zeeman state, and use the phase difference $\theta = \arg\left(\psi_{+1} \psi_{-1} \psi^*_0 \psi^*_0\right)$ which controls the degree of magnetization or nematicity.  The spin-dependent energy functional for the uniform spin-1 spinor gas given in Eq.\ (\ref{eq:spin1energyfunctional}) is then expressed as
\begin{eqnarray}
E^{(1)} &=& c_1^{(1)} n \rho_0 \left[(1-\rho_0) + \sqrt{(1-\rho_0)^2 - m^2} \, \cos\theta \right] \nonumber \\
& & + q (1 - \rho_0).
\label{eq:zhangfunctional}
\end{eqnarray}

\begin{figure}[tb]
\begin{center}
\includegraphics[width=0.8\columnwidth]{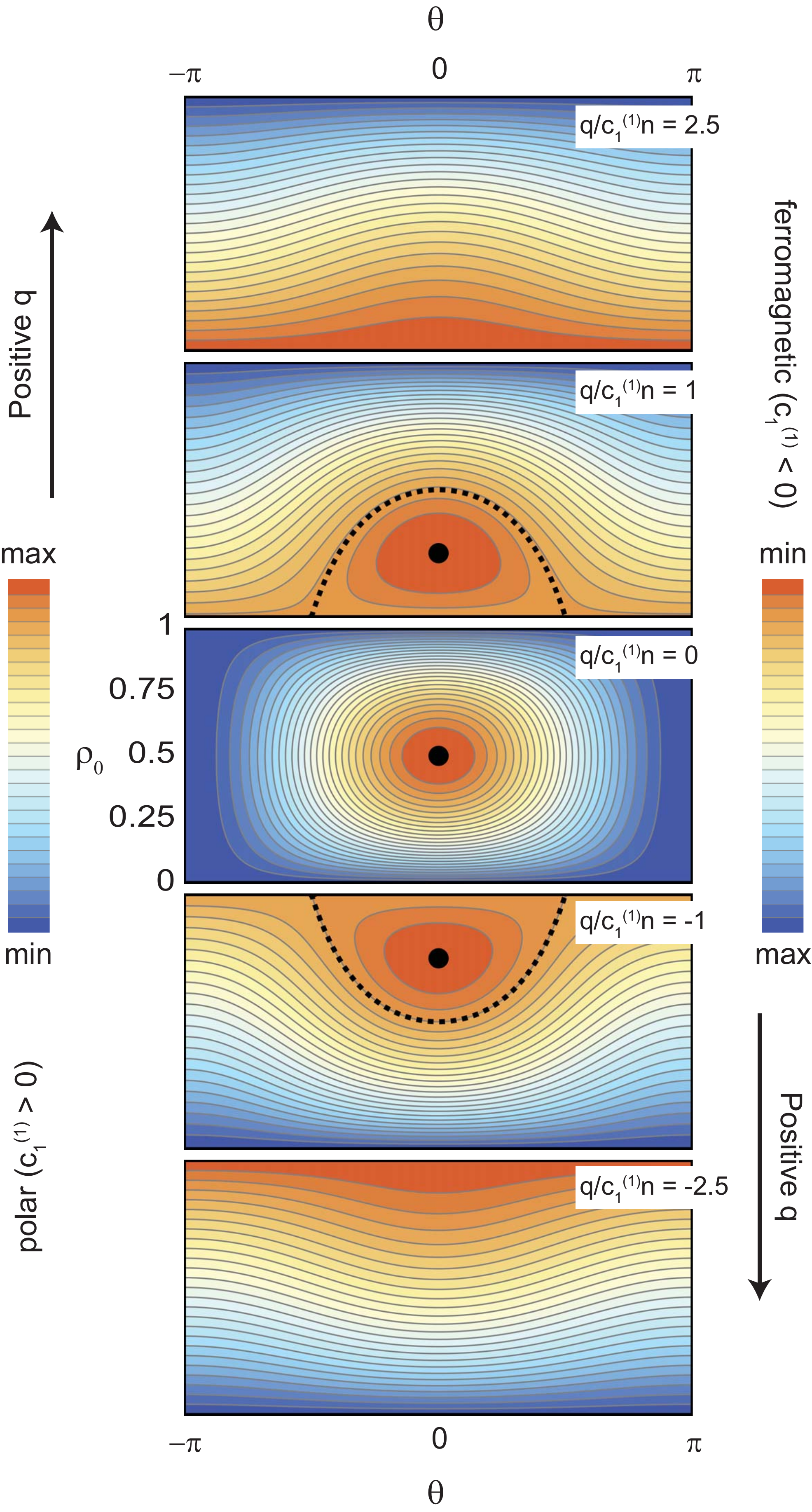}
\end{center}
\caption{(Color) Energy functional $E^{(1)}$ of an $F=1$ spinor Bose-Einstein condensate with zero longitudinal magnetization ($m=0$).  Here, $\rho_0 = \psi^\dagger_0 \psi_0$ and $\theta = \arg\left(\psi_{+1} \psi_{-1} \psi^*_0 \psi^*_0\right)$ for the spinor $(\psi_{+1}, \psi_0, \psi_{-1})^T$ written in the $\hat{F}_z$ eigenbasis.  For antiferromagnetic (ferromagnetic) interactions, minimum energies are shaded in blue (red).  Spin mixing causes the spinor gas to evolve along constant-energy contours (gray lines). In the interaction regime ($|q/c_1^{(1)}n| < 2$), a non-inert energy extremum emerges (marked by black dot) and also a separatrix (dashed curve), lying at $E^{(1)} = q$ ($E^{(1)} = 0$) for positive (negative) $q/c_1^{(1)}n$, separates the bounded and running spin mixing trajectories.}
 \label{fig:energyfunctional}
\end{figure}

This energy functional implicitly describes coherent spin dynamics in a single spin mode.  The dynamical variables $\rho_0$ and $\theta$ satisfy canonical relations, yielding equations of motion as $\dot{\theta} = (2/\hbar) \partial{E}/\partial \rho_0$ and $\dot{\rho_0} = -(2/\hbar) \partial{E}/\partial \theta$.  The energy $E$ is conserved under such dynamics; thus, the trajectory of $\rho_0$ and $\theta$  follows a constant energy contour (Fig.\ \ref{fig:energyfunctional}).  The energy extrema are thus identified as stationary states.  Below, we will discuss extensively the stability of the higher-energy stationary states.

Spin mixing occurs in two distinct dynamical regimes.  Consider the case of zero longitudinal magnetization ($m=0$).  For $|q| > 2 |c_1^{(1)}| n$, in the so-called ``Zeeman regime'' \cite{kron06tune}, all trajectories are running solutions in $\theta$, and the spin undergoes oscillatory orientation-to-alignment conversion -- alternating between magnetized and nematic states --  as expected for high-spin objects under quadratic spin Hamiltonians \cite{fano73}.  For $|q|<2 |c_1^{(1)}| n$, the so-called ``interaction regime,'' phase space is divided in two dynamical regions: one region contains oscillations that are bounded in $\theta$ (akin to oscillatory pendular motion) , and the other contains running solutions in $\theta$ (similar to over-the-top pendular motion).  Indeed, several works have drawn analogies between such spin-mixing dynamics and those of nonrigid pendulums and (internal) Josephson oscillations \cite{zhan05cohspin,chan05nphys,chan07fluc,jing09,yasu10internal,yasu10jltp}.  However, one should note the difference in topology between the phase space of a pendulum and that of the spin-1 spinor gas under this representation.  The phase space of a pendulum has the topology of an infinite cylinder: periodic in the angle and unbounded in the angular velocity.  In contrast, the $\rho_0-\theta$ phase space has the topology of a cone (for non-zero longitudinal magnetization) or a sphere (for zero longitudinal magnetization).

The transition between the Zeeman and interaction regimes is signified dramatically in the frequencies of small-amplitude spin oscillations.  Consider the case with $m=0$, and a state initialized near the transverse magnetized state ($\rho_0 = 1/2$, $\theta = 0$).  Deep in the Zeeman regime, $E^{(1)}$ is dominated by the term $q (1 - \rho_0)$, whence we obtain $\dot{\theta} \simeq -2 q / \hbar$ as expected for single-atom dynamics.  In the interaction regime, for $q=0$, the deviations $\delta\rho_0 = \rho_0 - 1/2$ and $\theta$ from the transverse magnetized state remain small.  To lowest order in these deviations, we find $\ddot{\theta} = - (2 c_1^{(1)} n/\hbar)^2 \theta$, so that spin mixing oscillations occur at the frequency $2 |c_1^{(1)}| n/\hbar$ determined by the spin-dependent contact interaction.  Between these regimes, the spin dynamics are slowed, reaching a ``spin-mixing resonance'' at $|q|=2|c_1^{(1)}|n$ where the spin-mixing period diverges \cite{chan05nphys,kron06tune,blac07naspinor}.  The resonance value of $q$ varies with magnetization $m$ \cite{blac07naspinor}.

Spin mixing oscillations in  both ferromagnetic ($^{87}$Rb) and antiferromagnetic ($^{23}$Na) $F=1$ spinor Bose-Einstein condensates have been observed.  The Chapman group initiated spin-mixing oscillations of $^{87}$Rb in the interaction regime by preparing condensates in spin states that were neither ferromagnetic nor polar, making use of microwave transitions between hyperfine levels.  From time-of-flight based measurements of $\rho_0$, the consequent trajectory in spin space was revealed to undergo several complete revolutions through $\rho_0$-$\theta$ phase space before the system relaxed to the predicted ground state \cite{chan05nphys}.  The Hamburg group focused on oscillations in the Zeeman regime, using $\pi/2$ rf pulses to rotate the atomic spin prior to Stern-Gerlach analysis, thereby permitting the observation of orientation-to-alignment oscillations \cite{kron06tune}.  Finally, working with $^{23}$Na condensates, the NIST group employed both Stern-Gerlach analysis and also dispersive optical detection (Sec.\ \ref{sec:dispersive}) to measure both $\rho_0$ and $\theta$, respectively \cite{blac07naspinor,liu09qpt}.  Altogether, these measurements enabled a determination of the spin-mixing energy from dynamics in the interaction regime \cite{chan05nphys,blac07naspinor}, and the observation of the spin-mixing resonance at nonzero quadratic Zeeman shifts \cite{kron06tune,blac07naspinor,liu09qpt}.

Spin mixing dynamics in $F=2$ spinor gases are richer owing to the larger phase space spanned by such dynamics.  As noted by \citet{kron08njp}, in addition to the aforementioned gauge and rotational symmetries, the equations of motion for the spin state are invariant under rotations by $\pi$ about a transverse spin axis, with the substitution $m \rightarrow -m$.  For the $F=2$ case, this additional non-trivial symmetry reduces the number of dynamical variables to four.  Deep in the interaction and Zeeman regimes -- the limits of $|q|$ being either much smaller or much larger than the spin-dependent interactions -- the spin-mixing again shows simple oscillatory behavior, as confirmed both by experiments and by theoretical approximations \cite{kuwa04spin2,kron06tune}.  In the interaction regime (small $|q|$), the specific trajectory of spin-mixing dynamics depends on the relative strengths of the spin-dependent interaction parameters $c_1^{(2)}$ and $c_2^{(2)}$, providing an experimental means to determine such relative strengths clearly  \cite{sait05diag}. Intriguingly, between the interaction and Zeeman regimes, where one obtains a spin-mixing resonance in the $F=1$ case, here the mean-field theory predicts the onset of chaotic motion at the mean-field level \cite{kron08njp}.

Finally, we mention several studies of spin-mixing dynamics in driven spinor gas systems.  \citet{pu00mag} and \citet{zhan01localization, zhan02spin2dynamics} considered theoretically the stability of spin mixing oscillations in the presence of external field noise to assess whether experimental limitations on such noise would bar the observation of coherent spin mixing dynamics.  It will be interesting to revisit such calculations in light of the prediction for chaotic dynamics in high-spin spinor gases.  They also assessed the response of the spinor gas to a periodic magnetic fields, finding regimes where the conventional Rabi oscillations of non-interacting spins give way to dynamical spin localization, akin to the physics of Josephson junctions (see also \citet{yasu10jltp}).  Dynamical localization is also predicted in the case that the spin-dependent scattering length is periodically modulated \cite{zhan10loc}.  Pulsed magnetic excitation is predicted to lead to chaotic spin dynamics, even in the $F=1$ case \cite{chen10chaos}, similar to quantum chaotic phenomena observed in other pulsed spin systems \cite{naka94chaosbook}.

\subsubsection{Many-body single-mode dynamics}

For $F=1$ spinor gases, a many-body single-spatial mode treatment of spin-mixing dynamics at $q=0$ is obtained directly from the exact energy spectrum derived by \citet{law98spin2}, as explained in Sec.~\ref{sec:fragmentation}. Indeed, an analytic Heisenberg-picture solution for spin-mixing dynamics based on this spectrum has been obtained \cite{chen08exactsol}. To compare with mean-field theory, let us consider the specific case of spin mixing dynamics for a state prepared initially near a ferromagnetic state.  As in experiments \cite{chan05nphys}, we consider the evolution of the Zeeman populations in the $|m_z=0\rangle$ state.  Such populations, considered above at the mean-field level, are determined from the many-body spin states as the expectation value of a sum of single-atom projection operators.  This sum is an operator of rank 2, so the temporal oscillations of the $|m_z=0\rangle$ state population occur at the frequency determined by the energy difference between many-body states of total angular momentum $L$ and $L - 2$:
 \begin{equation}
\Omega_{sm}^{N}= \frac{c^{(1)}_1 n}{2 \hbar N} \left(4 L - 2\right)
 \end{equation}
This result reproduces both the two-atom result (Sec.\ \ref{sec:twoatommixing}) and also, for large atom number $N$ and with $L=N$ near the ferromagnetic initial state, the mean-field result discussed above (Sec.\ \ref{sec:meanfieldmixing}).

At longer evolution times, the anharmonicity of the energy spectrum becomes evident \cite{law98spin2}.  Consider a state initialized in a coherent spin state that, if substantially different from the ferromagnetic state, represents a superposition of angular momentum eigenstates spanning a range $\Delta L \sim \sqrt{N}$.  Spin mixing dynamics now involve a range of Bohr frequencies spanning a bandwidth of $\Delta \omega \sim \Omega_{sm}^{N} N^{-1/2}$.  After an evolution time $t > (\Delta \omega)^{-1}$, the coherent spin-mixing oscillations predicted by mean-field theory are lost to ``quantum phase diffusion,'' similar to that discussed in the physics of scalar Bose-Einstein condensates \cite{lewe96phas,wrig96col1,pita97phen}.  However, given that all Bohr frequencies remain integer multiples of a common factor, even beyond the collapse of mean-field dynamics, the exact many-body dynamics nevertheless exhibit remarkable structure as seen in the predicted probability distribution for measurements of Zeeman-state populations \cite{dien06dynamics}.  The experimental observation of such many-body effects awaits the creation of well-controlled mesoscopic spinor Bose-Einstein condensates.

\subsection{Spin mixing instability}
\label{sec:spinmixinginstability}

In the interaction regime, the energy functional of Eq.\ (\ref{eq:zhangfunctional}) contains also energy maxima.  As a specific example, for a ferromagnetic $F=1$ system, with $0<q\leq 2|c_1^{(1)}|n$ the high-energy state with $\rho_0=1$ (inert state) is, at the mean-field level, stationary.  Whereas the mean-field ground state has a non-zero magnetization oriented in the transverse spin plane, the $\rho_0=1$ stationary state does not select a transverse spin orientation.

Experiments and theory have shown that this mean-field stationary state is, in fact, dynamically unstable.  Akin to a marble placed at the apex of a convex potential landscape, the collective spin state will diverge exponentially from the unstable stationary state on a trajectory initiated by even microscopic fluctuations.  Starting from an unbroken symmetric state, identified by zero-valued expectations of the relevant magnetic moments, the system evolves toward states of strong non-zero transverse magnetization and broken symmetry.  The relevance of such dynamics to topics such as quantum noise, quantum amplification, symmetry breaking and phase transitions has led to many studies of the spin mixing instability in both single-mode and spatially extended spinor Bose gases.

To discuss this full range of topics, let us consider the dynamics of the inert $|m_z = 0\rangle$ condensate under the Bogoliubov approximation, retaining terms up to second order in the fluctuations atop the condensate wavefunction.  We will find it convenient to work in the Cartesian or polar-state basis, with $\psi_\alpha$ being the $|m_\alpha=0\rangle$ eigenstate for the $\hat{F}_\alpha$ spin operator ($\alpha \in \{x, y, z\}$).  The spin-dependent Hamiltonian is approximated as follows:
\begin{eqnarray}
\hat{H} & = & \sum_{\beta = \{x, y\}} \int d^3 \mathbf{r}\,  \hat{\psi}^\dagger_\beta \left( \hami_0 + c^{(1)}_1 n(\mathbf{r}) \right) \hat{\psi}_\beta \nonumber \\
& & - \frac{c^{(1)}_1 n(\mathbf{r})}{2} \left( \left.\hat{\psi}_\beta^\dagger\right.^2(\mathbf{r}) + \left.\hat{\psi}_\beta\right.^2(\mathbf{r}) \right)
\label{eq:fromSauNJP}
\end{eqnarray}
Here we account for spatial variation of the spinor condensate: $n(\mathbf{r})$ is the density of the initial-state condensate, and $\hat{\psi}_\beta$ is the field operator for particles in the initially unoccupied polar states labeled by $\beta$.  Such particles experience the single-particle Hamiltonian of the form $\hami_0 = -\hbar^2 \nabla^2 / 2 m + V(\mathbf{r}) + q$ where, for simplicity, we assume all particles experience the same potential $V(\mathbf{r})$.

Now, let us simplify the problem by considering the dynamics of a single spin-excitation (magnon) mode, with two polarizations as labeled by $\beta$.  Such a mode can be identified either as a discrete magnon mode in the single-spatial-mode regime discussed above, as a linear combination of momentum eigenstates in the case of a homogeneous condensate, or by careful consideration of the appropriate Bogoliubov-de Gennes equations for this problem \cite{bara08anisotropic,sau09,sau10njp}.  Letting $\hat{\psi}_\beta$ now be the creation operator for particles in this magnon mode, the dynamics obey the following Hamiltonian:
\begin{equation}
\hat{\hami} = \sum_{\beta} \left(\epsilon + q + \bar{n} \right) \hat{\psi}_\beta^\dagger \hat{\psi}_\beta - \frac{c^{(1)}_1 \bar{n}}{2} \left( \left.\hat{\psi}_\beta^\dagger\right.^2 + \left.\hat{\psi}_\beta\right.^2\right)
\end{equation}
where the energy $\epsilon$ gives the kinetic and potential energy of the magnon mode.

Following the analogy of a marble rolling downhill, let us define Hermitian operators $\hat{Z}_\beta = (\hat{\psi}^\dagger_\beta + \hat{\psi}_\beta)/2$ and $\hat{P}_\beta = i (\hat{\psi}^\dagger_\beta - \hat{\psi}_\beta )/2$, and now obtain \cite{lama07quench}
\begin{equation}
\hami = \sum_{\beta} \left[ \left(\epsilon+q\right) \hat{Z}_\beta^2 + \left(\epsilon + q + 2 c^{(1)}_1 \bar{n}\right) \hat{P}_\beta^2 \right].
\label{eq:zandp}
\end{equation}
In the case that the coefficients of $\hat{Z}_\beta^2$ and $\hat{P}_\beta^2$ have the same sign, as obtained in the Zeeman regime when $|q|$ dominates the other energy scales, each polarization of the magnon mode evolves stably, precessing in the phase space spanned by the $\hat{Z}_\beta$ and $\hat{P}_\beta$ operators at a frequency given as $\omega = \sqrt{E^2}/\hbar$ with $E^2 =  (\epsilon+q)(\epsilon+q+2 c^{(1)}_1 \bar{n}) >0$.  In contrast, when the coefficients have opposite sign, which occurs when $\epsilon +q$ lies between $0$ and $-2 c^{(1)}_1 \bar{n}$, the magnon mode represents a dynamical instability of the initial-state condensate.  While $\langle \hat{Z}_\beta \rangle = \langle \hat{P}_\beta \rangle = 0$ in the initial state, the fluctuations of these operators become amplified parametrically: one quadrature of the $\hat{Z}_\beta-\hat{P}_\beta$ phase space grows exponentially at the rate $\omega = +\sqrt{-E^2}/\hbar$, while the other quadrature is attenuated.


This dynamical instability was noted first by a high sensitivity to noise in numerical simulations of spin mixing dynamics \cite{pu99dyn}, prompting investigations of the effects of field fluctuations on such dynamics \cite{pu00mag}.  A more focused study by \citet{robi01instab} was informed by the analogy between coherent matter-wave dynamics and optical four-wave mixing \cite{deng99fourwave}.  Robins \emph{et al.}\ examined the evolution of spatially uniform inert states of an $F=1$ condensate -- either the polar state (Case 1 in their work) or the magnetized state (Cases 2-4), in the absence of a quadratic Zeeman shift.  A modulational instability is observed only for ferromagnetic interactions and for the polar state, at modulation wavevectors $\mathbf{k}$ for which $\epsilon = \epsilon_k = \hbar^2 k^2 / 2 M$ gives $E^2<0$ as defined previously. The instability was predicted to cause an exponential rise in the population of previously unoccupied Zeeman states, and strong spatial modulation of the condensate spinor wavefunction.

The predicted transfer of Zeeman populations was observed in the first experiments on dynamics of ferromagnetic $F=1$ condensates \cite{chan04}.  A condensate prepared in the $|m_z = 0\rangle$ state at low $q$ evolved slowly for several hundred milliseconds before rapidly tending toward a mixture of Zeeman states.  Similar dynamics were observed also for antiferromagnetically interacting spinor gases ($c^{(1)}_1>0$) at negative quadratic shifts.  For example, the stability analysis for a $^{87}$Rb $F=2$ condensate prepared in the $|m_z = 0\rangle$ initial state is similar to that of the $F=1$ antiferromagnetic gas. Such condensates were found to undergo collision-induced spin mixing at low dc magnetic fields, for which $q<0$ in the $F=2$ $^{87}$Rb spin manifold.  The dynamics are substantially faster than that for the $F=1$ manifold, owing to the larger spin-dependent interaction strength \cite{schm04,kuwa04spin2}.  The spin-mixing instability was also demonstrated recently for the antiferromagnetic $F=1$ $^{23}$Na condensate, for which off-resonant microwaves were used to induce a negative quadratic Zeeman shift \cite{book11}.

\begin{figure}[tb]
\begin{center}
\includegraphics[width=0.9\columnwidth]{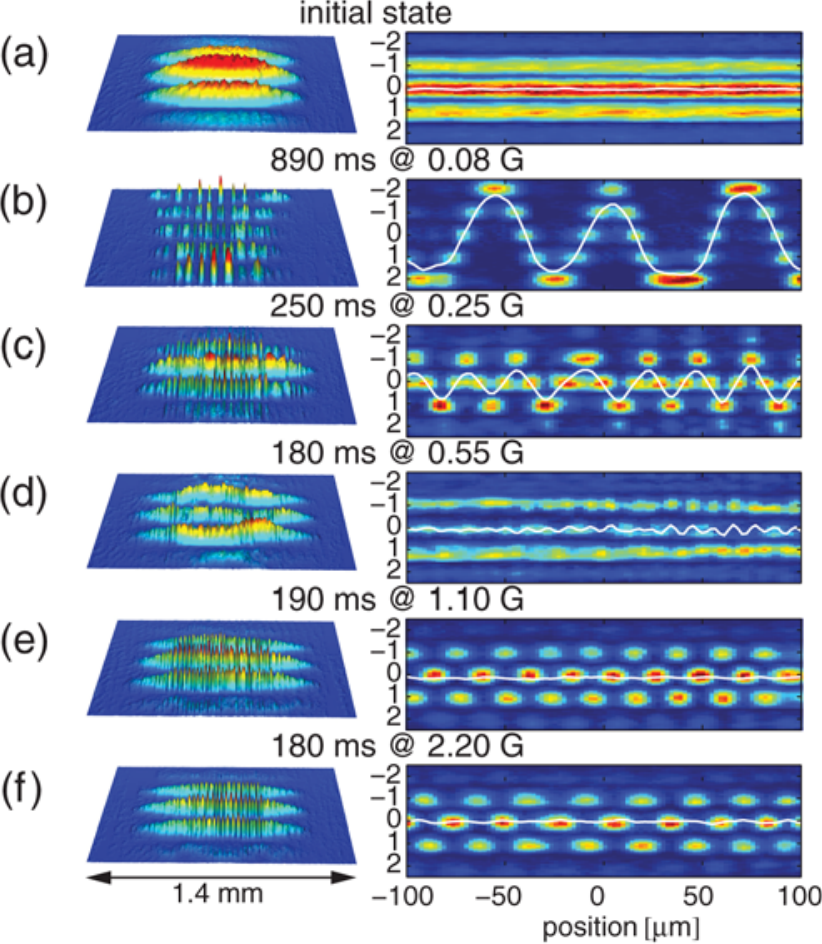}
\end{center}
\caption{(Color) Spontaneous pattern formation from modulational instability of a transversely magnetized $F=2$ $^{87}$Rb spinor condensate.  The condensate is prepared in a quasi one-dimensional optical trap.  The initial state (a) and saturated spin patterns arising for increasing magnetic field (b-f) are determined from Stern-Gerlach/time-of-flight images that reveal the in-trap axial density distributions of the five Zeeman-state populations.  Cases (b) and (c) represent the interaction regime (small $q$), while cases (e) and (f) represent the Zeeman regime (large $q$).  The characteristic length scale of the spin modulation varies strongly with $q$.  Figure reproduced from \citet{kron10spont}.}
\label{fig:f2pattern}
\end{figure}

Spinor Bose condensates display other instabilities as well.  Returning to the single-mode energy functional for the $F=1$ system (Eq.\ (\ref{eq:zhangfunctional})), the energy extremum defined by $\theta = 0$ (dubbed the ``phase-matched state'' in \citet{matu08}) and $\rho_0 = \left[1 - q/(2 c^{(1)}_1 n)\right]/2$, which exists for $|q|<|2 c^{(1)}_1 n|$, is an energy maximum under antiferromagnetic interactions (denoted by the black line in Fig.\ \ref{fig:spin1phasediagrams}(c)).  A wave-mixing analysis shows that this stationary state is also subject to modulational instability \cite{matu08} that can produce intricate spin domain structures \cite{matu10roton}.  Modulational instability was observed experimentally in a similar system -- the transversely magnetized $F=2$ $^{87}$Rb spinor condensate -- generating spin domain structures with a periodicity controlled by the quadratic Zeeman shift \cite{kron10spont} (Fig.\ \ref{fig:f2pattern}).

\subsection{Quantum quench dynamics and the Kibble-Zurek mechanism}
\label{sec:quantumquench}

Considering the $|m_z=0\rangle$ energy extremum for the ferromagnetic $F=1$ single-mode spin energy, we observe a transition from dynamical stability to instability coincident with the state's transition from an energy minimum to a maximum, i.e.\ with a second-order zero-temperature phase transition.  The evolution of systems prepared initially at the extremal state thereby exemplifies the dynamical response of a system quenched rapidly across a phase transition, evolving from a state of unbroken symmetry (here, $SO(2)$ axisymmetry) toward a manifold of broken-symmetry ground states.

In classical systems rapidly quenched across a finite-temperature phase transition, symmetry-breaking occurs inhomogeneously, with different portions of the system adopting different symmetry-broken phases.  Such a process was considered by \citet{kibb76}, in the context of symmetry breaking in the early universe, and by \citet{zure85cosmo}, in the context of low-energy laboratory experiments on material systems.  Their treatment, known as the ``Kibble-Zurek mechanism,'' translates the phenomenology of critical scaling of equilibrated systems near phase transitions into predictions regarding the typical size of the symmetry-broken phases.  Their theory also discusses the types of topological defects produced in such a quench, their initial density (related to the aforementioned typical size), and their subsequent evolution.  Laboratory experiments have explored aspects of the Kibble-Zurek mechanism, e.g.\ using liquid crystals \cite{chua91cosmo,bowi94}, pressure-quenched $^4$He \cite{hend94cosmo}, neutron-bombarded $^3$He \cite{baue96,ruut96}, and Josephson junctions \cite{carm00kz,mona02kz}.

Given theoretical progress in understanding quantum phase transitions, and experimental progress in preparing isolable low-temperature quantum systems, the Kibble-Zurek idea was naturally extended to the quench of a system across a quantum phase transition \cite{zure05qpt,dzia05,polk05univ}.  For an ideal experiment, the quench dynamics are now quantum mechanical, in that quantum fluctuations (rather than thermal or technical ones) are amplified into coherent superpositions of macroscopically distinct, broken symmetry states.  Ultracold atomic gases are unique materials with which to study such ideal quantum quenches.

\begin{figure}[tb]
\begin{center}
\includegraphics[width=0.9\columnwidth]{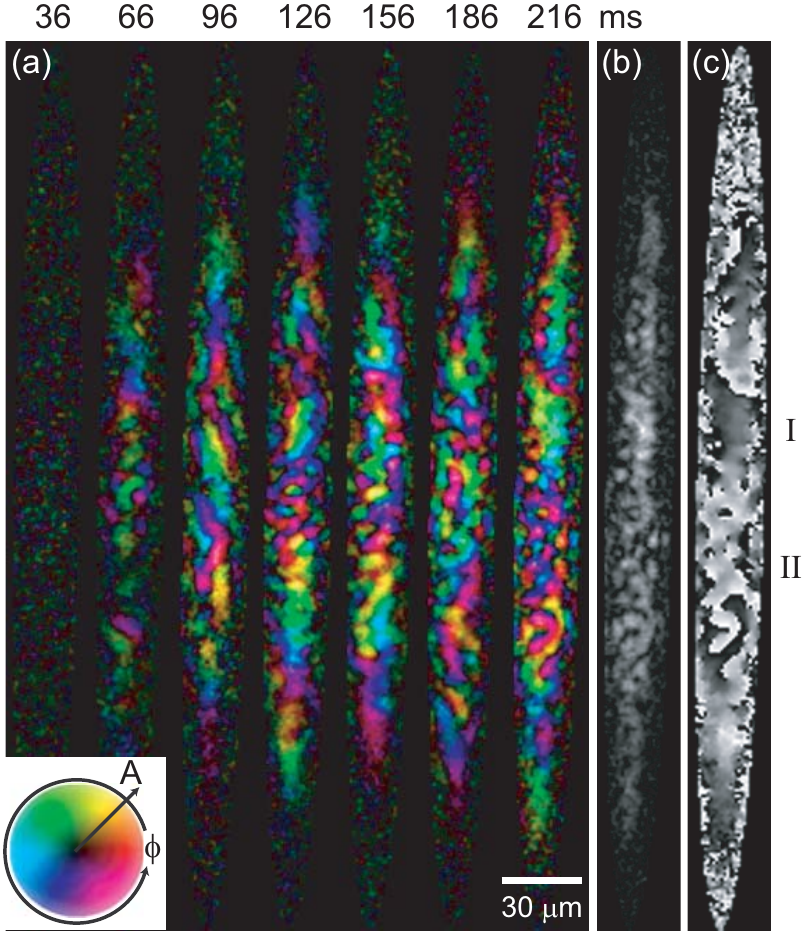}
\end{center}
\caption{(Color) Spontaneous symmetry breaking in a ferromagnetic $F=1$ $^{87}$Rb spinor Bose-Einstein condensate quenched across a phase transition.  A condensate is prepared in the $|m_z = 0\rangle$ state at high $q$, after which $q$ is suddenly reduced to $q < |c_1^{(1)}| n_0$ where spin-dependent interactions favor a transverse ferromagnetic state ($n_0$ is the central column-averaged condensate density).  (a) The transverse column-integrated magnetization, measured for condensates at variable times after the quench (indicated on top), is shown with the magnetization orientation ($\phi$) indicated by hue, and magnitude ($A$) indicated by brightness.  The maximum brightness, indicated by the color wheel at left, corresponds to full magnetization of the condensate center.  For the data at $216$ ms, the magnetization density (b) and orientation (c) are shown separately.  The spin texture at position I is characterized by an extended region with magnetization of non-zero magnitude and slowly varying orientation.  Smaller magnetic domains at position II are divided by domain walls with zero magnetization.  The gray scale in (c) covers the range 0 to $2 \pi$.  Regions outside the condensate are indicated in black.  Figure reproduced from \citet{sadl06symm}.} \label{fig:sadlerquench}
\end{figure}

The evolution of a quenched spinor Bose-Einstein condensate was studied experimentally by \citet{sadl06symm}.  A large $F=1$ $^{87}$Rb spinor condensate, spatially extended in two dimensions, was prepared in the $|m_z = 0\rangle$ initial state at a high quadratic Zeeman shift. Following a rapid reduction of the quadratic Zeeman shift, which initiated the spin-mixing instability, the gas evolved for a variable time before it was probed by dispersive magnetization-sensitive imaging.  The gas showed little magnetization for tens of milliseconds, after which an inhomogeneous magnetization landscape spontaneously emerged (Fig.\ \ref{fig:sadlerquench}).

For the sudden quench studied in this experiment, the spatial spin correlations of the inhomogeneously magnetized condensate are determined by, and are, therefore, reflective of, the spin-mixing amplification spectrum \cite{lama07quench}.  Considering the simplest case of a homogeneous initial state, at early times after the onset of the spin-mixing instability, the evolution of the initial spin fluctuations at each wavevector can be considered independently according to Eq.\ (\ref{eq:zandp}), where the temporal evolution is defined through the magnon dispersion relation $E^2(k,q) = (\epsilon_k + q)(\epsilon_k + q + 2 c^{(1)}_1 n)$.  For a quench to $q \simeq 0$, this highest temporal gain occurs at the wavevector $k_{\rm max} = 1/\xi_\mathrm{sp}$, leading to domains with dominant size $\pi \xi_\mathrm{sp}$ where $\xi_\mathrm{sp} = \hbar / \sqrt{2 m c^{(1)}_1 n}$ is the spin healing length for this system, and with time constant of $\tau = \hbar / (c^{(1)}_1 n)$.  These predictions were in rough agreement with experiment.  However, details of the spatial spin correlation function were unexplained by the purely-quantum, homogeneous theory of \citet{lama07quench}.   Refined theories were constructed that account explicitly for the anisotropy of the trapped condensate by defining the proper magnon modes for that case \cite{bara08anisotropic}.  Numerical simulations obtained agreement with measurements by varying the spatial Fourier spectrum of the initial noise \cite{sait07quench}, perhaps indicating technical imperfections in the initial state preparation.

Following this work, several theoretical papers considered the realization of the Kibble-Zurek mechanism in spin-ordering transitions of the ferromagnetic $F=1$ spinor Bose-Einstein condensate.  \citet{lama07quench} distinguished between deep quenches ($q < |c^{(1)}_1| n)$ in which spin domains will emerge at a length scale determined by the depth of the quench, and shallow quenches ($|c^{(1)}_1 n | < q < 2 |c^{(1)}_1| n$) for which spin correlations will show ``light-cone'' dynamics and expand linearly in time.  The latter behavior is reminiscent of the Lieb-Robinson expansion of quasiparticle correlations and entanglement predicted for quantum quenches that do not necessarily cross phase transitions \cite{cala06,brav06,cram08}.  Gradual linear quenches across the phase transition were considered, essentially by considering evolution under the magnon dispersion $E^2(k, q(t))$ with $q(t)$ varying linearly in time, and scaling laws for the characteristic domain length and spin-vortex density were obtained \cite{lama07quench,sait07kz,uhlm07quench,dams07quench}.  \citet{sait07kz} noted that the spin-mixing instability has the special feature of local spin conservation; while this conservation requires that regions of spin anti-correlation exist at short range (as observed by \citet{sadl06symm}), the standard Kibble-Zurek mechanism still applies over longer length scales.

\subsection{Parametric spin amplification}
\label{sec:spinsqueezing}

In describing the experiments discussed above as quantum quenches, what is meant specifically by the word ``quantum?''  Subsequent work by \citet{lesl09amp} explored this question by testing whether the macroscopic magnetization patterns observed after the quench could indeed be ascribed to the quantum-limited parametric amplification of quantum noise.  In the spatially extended systems which they studied, the spin-mixing instability is represented as a spatially broadband parametric amplifier, i.e.\ one with a wavelength dependent amplification, the spatial spectrum of which is defined by the magnon dispersion relation $E^2(k,q)$.  That amplification spectrum is detected experimentally by quenching to different values of $q$, feeding spatially white noise at the input of the amplifier and detecting the post-amplification output via the spatial Fourier spectrum of the magnetization.  The amplification spectrum was indeed observed to vary with $q$, with shallow quenches amplifying longer wavelength features, while deeper quenches produced small length scale spin domains \cite{lesl09amp,sau09}.  However, the experiments did not verify the quantum nature of the seeding and amplification due to uncertainties of the spin-dependent interaction energies for $^{87}$Rb \cite{klau01rbspin,vankemp02,chan05nphys,wide06precision}.

\begin{figure}[tb]
\begin{center}
\includegraphics[width=0.9\columnwidth]{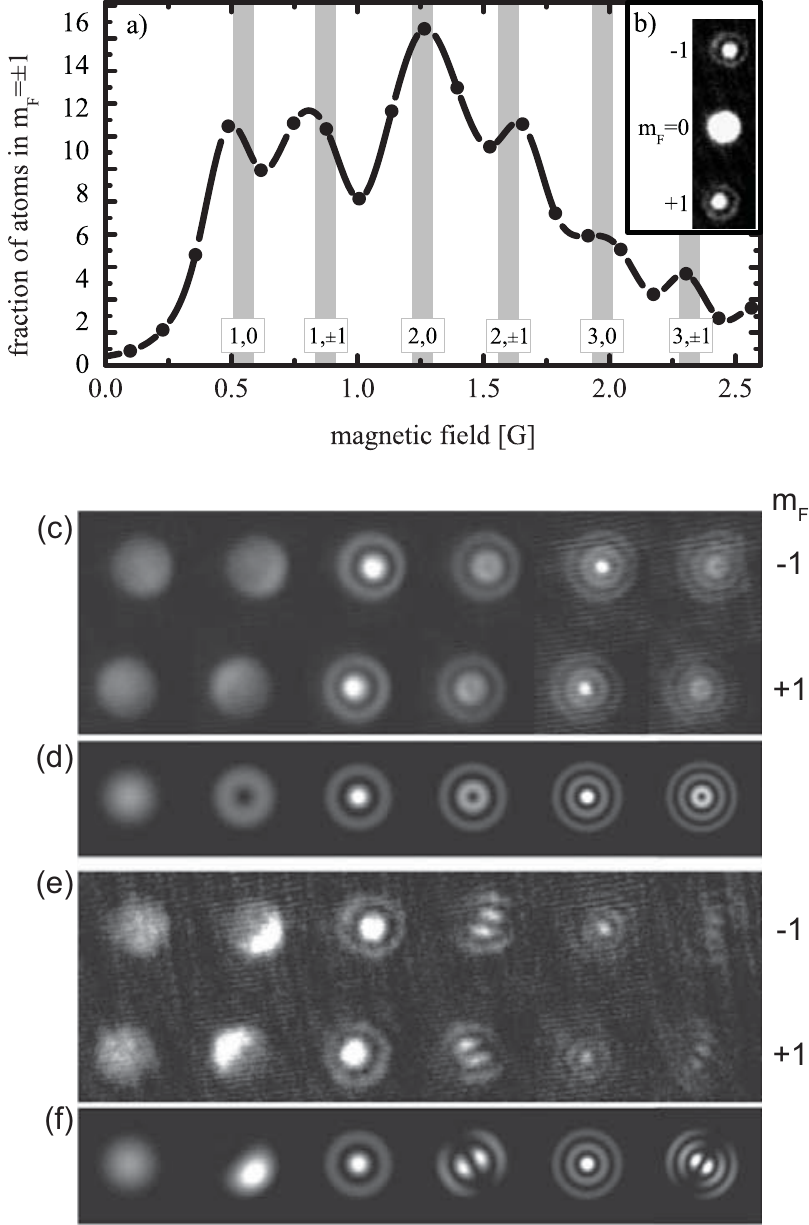}
\end{center}
\caption{For a spinor condensate trapped in the single spatial-mode regime, the spin mixing instability can be tuned to discrete spin excitation modes by varying the quadratic Zeeman shift.  (a) The fraction of atoms transferred into the $|m_z = \pm 1\rangle$ sublevels from the initial $|m_z = 0\rangle$ state, within 18.5 ms of the onset of dynamical instability (determined from images such as (b)), shows a multi-resonant structure; here, experiments are performed with $F=2$ $^{87}$Rb atoms for which $q$ is negative for dc magnetic fields.  The labels $(n,l)$ give the radial and angular quantum numbers of transverse spatial output modes.  Resonances occur when $q$ matches the output mode energy (theoretical predictions shown by gray bars).  The spatial  structure of the output modes matches the predicted patterns for averages over many realizations (c:experiment, d:theory), and for single realizations (e:experiment, f:theory). The single realizations show spontaneous breaking of both spin and spatial symmetry as particular superpositions of degenerate output modes are selected at random.  Figures reproduced from \citet{sche10}.} \label{fig:klempt}
\end{figure}

Stronger evidence for fully quantum mechanical evolution has been obtained in single-spatial-mode experiments, performed with spinor Bose gases in tightly confining optical traps.  The Hannover group studied the spin-mixing instability of the $|m_z = 0\rangle$ state for the $F=2$ $^{87}$Rb spinor condensate, in a situation where discrete spin-excitation modes were separated energetically by gaps, on the order of the transverse trapping energy, that were made larger than the spin-mixing energy.  By tuning the quadratic Zeeman shift to match the excitation energies $\epsilon$, the spin-mixing instability led to the amplification of discrete spin-excitation modes for which the amplification gain was largest.  Comparing the Zeeman populations generated after a fixed duration of spin mixing for the various magnon modes, they concluded that higher-order spatial modes were less contaminated by technical noise, such as that due to imperfect purification of the initial spin state, potentially to the level where the initial fluctuations of those higher-order modes was due to vacuum fluctuations alone \cite{klem09multi,klem10vacuum}.  In their cylindrically symmetric trap, spin-excitation modes can be labeled by radial and angular quantum numbers, with the states $(n, \pm l)$ being degenerate.  Remarkably, they observed the amplification of spin fluctuations selected a coherent superposition of such states which varied between runs of the experiment (Fig.\ \ref{fig:klempt}), representing a spontaneous breaking of spin and spatial rotational symmetries \cite{sche10}.

\subsection{Quantum spin-nematicity squeezing}

Distinctly quantum-mechanical evolution was observed in recent experiments on spin squeezing generated by the spin mixing instability.  The concept of spin squeezing has been derived most extensively with reference to the collective spin state of an ensemble of spin-1/2 particles.  In the absence of interparticle correlations, such an ensemble is described by the single-body density matrix and, thereby, defined by its net vector spin $\mathbf{F}_{\rm tot} = \langle \mathbf{\hat{F}}_{\rm tot}\rangle$ where $\mathbf{\hat{F}}_{\rm tot}$ is the symmetric sum of vector spin operators over the ensemble's particles.  Allowing the orientation of $\mathbf{F}_{\rm tot}$ to define the $\mathbf{z}$ axis, the uncertainty relation $\langle (\Delta F_x)^2\rangle \langle (\Delta F_y)^2 \rangle \geq \langle F_z \rangle^2/4$ is satisfied by partitioning the uncertainties equally between the two transverse spin components.  Correlations among the particles allow one to apportion the uncertainty unevenly, allowing the measurement uncertainty of one transverse spin component to be lower than the standard quantum limit $\langle (\Delta F_\perp)^2 \rangle_{SQL} = |\langle F_z \rangle|/2$ \cite{wine92squeeze,kita93}.  Such squeezed states can provide sub-quantum-limited performance for atom-based sensors such as atomic clocks \cite{appe09,schl10states}, magnetometers \cite{kosc10}, and interferometers \cite{gros10nonlinear}.

Spin squeezed states can be generated naturally by the evolution of an ensemble under a system Hamiltonian that contains terms quadratic in the spin operators \cite{kita93}.  Such terms are generated by binary collisional interactions, e.g.\ in Bose-Einstein condensates, as pointed out in \citet{sore01ent} and recently realized for a pseudo-spin-1/2 condensate in a state-dependent potential \cite{ried10entanglement}.  Naturally, researchers suggested that spin squeezing would be generated by the spin-dependent interactions in spinor Bose-Einstein condensates \cite{duan02entanglespinor}.  However, the spin-1/2 theory of \emph{vector} spin squeezing is clearly inadequate for describing squeezing in higher-spin spinor Bose gases, since the single-particle density matrix is characterized by more observables than just the vector spin.

Several frameworks have been suggested for describing spin squeezing in this context.  \citet{duan02entanglespinor} suggest that the definition of spin squeezing for spin-1/2 systems can be adapted to the higher-spin case by selecting any two orthogonal single-particle states, or, equivalently, any two orthogonal bosonic modes based on those single-particle states, and denoting them as pseudo-spin +1/2 and -1/2 states, respectively.

A similar approach is taken by \citet{sau10njp}, who point out that for any coherent spin state of a spin-$F$ ensemble, in which all particles occupy the single-particle spinor $|\psi\rangle$, one can select $2 F$ spinors $|\xi_\alpha\rangle$ that form an orthonormal set with $|\psi \rangle$.  Based on these, one defines spin fluctuation observables $M^{(1)}_\alpha = \sum_j \left(|\xi_\alpha\rangle \langle \psi | + |\psi\rangle \langle \xi_\alpha | \right)_j$ and $M^{(2)}_\alpha = \sum_j i \left(|\xi_\alpha\rangle \langle \psi | - |\psi\rangle \langle \xi_\alpha | \right)_j$ where the sum is taken over all particles in the ensemble.  These observables, each of which has an expectation value of zero and variance equal to the standard quantum limit for the coherent spin state $|\psi\rangle$, represent $2 F$ independent polarizations of spin fluctuations, and play the role of transverse-spin operators for independent pseudo-spin-1/2 systems.

For a spinor Bose condensate prepared initially in the $|\psi\rangle = |m_z=0\rangle$ spin state, choosing the orthogonal states $\{|\xi_\alpha\rangle\} = \{|\psi_x\rangle, |\psi_y\rangle\}$ from the polar-state basis, one finds the two polarizations of spin-fluctuation observables to be $\left(M^{(1)}_x = -N_{xz}, M^{(2)}_x = F_y \right)$ and $\left(M^{(1)}_y = -N_{yz}, M^{(2)}_y = -F_x \right)$.  That is, in contrast to the vector spin squeezing familiar for spin-1/2 particles, here, spin squeezing implies the reapportionment of fluctuations between conjugate spin-vector ($F_x$ and $F_y$) and spin-quadrupole components ($N_{yz}$ and $N_{xz}$) \cite{lesl09amp,sau10njp}.  More recently, a rigorous set of criteria for squeezing and entanglement in high-spin gases was provided by \citet{vita11arb}.

\begin{figure*}[tb]
\begin{center}
\includegraphics[width=6 in]{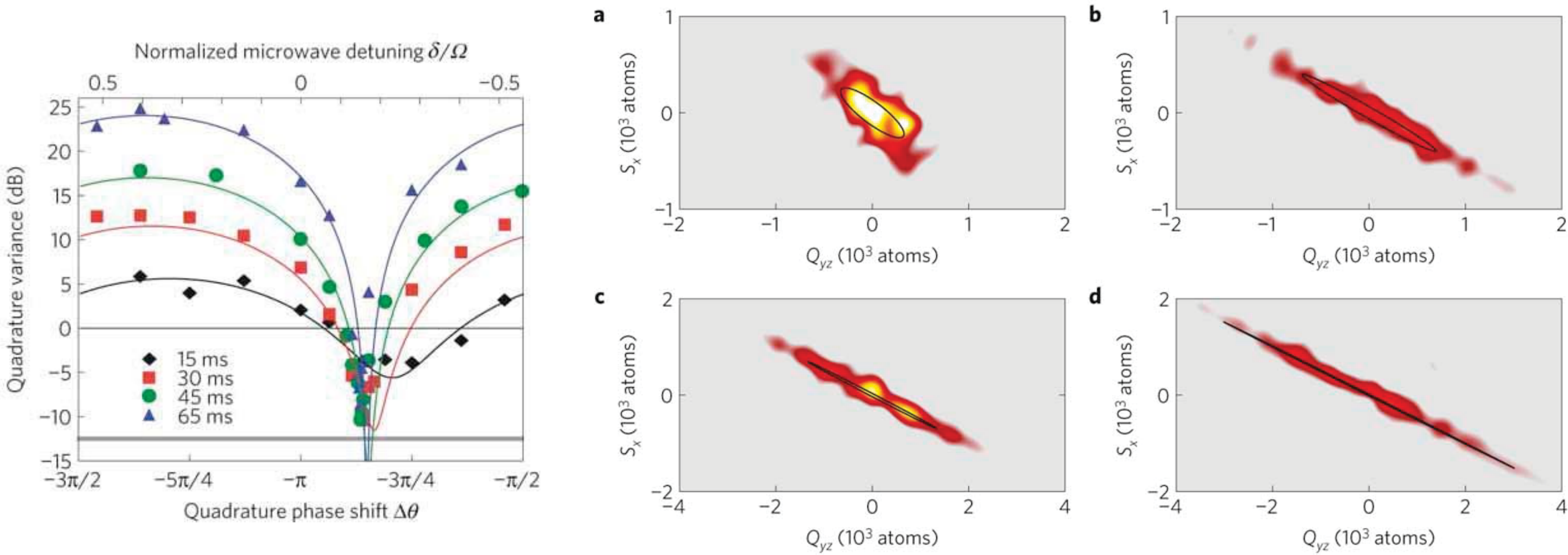}
\end{center}
\caption{(Color online) Observation of spin-nematicity squeezing.  An $F=1$ $^{87}$Rb condensate in the single-mode regime is prepared in the $|m_z=0\rangle$ initial state and allowed to evolve under a spin-mixing dynamical instability.  The gas is then probed by first using a brief pulse with variable $q$ (controlled by a microwave drive with detuning shown on the top axis) to rotate the spin state in the transverse spin-nematicity plane (by the quadrature phase shift $\Delta \theta$), and then using a $\pi/2$ spin rotation and Stern-Gerlach analysis to measure one component of the transverse spin.  (Left) Choosing $\Delta \theta$ to measure the quadrature that is squeezed by the dynamical instability, the measurement variance is found to be below the atomic shot-noise level by as much as 8.6 dB.  (Right) From measurements at all quadrature angles one reconstructs the Wigner function in the spin-nematicity phase space produced following (a) 15, (b) 30, (c) 45, and (d) 65 ms of parametric amplification by the spin-mixing instability.  The squeezed and anti-squeezed quadratures are clearly observed.  Black ellipses indicate the $1/\sqrt{e}$ measurement width calculated from theory.  Figures reproduced from \citet{haml12squeezing}.}
\label{fig:chapmansqueezing}
\end{figure*}

Such spin-nematicity squeezing has been recently observed (Fig.\ \ref{fig:chapmansqueezing}).  An $|m_z=0\rangle$ condensate prepared in a tight optical trap was evolved under the spin mixing instability for several spin-mixing times.  To observe the squeezed quadrature, the gas was first exposed to a pulsed quadratic Zeeman shift, which rotates spin fluctuations in the spin-nematicity phase space, and then the transverse spin was measured by applying a $\pi/2$ spin rotation and then measuring $F_z$ via Stern-Gerlach analysis.  Significant squeezing, at the level of around 8 dB below the standard quantum limit, was observed when probing the squeezed quadrature \cite{gros11spinor}.

Similar experiments were conducted by \citet{gros11spinor}, working with a sample of several hundred condensed atoms in the $F=2$ manifold of $^{87}$Rb.  They interpret the measurement of coherences between the initial $|m_z = 0\rangle$ state and the $|m_z = \pm 1\rangle$ states produced by the dynamical instability as a form of homodyne measurement akin to those performed in quantum optics.  However, sub-shot-noise squeezing was not detected in this work due to technical limitations.

Spin-nematicity squeezing records the coherence generated between the particles emitted by spin mixing (in the $|m_z \pm 1\rangle$ states) and the condensate from which they are produced.  Alternately, one may focus on the correlations generated between the emitted particles themselves.  By spin conservation, the populations generated by spin mixing in the $|m_z = \pm 1\rangle$ states should be strictly identical; indeed, experimental measurements of such populations have shown sub-Poissonian, resolution-limited differences in their populations \cite{chan04,book11correlations,luck11squeeze}.  However, such number correlations do not, on their own, imply any quantum correlations or entanglement.  Rather, to detect such correlations, one probes whether the particles emitted by spin mixing inhabit two (one for each spin state) common bosonic modes.  Under such conditions, the $|m_z = \pm 1\rangle$ atomic populations are represented as a superposition of twin Fock states \cite{pu00,duan00squeeze,mias08}, $|N_+, N_-\rangle$, with $N_+ = N_-$, similar to the photonic states produced by parametric down-conversion.

One feature of this twin Fock state is that the variance in the population difference between the two modes is initially minimal, but grows rapidly if one first applies a coherent rotation by an angle $\phi$ between the two modes (similar to the action of an optical beam splitter on light).  This feature was observed in recent experiments by the Hannover group, who point out that the rapid increase in measurement variance gives a quantum-enhanced sensitivity to the rotation angle $\phi$ \cite{luck11squeeze}.  An additional feature, pointed out by \citet{mias08}, is that, in the twin-Fock-state superposition, the populations $N_+ = N_-$ should show thermal-like fluctuations even though the emitted particles inhabit a zero-entropy quantum state.  Such fluctuations, analogous to Hawking radiation, have not yet been studied experimentally.




%% file: rmp_ch_equilibration.tex
\section{Coarsening and equilibration}
\label{sec:equilibration}

The equilibration of a quenched system, such as a ferromagnet quenched suddenly below its Curie point or a binary alloy undergoing phase separation, consists generally of two stages of evolution: an initial stage dominated by the fastest timescales for establishing locally ordered regions, and an ensuing coarsening stage during which the order parameter becomes ordered on length and time scales that grow ever longer.  In degenerate spinor Bose gases, the former evolution is described well in terms of the coherent quantum dynamics of collective spin mixing following a quantum quench.  The nature of the latter stage of evolution, the long-time relaxation of spin-mixing oscillations and the phase-ordering kinetics of spin textures, remains largely unexplored.  Here, we describe several aspects of relaxation and equilibration.

\subsection{Relaxation of spin-mixing oscillations}

Spinor Bose-Einstein condensates prepared in non-equilibrium spin states undergo coherent spin-mixing dynamics, marked by the deterministic initial evolution both of the mean-field spinor wavefunction and its many-body quantum fluctuations.  At longer times, however, the signatures of this coherent evolution are lost.  This behavior is exhibited most clearly in two experimental reports.  Studying spin-mixing dynamics of an $F=1$ $^{87}$Rb condensate, the Chapman group observed coherent oscillations of the Zeeman populations that were damped out after a few oscillations \cite{chan05nphys}.  The observed damping coincided with the appearance of spatial structure along the long axis of their prolate gas samples.  In those experiments, the comparison between the condensate radii of $(3.2, 3.6, 36)\, \mu$m and the expected size of spin domains $\pi \xi_\mathrm{sp} = 9\, \mu$m supports the notion that the observed damping can be explained as resulting from inhomogeneous broadening of the spin-mixing frequency along the long axis of the gas \cite{murp06finiteT}.

A second experiment, by the NIST group, probed the spin-mixing oscillations of $F=1$ $^{23}$Na condensates \cite{liu09fluctuations}.  For that work, the condensate was prepared with spatial dimensions (radii ranging from 6 to 8 $\mu$m) roughly equal to the expected size of spin domains ($\pi \xi_\mathrm{sp} = 8 \, \mu$m), so that the single-mode approximation should apply.  Experiments were conducted with a small quadratic Zeeman shift, for which the mean-field spin-mixing dynamics divided between the closed-orbit trajectories (in $\rho_0$-$\theta$ space; see Sec.\ \ref{sec:meanfieldmixing}) of the higher-energy spin states, and the running trajectories of the lower-energy spin states.  These two dynamical regions are divided by a separatrix marked by large, slow variations in the $|m_z=0\rangle$ Zeeman population \cite{blac07naspinor}.

The spinor gases were prepared initially in the high-energy, transverse ferromagnetic state.  The ensuing spin-mixing dynamics, monitored by observing the Zeeman populations, indicated that the spin system underwent ``orbital decay,'' evolving from high- to low-energy spin-mixing trajectories.  The authors modeled this decay phenomenologically by adding a damping term to the mean-field dynamical equations.  However, the true nature of this damping remains unclear.  \citet{endo11kineticjltp} have provided a framework for describing dissipation of spin mixing dynamics through damped collisional dynamics of the non-condensate fraction.  However, a quantitative prediction has not yet been derived from this framework to test its applicability to the experimental situation.

\subsection{Elements of coarsening dynamics}

Far outside the single-mode regime, the equilibration dynamics of a spinor gas that is quenched across either a quantum or thermal phase transition may be explained using the theories of phase-ordering kinetics and dynamical critical phenomena that have been developed to describe a wide variety of systems \cite{hohe77,bray02}.  Elements of such theories include the hypothesis that domain structures and correlation functions at different times in the equilibration process differ only in the overall length scale, and that this length scale grows in time by the power law $l(t)\propto t^{1/z}$, with $z$ being a dynamical critical exponent.  Such scaling has been examined numerically, and, to a limited extent, experimentally in several models of relaxation dynamics that differ in the symmetry of the order parameter and in the presence of conserved quantities.  Scaling dynamics can often be related to the behavior of topological structures, such as vortices and domain walls.  In this sense, studies of topological structures in spinor Bose gases should shed light on their coarsening dynamics.

\begin{figure*}[tb]
\begin{center}
\includegraphics[width=5 in]{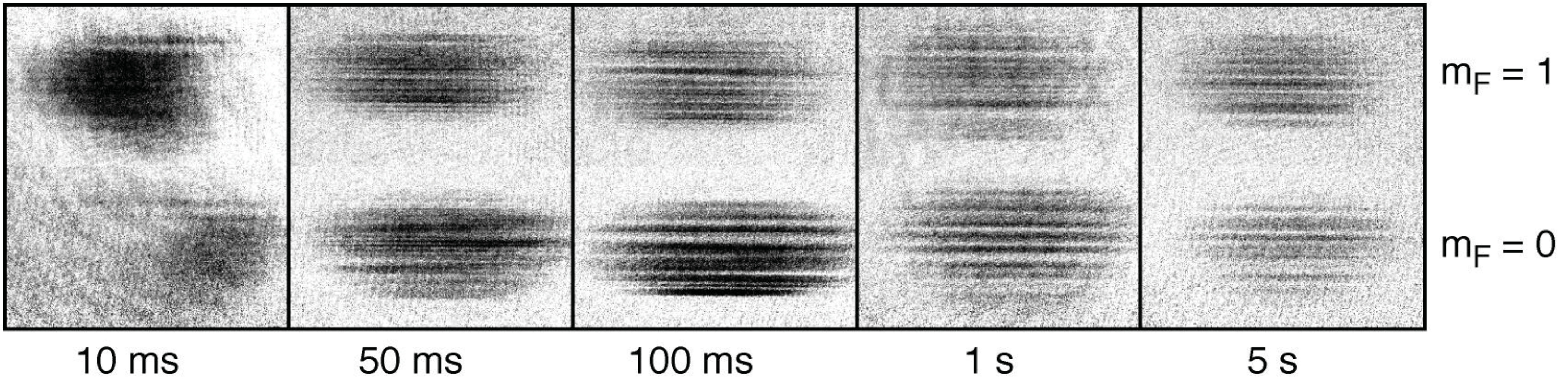}
\end{center}
\caption{A two component $^{23}$Na spinor Bose-Einstein condensate is quenched by preparing it in a uniform superposition of the $|F=1, m_z = 0, +1\rangle$ Zeeman sublevels in a narrow, quasi-one-dimensional optical trap, and at a high quadratic Zeeman shift for which spin mixing into the unoccupied $|m_z = -1\rangle$ state is strongly  suppressed.  Stern-Gerlach imaging is used to determine the spin-state distribution at a variable time after the state preparation.  The gas undergoes phase separation, initially in the narrow radial direction (10 ms) but soon settling into a distribution of pure-spin domains along the long axis of the gas (vertical in image), with domain lengths of 40 $\mu$m or less.  At the low temperatures of this experiment (no non-condensed gas is discernible) and without a magnetic-field gradient to induce directed diffusion or quantum tunneling, no significant coarsening occurs over the few-second lifetime of the trapped gas.  Figure reproduced from \citet{mies99meta}. }
\label{fig:miesner}
\end{figure*}

The mechanisms of equilibration in spinor Bose gases were explored in  experiments on the appearance of sharp domain walls in $F=1$ spinor condensates of sodium, and on their role in impeding phase ordering of the gas \cite{mies99meta,stam99tunprl}.  Rf pulses were used to prepare a condensate in a superposition of the $|m_z=0\rangle$ and $|m_z=1\rangle$ Zeeman sublevels, at a high value of the quadratic Zeeman shift where spin mixing into the $|m_z=-1\rangle$ state is energetically suppressed so that the number of atoms in each Zeeman state is conserved.  In accordance with the phase diagram of Fig.\ \ref{fig:spin1phasediagrams}, with the relation $a_{01} > \sqrt{a_{00} \, a_{11}}$ between s-wave scattering lengths in the case of antiferromagnetic $F=1$ interactions, the initially overlapping states underwent rapid phase separation (Fig.\ \ref{fig:miesner}).  Such dynamics can be described as due to a dynamical instability that is most dominant at the wavevector equal to the inverse spin healing length \cite{gold97,timm98phas,grah98binary,isos99spin}.  This experiment represents another ``quantum quench,'' in which a system is prepared in the non-equilibrium admixed state under conditions that favor the breaking of an Ising symmetry by the process of phase separation.  The observed dynamics resemble the classical spinodal decomposition of binary alloys \cite{ao98,ao00}.

The condensates in these experiments were extremely prolate, so as to be effectively one-dimensional with respect to spin dynamics.  Accordingly, the textures formed after the quench showed only axial variations in spin composition, comprised of single-component domains of length $\sim 40 \, \mu$m separated by sharp domain walls.

In the case of a conserved order parameter, classical phase ordering proceeds by diffusion, which drives particles from the surface-tension-induced high chemical potential of small domains to the lower chemical potential of larger domains \cite{bray02}.  In the phase-separated, quantum-degenerate Bose gas, such diffusion describes the evaporation of condensed atoms from one portion of the gas, and the recondensation of the normal fluid into another portion of the gas (where the chemical potential is lower).  The quantum-degenerate gas also features a second means of transport: quantum tunneling of the superfluid across the potential barriers erected by domains and domain walls \cite{stam99tunprl}.

Both these modes of coarsening were observed by preparing gases initially in the simpler starting condition of having just one large domain of each component.  A magnetic field gradient was then applied, which favored the spatially reversed rearrangement of spin domains along the long-axis of the trap.  In this case, the chemical potential difference that drives particle motion is controlled by the strength of the field gradient, rather than by surface tension.

The quantum tunneling current across the spin domain wall is regulated by the height of the potential barrier, which is proportional to the spin-dependent mean-field energy (and hence to the condensate density), and by its width, which is inversely proportional to the magnetic field gradient.  At high condensate density and weak magnetic field gradients, particle transport was dominated by the diffusion and recondensation of the thermal gas.  The particle current due to this recondensation was slow and was relatively insensitive to the field gradient.  At lower condensate densities and stronger field gradients, transport was dominated by the direct quantum tunneling of the superfluid.  The rapid (exponential) variation of the tunneling rate led to a rapid rearrangement of the gas once the gradually decaying condensate density fell below a threshold value.

Upon allowing for spin-mixing among all substates of the spinor gas, one expects transport and coarsening mechanisms to differ from those studied in the aforementioned experiments.  For example, in the $F=1$ spinor gas at low quadratic Zeeman shift, the $|m_z=0\rangle$ population is no longer a locally conserved quantity.  Thus, domains of $|m_z=0\rangle$ atoms can be transformed by spin mixing into the $|m_z = \pm 1\rangle$ states.  This transformation is seen in quench experiments with both ferromagnetic  (\citet{sadl06symm}; see Sec.\ \ref{sec:quantumquench}) and antiferromagnetic (\citet{book11}; see Fig.\ \ref{fig:sgtof}) $F=1$ condensates when  the $|m_z=0\rangle$ gas is made locally unstable.  The MIT group studied the fate of $|m_z=0\rangle$ domains under conditions where the domains are locally stable (due to a positive quadratic Zeeman shift), but globally energetically unstable (due to a magnetic field gradient) against spin mixing \cite{mies99meta}.  The $|m_z=0\rangle$ domain evaporated via thermally activated spin mixing, with the spin mixing products eventually recondensing into the $|m_z=\pm 1\rangle$ spin domains that flank the $|m_z=0\rangle$ domain on either side.

\subsection{Long-time equilibration dynamics}

While some of the elemental processes in coarsening have been studied, our understanding of the mechanisms and timescales for equilibration remains incomplete and poses vexing problems for future research.  Indeed, recent experiments on relaxation and coarsening in spatially extended $F=1$ $^{87}$Rb condensates \cite{veng10periodic,guzm11} raise the question of whether spinor Bose gases can be reliably equilibrated within experimentally accessible timescales.  These experiments were motivated by observations made on degenerate spinor gases that had undergone dynamical instabilities, brought about either by rapid sweeps of the quadratic Zeeman shift \cite{sadl06symm,lesl09amp} or by the imposition of a helical spin texture \cite{veng08helix}, and allowed to evolve for several times the dynamical instability timescale.  Under a variety of initial conditions, such gases developed long-lasting spatial modulations of the spin texture, comprised typically of $\sim 10\,\mu$m long domains of alternating magnetization.  Following the realization that the MDDI should significantly influence the structure of low-energy spin textures in rubidium spinor condensates, the question arose whether these spatial modulations could be characteristic of the thermal equilibrium state.

This question was explored in two works in which un-magnetized, nondegenerate spinor gases were produced, cooled gradually by evaporative cooling to below the degeneracy temperature, and then held at constant temperature for a variable equilibration time before being probed either by time-of-flight or \emph{in situ} imaging.  The first study \cite{veng10periodic}, which was limited to moderate equilibration times of several 100 ms, explored the textures produced at variable temperature and for compositions of the nondegenerate gas that differed in the fractional population of the $|m_z=0\rangle$ state.  For atoms prepared in an initial $(1/3, 1/3, 1/3)$ population mixture of the $m_z = (+1, 0, -1)$ Zeeman substates, the fine-scale magnetization modulations were observed at all temperatures below the degeneracy temperature.  In contrast, for atoms prepared in the $(1/4, 1/2, 1/4)$ initial state, the modulated spin textures were observed at the lowest temperatures, while at higher temperatures the degenerate gases were nearly uniformly transversely magnetized.

The persistence of the observed spin textures was examined theoretically, primarily by computing either the ground-state or the temporal evolution determined by the zero-temperature Gross-Pitaevskii equation, accounting for both the spin-dependent s-wave interactions and the MDDI.  \citet{cher09roton} found that uniformly magnetized rubidium condensates were subject to a roton-like instability with particular wavevectors and spatial anisotropy, suggesting an instability toward the formation of a spatially periodic ground-state texture.  The possibility that the observed structures were long-lived spin-vortex lattices was suggested by an imaginary-time calculation for the ground-state spin texture \cite{zhan10jltp}, but belied by computations of the real-time dynamics \cite{kawa10spont}.  Equilibrium phases with striped magnetization patterns were predicted to be generated by the competition of the kinetic-energy stiffness of the condensate magnetic order parameter and the MDDI \cite{kjal09}; however, as with the other theory works, the predicted length scale of the ground-state modulation was several times larger than that observed experimentally.

\begin{figure}[tb]
\begin{center}
\includegraphics[width=\columnwidth]{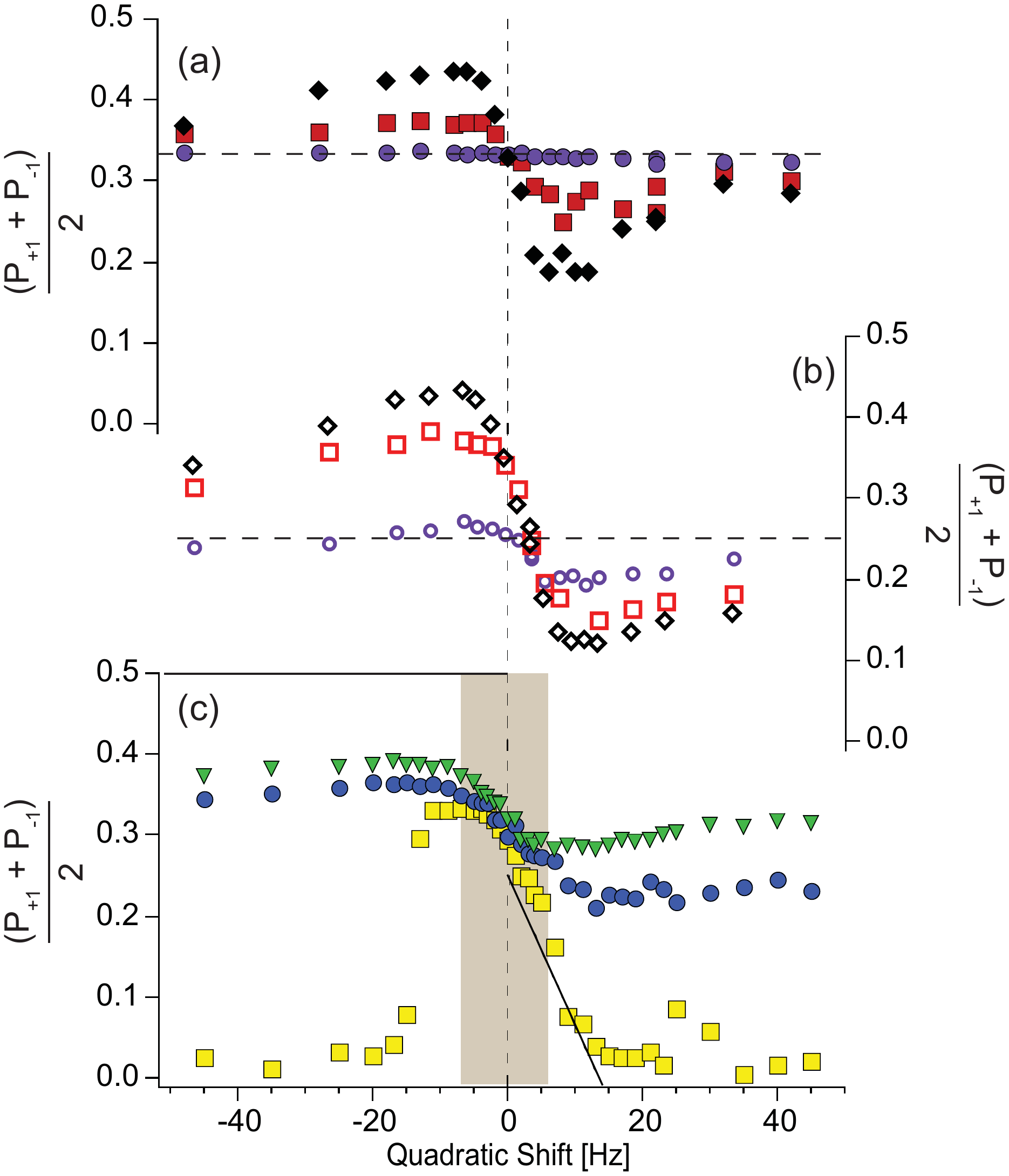}
\end{center}
\caption{(Color) Equilibration of condensate Zeeman populations following the preparation of an unmagnetized thermal gas, and evaporative cooling to below the quantum degeneracy temperature.  The $F=1$ $^{87}$Rb gas is held in a quasi two-dimensional optical potential, well outside the single spatial-mode regime.  The average population in the $|m_z = \pm 1\rangle$ states, $(P_{+1} + P_{-1})/2$, is determined  at various evolution times $t=$ 200 ms (purple circles), 1 s (red squares), and 3 s (black diamonds) for initial thermal spin populations of (a) $(1/3, 1/3, 1/3)$ or (b) $(1/4, 1/2, 1/4)$ in the $m_z = (1, 0, -1)$ states.   In (c), the long-time (2 seconds) Zeeman populations are shown for the initial spin mixtures of figure (a) (green triangles), figure (b) (blue circles), and for the populations $(0,1,0)$ (yellow squares).  For a narrow range $|q|/h\lesssim 10$ Hz (shaded), Zeeman populations evolve toward a common steady state that is qualitatively in accord with mean-field predictions (solid line in c) based on the spin-dependent contact interaction.  Outside that range, initial population differences persist throughout the experimentally accessed equilibration times.  Figure reproduced from \citet{guzm11}.}
\label{fig:slowequil}
\end{figure}

The mismatch between experiment and theory was partly explained by the fact that these finely modulated textures were non-equilibrium states, which coarsen over several seconds to form larger-scale magnetic domains (Fig.\ \ref{fig:equilibriumrb}).  Studying spinor condensates produced at the lowest temperatures, \citet{guzm11} found that the equilibration time for condensates with different initial spin compositions to reach the same final steady-state spin compositions depended strongly on the quadratic Zeeman energy: for small $|q| \lesssim h \times$ 10 Hz, condensates arrived at a common steady-state spin distribution within several seconds, while for larger $|q|$, their equilibration was dramatically slowed (Fig.\ \ref{fig:slowequil}).  The equilibration dynamics at small $|q|$ was marked by four trends:  First, the Zeeman populations were observed to evolve from their initial values to a common final steady state.  Second, the degenerate gas became increasingly magnetized as measured by the mean value of $|\mathbf{F}^2(\mathbf{r})|$.  Third, the spin texture developed a spin-space anisotropy, marked by the difference between the longitudinal ($|\mathbf{F}^2_z(\mathbf{r})|$) and transverse ($(|\mathbf{F}^2_x(\mathbf{r})| + |\mathbf{F}^2_y(\mathbf{r})|)/2$) magnetization where $\mathbf{z}$ denotes the axis defined by the quadratic Zeeman shift.  As predicted by mean-field theory, the textures showed predominantly transverse magnetization (easy-plane) for $q>0$ and longitudinal magnetization (easy-axis) for $q<0$.  Fourth, the size of the commonly magnetized regions coarsened over time from the $l \sim 10\, \mu$m domain length seen at short equilibration times to a final length comparable to the overall size of the trapped samples.  All four trends occurred on a similar time scale of a few seconds, although the mechanisms that determine the time scales for each of these trends are unknown.

An ambitious target for future experiments is to relate these observed trends quantitatively to the predictions for phase-ordering kinetics.  The presence of a Bose-Einstein condensate and the condition of locally conserved magnetization from the rotationally symmetric s-wave interactions (ignoring the MDDI) qualify the $F=1$ spinor Bose gas as obeying the ``Model F'' description of coarsening, as defined by \citet{hohe77}.  This assignment is supported by numerical comparisons between the coarsening dynamics predicted by the stochastic time-dependent Landau-Ginzburg and the energy-conserving Gross-Pitaevskii equation \cite{muke07dyn}.  Such a comparison is valid in the $q<0$ region, where a ferromagnetic $F=1$ condensate should develop longitudinal magnetization with long-range Ising-type order.  The growth of the characteristic domain size $l(t)$ was reported by \citet{guzm11}, but it is doubtful whether a meaningful comparison to the predicted power-law growth ($l(t) \propto t^{1/3}$) can be made over the limited time- and length-scales accessed in that work.

For ferromagnetic condensates evolving in two dimensions with $0 \leq q < 2 |c_1^{(1)} n|$, one does not expect long-range order to develop at thermal equilibrium.  Numerical studies suggest that the dynamics of such gases following a quantum (or thermal) quench may show a long-lived ``prethermalized'' state with well-characterized correlations that differ from those at thermal equilibrium \cite{barn11pretherm}.

%% file: rmp_ch_future.tex
\section{Application of spinor gases to magnetometry}

\label{sec:magnetometry}

Aside from being objects of fundamental interest, spinor Bose gases also lend themselves to the application of magnetic field sensing with high sensitivity and high spatial resolution.   Atomic Bohr frequencies $\omega$ may generally be measured using an atomic ensemble by the following procedure: First, the ensemble is prepared in a common polarized state, e.g.\ through optical pumping or Stern-Gerlach separation.  Next, to initiate the measurement, all atoms are placed in a coherent superposition of internal states, with a  relative phase $\phi$ between the internal states.  During the measurement time $\tau$, $\phi$ advances by an amount $\phi = \omega \tau$.  To conclude the measurement, the phase $\phi$ is estimated, by a single projective measurement or several weak measurements.

Subdividing the ensemble spatially allows one to measure the spatial inhomogeneity of the Bohr frequency, which may come about from an inhomogeneous environmental variable, such as the magnetic field.  For an uncorrelated initial state, the measurement within a single measurement pixel is limited by atomic shot noise to the imprecision $\delta \omega(l) = \left[N(l) D t\right]^{-1/2}$ where $N(l) \propto l^d n$ is the number of atoms within a pixel of length $l$ and dimension $d$, $n$ is the relevant (volume, column or line) density, $D$ is the measurement duty cycle and $t$ the total time allotted to the measurement.

Several features of spinor Bose-Einstein condensates qualify them for use in such spatially resolving sensors.  The Bose-Einstein condensate is a high-density, low-energy medium.  The high density reduces the atomic shot noise as defined above.  The low-energy permits the use of shallow optical traps, which are benign to the atomic spin and from which spontaneous scattering is minimal.  The absence of thermal diffusion implies that atoms remain well localized for long times.  Finally, the rotational symmetry of the s-wave interactions ensures that Larmor precession of a fully magnetized gas under a linear Zeeman shift is unaffected by interactions.  Altogether, these qualities allow for small measurement imprecision and systematic error.

\begin{figure}[tb]
\centering
\includegraphics[width=0.8\columnwidth]{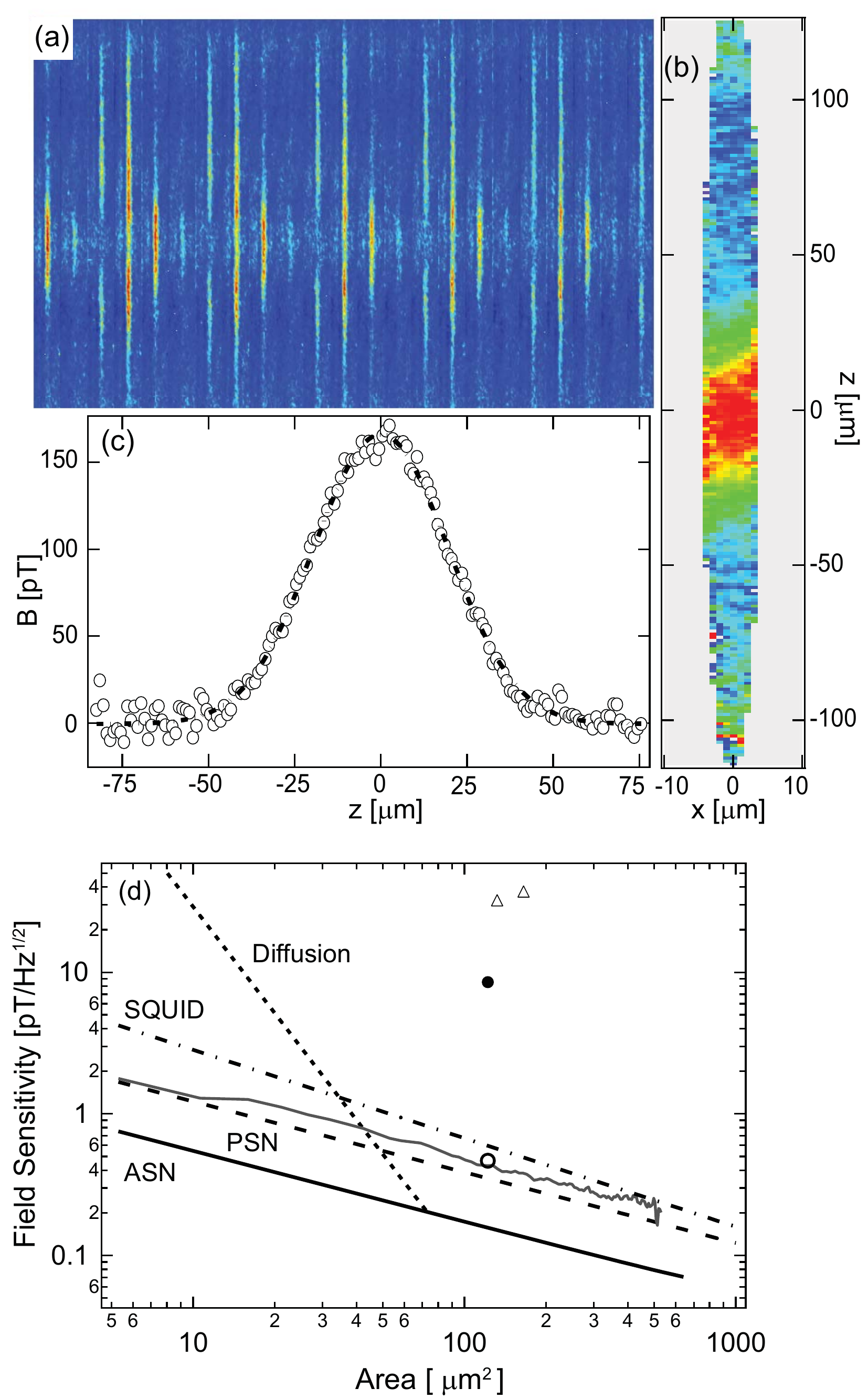}
\caption{(Color) Spatially resolved magnetometry with a spinor Bose-Einstein condensate.  A transversely magnetized condensate undergoes Larmor precession before the precession phase is measured by magnetization sensitive imaging.  (a) Repeated images of one component of the transverse magnetization, from which (b) the Larmor precession phase (indicated by color) and (c) the magnetic field is spatially resolved across the gas. (d) The field sensitivity determined from imaging noise (grey line, assuming unity duty cycle) agrees well with the  photon shot noise limit (PSN) on magnetization  imaging, and remains above the atomic shot-noise limit (ASN).  The sensitivity determined also from measurements of the light-induced magnetic field (shown in (a-c)) is plotted for the actual (filled circle) or for unity duty cycle (open circle).  Also shown are the quantum noise limit for SQUID magnetometers (dot dashed), and the sensitivities achieved in dc SQUID magnetometers (triangles) of  \citet{kirt95mag} and \citet{lee96mag}.  Dotted line indicates the worsening of sensitivity due to atomic motion, an effect that can be avoided by use of an optical lattice.  Figure reproduced from \citet{veng07mag}.}
\label{fig:magnetometer}
\end{figure}

High-resolution magnetometry with two-dimensional spatial resolution was demonstrated using an $F=1$ $^{87}$Rb spinor Bose-Einstein condensate \cite{veng07mag}.  The gas was prepared in a transversely magnetized state and allowed to Larmor precess for $\tau = 250$ ms before the precession phase was read out using magnetization-sensitive phase contrast imaging.  The sensor was tested by measuring the measurement noise both in the absence and the presence of a short length-scale magnetic signal, which was generated via the vector ac Stark shift of a focused, circularly polarized, off-resonant light beam (Fig.\ \ref{fig:magnetometer}).  The Allan deviation, measured for measurement areas $A = l^2$ in the range 6 - 400 $\mu\mbox{m}^2$, was limited by photon shot noise at a level slightly higher than the atomic shot-noise limit.  Even in this proof-of-principle experiment, the achieved measurement sensitivity was comparable to that achieved with low-noise SQUID-based magnetic microscopes at similar spatial resolution\footnote{Hot-atom magnetometers achieve much better field sensitivity \cite{budk07nphys}, but are limited in spatial resolution at the millimeter scale}.

Magnetic field sensing has also been demonstrated using scalar Bose-Einstein condensates, based on the observation that spatial variations of the magnetic field on the order of 1 nT produce variations in the magnetic potential energy, which then lead to visible density variations in the compressible condensed gas \cite{wild05magnature}.  However, the detection of Larmor precession allowed for measurements of far smaller field variations that were in the 1 pT range.  Spatially resolved Larmor precession in a cold, optically trapped, non-condensed gas was recently demonstrated, but was limited to poorer spatial resolution ($l^2 > 1000\, \mu\mbox{m}^2$) and short measurement times ($\tau < 1 \, \mbox{ms}$) \cite{kosc11}.

Several measurement targets for such a cold-atom sensor have been suggested.  The aforementioned scalar condensate magnetometer has been used to image the magnetic field corrugations generated by current flow through a disordered thin conductor \cite{aign08}.  This measurement established that cold atoms can be used as a direct probe of solid-state science.  Cold atomic gases have also been used as a ``scanning-probe microscope'' to visualize three-dimensional micro- and nano-structures (at micron-scale resolution) \cite{hung10,gier11}.  Mechanically vibrating structures that are either ferromagnetic or carry current will produce time-varying magnetic fields that can induce spin flip transitions in nearby spinor gases, allowing for detection of small scale motion and, perhaps, for coherent coupling between the two systems \cite{treu07proposal,kalm11galvo,stei11backaction}.  Alternately, sensing temporal fluctuations in the Larmor precession frequency may provide a means to sense, and, indeed, to image vortices migrating through thin superconducting films \cite{sche07vortex}.

Several extant or foreseeable technological advances may make such spinor-based magnetometry more practical and powerful.  The duty cycle of the sensor, which was just $D = 0.003$ in the work of \citet{veng07mag}, can be increased by shortening the time required to form the spinor condensate (to around 1 s) and by increasing the measurement time;  indeed, the Berkeley group has recently observed coherent Larmor precession in a spinor condensate lasting over 3 s.  The apparatus required for producing large spinor condensates can be shrunk and simplified significantly, e.g.\ as demonstrated by \citet{sali11qip}.  Thin membranes can be employed to separate the spinor gas from its measurement target, similar to methods used to probe room-temperature objects using cryogenic-temperature SQUID microscopes \cite{lee96mag}, so that the measurement target need not be UHV compatible and can be rapidly exchanged.  Additional improvements may stem from utilizing the physical properties of the spinor condensates themselves.  For example, spin-mixing instabilities can be used to amplify weak spatially varying signals \cite{lesl09amp}, allowing higher sensitivity measurement even at low optical-detection efficiency.  Spatially multimode spin-nematic squeezing, generated either by atomic interactions or by measurement, may provide measurement resolution below the atomic shot-noise limits \cite{lesl09amp,sau10njp}.

%% file: rmp_conclusions.tex
\section{Conclusions}
\label{sec:conclusions}

Several concepts regarding spinor Bose gases were emphasized in this Review.  First, we have emphasized the properties of spinor gases that are necessitated by rotational symmetry.  In contrast to generic multi-component quantum fluids, which can be represented formally in terms of a pseudo-spin internal-state wavefunction, spinor gases are comprised of particles whose internal-state wavefunction responds under geometric rotations as a spherical tensor of rank $2 F$.  If it were not for experimental conditions that break symmetry, the Hamiltonian of a self-interacting spinor gas would necessarily be invariant under rotations.

In Sec.\ \ref{sec:rotsymm}, we have discussed how this rotational invariance constrains the interactions among spinor gas atoms.  If we first neglect magnetic dipolar interactions, the ultracold nature of spinor gases indicates the use of partial waves to describe interparticle scattering, with only the lowest partial wave contributing to low-energy collisions. By rotational symmetry, such s-wave binary collisions are distinguished only by the total spin of the colliding pair, not by its orientation.  By Bose symmetry, s-wave collisions occur only when this total spin is even.  Altogether, the interaction Hamiltonian is enormously simplified by these symmetries.

The symmetries of this interaction Hamiltonian are reflected also in its mean-field stationary solutions and its exact many-body ground states.  In Sec.\ \ref{sec:quantumstates}, we presented a scheme to visualize these solutions graphically.  Rotational invariance of the Hamiltonian implies that these stationary states reside in a degenerate manifold of states, and we outlined consequences of this fact in describing some of the unusual textures and topological defects that occur in spinor Bose-Einstein condensates (Sec.\ \ref{sec:topological}) and in treating their low-energy dynamics (Sec.\ \ref{sec:Hydrodynamics}).  Finally, the rotational symmetry of interactions implies that, for certain internal states, the dynamics of interacting spinor Bose gases under applied magnetic fields are the same as those of free particles.  On the one hand, in the absence of dipolar relaxation, this fact allows one to gauge away the effects of uniform magnetic fields and study low-energy dynamics of certain spinor gases even in the presence of substantial and uncontrolled magnetic fields (discussed in Sec.\ \ref{sec:linearzeeman}).  On the other hand, the immunity of Larmor precession to interaction effects makes the spinor Bose gas suitable for magnetic field sensing (Sec.\ \ref{sec:magnetometry}).

A second concept central to this Review is the interplay between magnetic order and superfluidity.  Both phenomena signify phase coherence among atoms in the fluid: magnetic order stems from the phases controlling internal-state coherences, while superfluidity, at least in three dimensions, stems from coherence of the global phase of the gas-particle wavefunctions.  The long-range phase coherence of a superfluid implies that a spinor Bose gas display also magnetic order, though the converse implication does not necessarily hold.  The question of whether Bose-Einstein condensation and magnetic ordering must occur simultaneously was taken up in Sec.\ \ref{sec:magneticorder} where we considered three scenarios in which long-range coherence and the influence of Bose statistics are weakened, namely non-zero temperature, restriction of the spinor gas to lower dimensions, and the effects of optical lattice potentials.

The connection between superfluid and magnetic order was discussed also in the context of spin textures in Sec.\ \ref{sec:textures}.  We have highlighted the relation between superflow and the spin-rotation symmetry of the ground-state manifolds of a spinor Bose-Einstein condensate (Sec.\ \ref{sec:superflow}), a relation familiar from studies of superfluid $^3$He that is manifest in the properties of the varied textures and defects predicted for spinor condensates.  This relation also arose in the description of low-energy hydrodynamics of spinor Bose-condensates, as discussed in Sec.\ \ref{sec:Hydrodynamics}.

A third focus of this Review was non-equilibrium quantum dynamics.  Quantum degenerate atomic gases are a new class of materials, and many of the scientific concepts explored in research on such gases are similar to those explored with condensed matter.  On the one hand, the short lifetime of the gas, its isolation from a thermal environment, and the presence of conserved quantities make it challenging to associate the observed behavior of the gas with a state of thermal equilibrium.  On the other hand, these same properties make the gaseous systems particularly appealing for studies of non-equilibrium dynamical effects that would be hard to simulate in denser solids and liquids.  Focusing on non-equilibrium spin dynamics is particularly appealing because the spin state of the gas can be initialized with very high fidelity, at noise levels much smaller than might be expected at thermal equilibrium given the nK-scale kinetic temperatures of the gas.

In Sec.\ \ref{sec:spinmixing} we discussed the non-equilibrium phenomena of coherent spin-mixing oscillations and instabilities.  We highlighted the connection between few-body, many-body, and mean-field treatments of such dynamics, illustrated by experimental examples.  In discussing the rich area of spin-mixing instabilities, we drew connections to four-wave mixing of optical and matter waves, the dynamics of systems near quantum phase transitions, symmetry-breaking in closed quantum systems, and the dynamical generation of correlated quantum spin states of practical use for metrology.

With these far-reaching developments, the authors of this review consider the field of spinor Bose gases still in its infancy. We conclude this review by enumerating some outstanding issues. One important goal is to make quantitative test of mean-field and Bogoliubov theories on which so many theoretical studies are based. This can be done most straightforwardly by measuring the collective modes. An equally pressing issue is to understand why non-equilibrium states persist so long in spinor gases. This persistence marks a sharp contrast with scalar gases which relax to their equilibrium states quickly, and also raises the question of whether it is experimentally feasible to explore equilibrium properties of spinor Bose gases, and, if so, under what circumstances.  The question of equilibration also hinges on the properties of spinor Bose gases at non-zero temperature, which remain poorly understood and unexplored.

Another inviting research target is the characterization of correlated spin states produced either dynamically or at equilibrium in spinor Bose-Einstein condensates.  Recent studies on spin-nematicity squeezing \cite{luck11squeeze,haml12squeezing} have opened this line of investigation. The beautiful prediction of a fragmented ground state for antiferromagnetic spinor condensates by \citet{law98spin2} may be confirmed by many of the same techniques used in recent experiments.

Topological aspects are also worthwhile to investigate because spinor condensates are among the rare systems for which the needed manipulations to prepare topological objects are experimentally available.  The evolution of topological structures in spin textures, and their role in determining the character of a spinor Bose gas in restricted dimensions, will offer important examples and fresh perspectives on the dynamics of quantum fields.  New topological structures may also be found at interfaces between different phases of the spinor Bose gas; for example, it would be interesting to ask whether such interfaces could accommodate topological defects such as a boojum, which is a point singularity on a surface in superfluid helium-3~\cite{bhat77}.

The topic of spinor gases in optical lattices is especially rich.  Such lattices add new twists to the physics of spinor gases because the dimensions, filling fraction, and geometry of the lattices can introduce strong correlations, phase transitions, and spin frustration in the same system.  Together with newly developed techniques for synthesizing gauge fields and for generating more complex, spin-dependent, lattice potentials, such lattice-trapped spinor gases offer a new paradigm for exploring quantum phenomena and simulations.